\documentclass{emulateapj}

\def\comm#1           {{\tt (COMMENT: #1)}}
\def\kms   {\hbox{km s$^{-1}$}}

\begin{document}

\title{ The Milky Way Tomography with SDSS: II. Stellar Metallicity}

\author{
\v{Z}eljko Ivezi\'{c}\altaffilmark{1},
Branimir Sesar\altaffilmark{1},
Mario Juri\'{c}\altaffilmark{2},
Nicholas Bond\altaffilmark{3},
Julianne Dalcanton\altaffilmark{1},
Constance M. Rockosi\altaffilmark{4},
Brian Yanny\altaffilmark{5},
Heidi J. Newberg\altaffilmark{6},
Timothy C. Beers\altaffilmark{7},
Carlos Allende Prieto\altaffilmark{8},
Ron Wilhelm\altaffilmark{9},
Young Sun Lee\altaffilmark{7},
Thirupathi Sivarani\altaffilmark{7},
John E. Norris\altaffilmark{10},
Coryn A.L. Bailer-Jones\altaffilmark{11},
Paola Re Fiorentin\altaffilmark{11,12},
David Schlegel\altaffilmark{13},
Alan Uomoto\altaffilmark{14},
Robert H. Lupton\altaffilmark{3},
Gillian R. Knapp\altaffilmark{3},
James E. Gunn\altaffilmark{3},
Kevin R. Covey\altaffilmark{15},
J. Allyn Smith\altaffilmark{16},
Gajus Miknaitis\altaffilmark{5},
Mamoru Doi,\altaffilmark{17},
Masayuki Tanaka\altaffilmark{18},
Masataka Fukugita\altaffilmark{19},
Steve Kent\altaffilmark{5},
Douglas Finkbeiner\altaffilmark{15},
Jeffrey A. Munn\altaffilmark{20},
Jeffrey R. Pier\altaffilmark{20},
Tom Quinn\altaffilmark{1},
Suzanne Hawley\altaffilmark{1},
Scott Anderson\altaffilmark{1},
Furea Kiuchi\altaffilmark{1},
Alex Chen\altaffilmark{1},
James Bushong\altaffilmark{1},
Harkirat Sohi\altaffilmark{1},
Daryl Haggard\altaffilmark{1},
Amy Kimball\altaffilmark{1},
John Barentine\altaffilmark{21},
Howard Brewington\altaffilmark{21},
Mike Harvanek\altaffilmark{21},
Scott Kleinman\altaffilmark{21},
Jurek Krzesinski\altaffilmark{21},
Dan Long\altaffilmark{21},
Atsuko Nitta\altaffilmark{21},
Stephanie Snedden\altaffilmark{21},
Brian Lee\altaffilmark{13},
Hugh Harris\altaffilmark{20},
Jonathan Brinkmann\altaffilmark{21},
Donald P. Schneider\altaffilmark{22},
Donald G. York\altaffilmark{23}
}

\altaffiltext{1}{Department of Astronomy, University of Washington, Box 351580, Seattle, WA 98195
\label{Washington}}
\altaffiltext{2}{Institute for Advanced Study, 1 Einstein Drive, Princeton, NJ 08540
\label{IAS}}
\altaffiltext{3}{Princeton University Observatory, Princeton, NJ 08544
\label{Princeton}}
\altaffiltext{4}{University of California--Santa Cruz, 1156 High St., Santa Cruz, CA 95060
\label{UCSC}}
\altaffiltext{5}{Fermi National Accelerator Laboratory, P.O. Box 500, Batavia, IL 60510
\label{FNAL}}
\altaffiltext{6}{Department of Physics, Applied Physics, and Astronomy,
Rensselaer Polytechnic Institute, 110 8th St., Troy, NY 12180
\label{Rensselaer}}
\altaffiltext{7}{Dept. of Physics \& Astronomy, CSCE: Center for the Study of Cosmic 
Evolution, and JINA: Joint Institute for Nuclear Astrophysics, Michigan State University, 
East Lansing, MI  48824, USA 
\label{JINA}}
\altaffiltext{8}{McDonald Observatory and Department of Astronomy, 
University of Texas, Austin, TX 78712
\label{Texas}}
\altaffiltext{9}{Department of Physics, Texas Tech University, Box 41051, Lubbock, TX 79409
\label{TexasTech}}
\altaffiltext{10}{Research School of Astronomy \& Astrophysics, The Australian National 
University, Cotter Road, Weston, ACT 2611, Australia 
\label{RSAA}}
\altaffiltext{11}{Max Planck Institut f\"{u}r Astronomie, K\"{o}nigstuhl 17,
69117 Heidelberg, Germany
\label{MPAstro}}
\altaffiltext{12}{Department of Physics, University of Ljubljana, Jadranska 19, 
1000 Ljubljana, Slovenia
\label{Ljubljana}}
\altaffiltext{13}{Lawrence Berkeley National Laboratory, One Cyclotron Road, 
MS 50R5032, Berkeley, CA, 94720 
\label{LBNL}}
\altaffiltext{14}{Department of Physics and Astronomy, 
The John Hopkins University, 3701 San Martin Drive, Baltimore, MD 21218
\label{JHU}}
\altaffiltext{15}{Harvard-Smithsonian Center for Astrophysics, 60 Garden Street,
                          Cambridge, MA 02138
\label{Harvard}}
\altaffiltext{16}{Dept. of Physics \& Astronomy, Austin Peay State University, 
Clarksville, TN 37044
\label{AustinPeay}}
\altaffiltext{17}{Institute of Astronomy, University of Tokyo, 2-21-1 Osawa,
Mitaka, Tokyo 181-0015, Japan
\label{UT}}
\altaffiltext{18}{Dept. of Astronomy, Graduate School of Science, University of Tokyo,
Hongo 7-3-1, Bunkyo-ku, Tokyo, 113-0033, Japan
\label{UT2}}
\altaffiltext{19}{
Institute for Cosmic Ray Research, University of Tokyo, Kashiwa, Chiba, Japan
\label{UT3}}
\altaffiltext{20}{U.S. Naval Observatory, Flagstaff Station, P.O. Box 1149, Flagstaff, AZ 86002
\label{USNOFlagstaff}}
\altaffiltext{21}{Apache Point Observatory, 2001 Apache Point Road, P.O. Box 59, 
Sunspot, NM 88349-0059
\label{APO}}
\altaffiltext{22}{Department of Astronomy
and Astrophysics, The Pennsylvania State University, University Park, PA 16802
\label{PennState}}
\altaffiltext{23}{University of Chicago, Astronomy \& Astrophysics Center
and The Enrico Fermi Institute, 5640 S. Ellis Ave., Chicago, IL 60637
\label{Chicago}}

\begin{abstract}
In addition to optical photometry of unprecedented quality, the Sloan 
Digital Sky Survey (SDSS) is producing a massive spectroscopic 
database which already contains over 280,000 stellar spectra. Using 
effective temperature and metallicity derived from SDSS 
spectra for $\sim$60,000 F and G type main sequence stars
($0.2<g-r<0.6$), we develop polynomial models, reminiscent of traditional 
methods based on the $UBV$ photometry, for estimating these parameters from 
the SDSS $u-g$ and $g-r$ colors. These estimators reproduce SDSS spectroscopic 
parameters with a root-mean-square scatter of 100 K for effective 
temperature, and 0.2 dex for metallicity (limited by photometric errors),
which are similar to random and systematic uncertainties in spectroscopic
determinations. We apply this method to a photometric catalog of coadded SDSS observations 
and study the photometric metallicity distribution of $\sim$200,000 F 
and G type stars observed in 300 deg$^2$ of high Galactic latitude sky. 
These deeper ($g<20.5$) and photometrically precise ($\sim$0.01 mag) coadded 
data enable an accurate measurement of the unbiased metallicity distribution for a 
complete volume-limited sample of stars at distances between 500 pc and 8 kpc.
The metallicity distribution can be 
exquisitely modeled using two components with a spatially varying number 
ratio, that correspond to disk and halo. The best-fit number ratio of the 
two components is consistent with that implied by the decomposition of 
stellar counts profiles into exponential disk and power-law halo components
by Juri\'{c} et al. (2008).
The two components also possess the kinematics expected for disk and halo stars. 
The metallicity of the halo component can be modeled as a spatially 
invariant Gaussian distribution with a mean of $[Fe/H]=-1.46$ and a standard 
deviation of $\sim$0.3 dex. The disk metallicity distribution is non-Gaussian,
with a remarkably small scatter (rms$\sim$0.16 dex) and the median smoothly 
decreasing with distance from the plane from $-0.6$ at 
500 pc to $-0.8$ beyond several kpc. Similarly, we find using proper motion 
measurements that a non-Gaussian rotational velocity distribution of disk 
stars shifts by $\sim$50 km/s as the distance from the plane increases from 
500 pc to several kpc. Despite this similarity, the metallicity and rotational 
velocity distributions of disk stars are not correlated (Kendall's 
$\tau=0.017\pm0.018$). This absence of a correlation between metallicity 
and kinematics for disk stars is in a 
conflict with the traditional decomposition in terms of thin and thick disks, 
which predicts a strong correlation ($\tau=-0.30\pm0.04$) at $\sim$1 kpc 
from the mid-plane. Instead, the variation of the metallicity and rotational 
velocity distributions can be modeled using non-Gaussian functions that 
retain their shapes and only shift as the distance from the mid-plane 
increases. We also study the metallicity distribution using a shallower ($g<19.5$) 
but much larger sample of close to three million stars in 8500 sq. deg. of sky
included in SDSS Data Release 6. The large sky coverage enables the detection 
of coherent substructures in the kinematics--metallicity 
space, such as the Monoceros stream, which rotates faster than the LSR, and 
has a median metallicity of $[Fe/H]=-0.95$, with an rms scatter of only 
$\sim$0.15 dex. We extrapolate our results to the performance expected from the
Large Synoptic Survey Telescope (LSST) and estimate that the LSST will obtain 
metallicity measurements accurate to 0.2 dex or better, with proper motion 
measurements accurate to $\sim$0.2-0.5 mas/yr, for about 200 million F/G dwarf 
stars within a distance limit of $\sim$100 kpc ($g<23.5$). 
\end{abstract}

\section{                        Introduction                             }

A major objective of modern astrophysics is to understand when and how 
galaxies formed, and how they have evolved since then. Our own galaxy, 
the Milky Way, provides a unique opportunity to study a galaxy in great 
detail by measuring and analyzing the properties of a large number of 
individual stars.

The formation of galaxies like the Milky Way was long thought to be a
steady process leading to a smooth distribution of stars, with this
standard view exemplified by the Bahcall \& Soneira (1980) and 
Gilmore, Wyse \& Kuijken (1989) models, and described in detail by 
e.g. Majewski (1993). In these smooth models, the spatial distribution 
of stars in the Milky Way is usually\footnote{Infrared data towards the
Galactic center require addition of a bulge and a stellar bar
(e.g. Weinberg 1992; Jackson, Ivezi\'{c} \& Knapp 2002; and references
therein).}
modeled by three discrete components described using relatively simple 
analytic expressions: the thin disk, the thick disk, and the halo. 
However, recent discoveries of complex substructure in the distribution 
and kinematics of the Milky Way's stars (e.g. Ivezi\'{c} et al. 2000; Yanny 
et al. 2000; Vivas et al. 2001; Newberg et al. 2002; Gilmore, Wyse \& Norris 
2002; Majewski et al. 2003; Duffau, Zinn \& Vivas 2006; Vivas \& Zinn 2006; 
Grillmair 2006ab; Belokurov et al. 2006, 2007; Bell et al. 2007; Juri\'{c} et al. 2008) 
have deeply shaken this standard view. Unlike those smooth models that 
involve simple components, the new data indicate many irregular
structures, such as the Sgr dwarf tidal stream in the halo and the 
Monoceros stream closer to the Galactic plane. These recent developments, 
based on accurate large-area surveys, have made it abundantly clear that 
the Milky Way is a complex and dynamical structure that is still being shaped 
by the infall (merging) of neighboring smaller galaxies.

Numerical simulations suggest that 
this merger process plays a crucial role in setting the structure and 
motions of stars within galaxies, and is a generic feature of current 
cosmological models (e.g., Helmi et al. 1999; Springel \& Hernquist 2003;
Bullock \& Johnston 2005). Since the individual stars that make up the stellar 
populations in the Milky Way can be studied in great detail, their 
characterization will provide clues about the galaxy merging process that 
cannot be extracted from observations of distant galaxies (e.g., Abadi et al.
2003; Brook et al. 2004; and references therein). 

The three presumably discrete Milky 
Way components differ not only in their spatial profiles, but also in 
the detailed distributions of their kinematics and metallicity (e.g., Majewski 
1993; Ojha et al. 1996; Freeman \& Bland-Hawthorn 2002; Robin et al. 2003; 
Wyse 2006; and references therein). The thin disk,
with a scale height of $\sim$$300$~pc, has a vertical velocity dispersion 
of $\sigma_z\sim 20$ \kms, while the thick disk, with a scale height of 
$\sim$$1000$~pc, is somewhat warmer ($\sigma_z\sim 40$ \kms) and older, has a 
lower average metallicity ($[Z/Z_\odot]\sim -0.7$, e.g., Gilmore \& Wyse 1985), 
and has enhanced $\alpha$-element abundances (e.g., Fuhrmann 2004; Bensby 
et al. 2004; Feltzing 2006; Reddy et al. 2006; Ram\'{i}rez et al. 2007). 
In contrast, the halo 
is composed mainly of low-metallicity stars ($[Z/Z_\odot]<-1.0$,
e.g., Ryan \& Norris 1991), and has little or no net rotation. Hence, in
addition to their spatial profiles, the main differences between these 
components are in their rotational velocity distributions, velocity 
dispersions, and metallicity distributions. 

We note that a recent study by Carollo et al. (2007), based on
a sample of over 20,000 calibration stars with available spectra from SDSS Data Release 5, 
has demonstrated that ``the halo" of the Galaxy is likely to comprise two 
distinct components. According to these authors, the inner-halo component 
dominates the population of halo stars found at distances up to 10-15 kpc from 
the Galactic center (including the Solar neighborhood), and an outer-halo 
component dominates in the regions beyond 15-20 kpc. 
The inner halo stars are non-spherically distributed about the center
of the Galaxy, with an inferred axes ratio of $\sim$0.6, while
the outer halo comprises stars that exhibit a much more spherical spatial
distribution. Our present study only reaches to 8 kpc from the Sun, and hence
is likely to be dominated by inner-halo stars. Therefore, for the purpose of 
the present paper, we assume a single-component halo.

Despite the significant progress that has been made over the years, we still 
cannot answer some simple questions such as: Are the exponential profiles used 
to describe the spatial profiles of thin and thick disks an
oversimplification? Why do estimates for thick disk scale height differ 
by a factor of several between different studies (for a discussion see 
Siegel et al. 2002 and Juri\'{c} et al. 2008)?  Is the transition between thin and thick disks in 
metallicity and kinematics abrupt or continuous? Is there a large-scale 
metallicity gradient in the thick disk and halo? Does the disk scale length 
depend on metallicity? Can large spatial substructures be traced in kinematic and 
metallicity spaces?

To reliably answer these and similar questions, a data set needs to be 
both voluminous (to enable sufficient spatial, kinematic and metallicity 
resolution), diverse (accurate distance and metallicity estimates, 
as well as radial velocity and proper motion measurements are required), 
and faint (to probe a significant fraction of the Galaxy). Modern sky
surveys, such as the Sloan Digital Sky Survey (hereafter SDSS, York et al. 
2000), with its imaging and spectroscopic components, and the Two Micron All 
Sky Survey (Skrutskie et al. 2006) with its all-sky coverage, have 
recently provided such data sets.

Most studies of the Milky Way structure can be
described as investigations of the stellar distribution in the 
nine-dimensional space spanned by the three spatial coordinates, 
three velocity components, and three main stellar parameters 
(luminosity, effective temperature, and metallicity). Depending on the quality,
diversity and quantity of data, such studies typically concentrate
on only a limited region of this space (e.g., the nearby solar neighborhood, 
pencil beam surveys, kinematically biased surveys), or consider only 
marginal distributions of selected quantities (e.g., number density of 
stars irrespective of their metallicity or kinematics, luminosity function
determinations, proper motion surveys without metallicity or radial velocity 
information).
We use the SDSS data to study in detail the stellar distribution in this 
multi-dimensional space. We focus on 
the metallicity distribution of disk and halo stars in this contribution. 
In companion papers we discuss the spatial distribution of stars 
(Juri\'{c} et al. 2008, hereafter J08) and their kinematics 
(Bond et al. 2008, in prep., hereafter B08). 

In \S 2, we use the data for $\sim$60,000 probable F and G type main 
sequence stars provided by the SDSS spectroscopic survey to calibrate a method 
for estimating metallicity from the $u-g$ and $g-r$ colors measured by the
SDSS photometric survey. Readers who are not interested in technical aspects 
of this method may want to skip directly to \S 3, where we apply this 
method to two photometric catalogs constructed using SDSS data. One catalog 
contains averaged repeated observations, and provides sufficiently improved photometric 
accuracy and depth to study the metallicity distribution all the way to the 
disk-halo interface at several kpc from the Galactic plane. The second catalog, 
based on all SDSS photometric observations to date, covers a wide area and probes a 
significant fraction of the Galaxy. We summarize and discuss our results 
in \S 4.

\section{  Determination  of Stellar Metallicity  from 
                          SDSS Photometric Data                      }

The most accurate measurements of stellar metallicity are based on spectroscopic
observations. Despite the recent progress in the availability of stellar spectra
(e.g., SDSS has recently made publicly available\footnote{See http:
//www.sdss.org/dr6} over 280,000 stellar spectra as a part of its Data Release
6; the proposed extension of SDSS, known as SDSS-III, is capable of providing
another several hundred thousand stars with available spectra in the next few
years; RAVE\footnote{See http://www.rave-survey.aip.de/rave} may provide up to a
million spectra, primarily thin- and thick-disk stars, over the next few years),
the number of stars detected in imaging surveys is vastly larger. In addition to
generally providing better sky and depth coverage than spectroscopic surveys,
imaging surveys obtain essentially complete flux-limited samples of stars. The
simple selection criteria used for the photometric surveys are advantageous when
studying Galactic structure, compared to the complex targeting criteria that are
used for SDSS stellar spectra (see 
\S~\ref{SDSSspec} below). Hence, we use the extant SDSS spectroscopic data to calibrate a 
method for estimating metallicity from the SDSS imaging data, and
use this calibration to study the metallicity distribution of several
million disk and halo stars of the Milky Way. 

Stellar metallicity has long been estimated using photometric methods such 
as the traditional UV excess based $\delta (U-B)_{0.6}$ method (Wallerstein 1962; 
Sandage 1969). A blue main sequence (F and G type) star's metallicity
is correlated with the difference between the star's $U-B$ color and 
that which would be measured for a metal-rich star with the {\it same} 
$B-V$ color. This correlation is seen in both data (e.g. Carney 1979 and 
references therein) and detailed stellar models (Kurucz 1979). The Johnson
$UBV$ bands are similar to SDSS's $ugr$ bands and thus it should be possible 
to derive an analogous method applicable to SDSS photometric system, as recently
attempted by Karaali, Bilir \& Tuncel (2005). However, as they pointed 
out, their study did not utilize SDSS data, but a somewhat different 
photometric system. Unfortunately, even small photometric offsets and color 
terms between different photometric systems may have significant systematic
effects on derived metallicities. For example, the SDSS $u$-band 
measurements are offset from the AB system by $\sim$0.04 mag (Eisenstein 
et al. 2006; Holberg \& Bergeron 2006), leading to a metallicity bias of up 
to 0.2 dex. Here we derive photometric metallicity estimators for the SDSS 
filter set using SDSS Data Release 6 data. This calibration relies on the 
large number of stars ($\sim$287,000) with a homogeneous set of stellar 
parameters (effective temperature, metallicity and gravity) derived from 
moderate-resolution SDSS spectra (Beers et al. 2006; Allende Prieto et al. 
2006; Lee et al. 2007ab; Allende Prieto et al. 2007).

\subsection{     An Overview of the Sloan Digital Sky Survey   }
\label{overview}

The SDSS is a digital photometric and spectroscopic survey which 
covers about one quarter of the Celestial Sphere in the North Galactic
cap, as well as a smaller area ($\sim$300 deg$^{2}$) but much deeper
survey in the Southern Galactic hemisphere (Stoughton et al. 2002; 
Abazajian et al. 2003, 2004, 2005; Adelman-McCarthy et al. 2006). 
SDSS is using a dedicated 2.5m telescope (Gunn et al. 2006) to provide 
homogeneous and deep ($r < 22.5$) photometry in five bandpasses 
(Fukugita et al.~1996; Gunn et al.~1998; Smith et al.~2002; Hogg et al.~2002; 
Tucker et al.~2006) repeatable to 0.02 mag (root-mean-square scatter,
hereafter rms, for sources not limited by photon statistics, Ivezi\'{c} et al.~2003) 
and with a zeropoint uncertainty of $\sim$0.02-0.03 mag (Ivezi\'{c} et
al.~2004). The flux densities of detected objects are measured almost
simultaneously in five bands ($u$, $g$, $r$, $i$, and $z$) 
with effective wavelengths of 3540 \AA, 4760 \AA, 6280 \AA, 7690 \AA, 
and 9250 \AA. 
The large survey sky coverage will result in photometric measurements for well 
over 100 million stars and a similar number of galaxies\footnote{The recent
Data Release 6 lists photometric data for 287 million unique objects observed
in 9583 deg$^2$ of sky; Adelman-McCarthy et al. 2008; see http://www.sdss.org/dr6/.}.
The completeness of SDSS catalogs for point sources is $\sim$99.3\% 
at the bright end and drops to 95\% at magnitudes of 22.1, 22.4, 22.1, 
21.2, and 20.3 in $u$, $g$, $r$, $i$ and $z$, respectively. 
Astrometric positions are accurate to better than 
0.1 arcsec per coordinate (rms) for sources with $r<20.5$ (Pier et al.~2003), and 
the morphological information from the images allows reliable star-galaxy separation 
to $r \sim$ 21.5 (Lupton et al.~2002; Scranton et al. 2002). 
A compendium of other technical details about SDSS can be found 
on the SDSS web site (http://www.sdss.org), which also provides 
interface for the public data access.

\subsection{             SDSS spectroscopic survey of stars          }
\label{SDSSspec}
SDSS spectra are obtained with a pair of dual multi-object fiber-fed
spectrographs on the same telescope used for the imaging survey (Uomoto
et al., in prep). 
Spectroscopic plates have a radius of 1.49 degrees and take 640 
simultaneous spectra, each with wavelength coverage 3800--9200~\AA~and 
spectral resolution of $R \sim 2000$. The signal-to-noise ratio is
typically $>$4 per pixel at $g$=20, but is substantially higher for brighter
point sources, such as considered herein.

Targets for the spectroscopic survey are chosen from the SDSS imaging data, 
described above, based on their colors and morphological 
properties\footnote{The recent extension of SDSS survey, known as SDSS-II,
has different targeting priorities. In particular, the sub-survey known
as SEGUE (Sloan Extension for Galactic Understanding and Exploration), is optimized for Galactic structure studies.}. The targets include
\begin{itemize}
\item {\bf Galaxies:} a simple flux limit for ``main'' galaxies, flux-color cut
       for luminous red galaxies (Strauss et al. 2002; Eisenstein et al. 2001)
\item {\bf Quasars:} flux-color cut, matches to FIRST survey (Richards et al. 2002) 
\item {\bf Non-tiled objects (color-selected):} calibration stars (16 per
           plate), ``interesting'' stars (hot white dwarfs, brown dwarfs, 
           red dwarfs, red giants, blue horizontal branch stars, carbon stars, 
           cataclysmic variables, central stars of planetary nebulae), sky
\end{itemize}
Here, {\it (non)-tiled objects} refers to objects that are not guaranteed a
fiber assignment. As an illustration of the fiber assignments, SDSS Data 
Release 6 contains spectra of 791,000 galaxies, 104,000 quasars, and 287,000 
stars. 

The spectra are targeted and automatically processed by three pipelines:
\begin{itemize}
       \item {\bf target:} Target selection and tiling
       \item {\bf spectro2d:} Extraction of spectra, sky subtraction, wavelength 
                      and flux calibration, combination of multiple exposures 
       \item {\bf spectro1d:} Object classification, redshifts determination,
                        measurement of line strengths and line indices 
\end{itemize}  

For each object in the spectroscopic survey, a spectral type, 
redshift (or radial velocity), and redshift error is determined by matching the
measured spectrum to a set of templates.  The stellar templates are
calibrated using the ELODIE stellar library. Random errors for the radial
velocity measurements are a strong function of spectral type and 
signal-to-noise ratio, but are usually
$< 5$~{\kms} for stars brighter than $g\sim18$, rising sharply to
$\sim$$25$~{\kms} for stars with $g=20$.  Using a sample of multiply-observed stars, 
Pourbaix et al. (2005) estimate that these errors may be underestimated by a
factor of $\sim$$1.5$. Further technical details about SDSS spectroscopic
survey are available from www.sdss.org.

\subsection{               Stellar Atmospheric Parameter Estimation                   }

SDSS stellar spectra are of sufficient quality to provide robust and
accurate stellar parameters, such as effective temperature, surface gravity, and
metallicity (parameterized as [Fe/H]). These parameters
are estimated using a variety of methods implemented in an 
automated pipeline (the SEGUE Stellar Parameters Pipeline, SSPP; Beers et al. 2006). 
A detailed discussion of these methods and their performance can be found 
in Allende Prieto et al. (2006, 2007) and Lee et al. (2007a,b). 
Based on a comparison with high-resolution abundance 
determinations, they demonstrate that the combination of spectroscopy and 
photometry from SDSS is capable of delivering estimates of $T_{\rm eff}$, 
$log(g)$, and $[Fe/H]$ with {\it external accuracies} of 190~K (3.2 \%), 0.28 dex, 
and 0.17 dex, respectively. These tests indicate that {\it mean systematic errors}
for $[Fe/H]$ and $T_{\rm eff}$ should not be larger than about 0.2 dex and 100 K, 
and may be below 0.1 dex and 75 K (Lee et al. 2007b). Note that these estimates 
apply to stars with a wider range of temperatures than we consider in this
study. 

We use the final adopted values, called $teffa$ and $feha$ in the SDSS
{\it sppParams} table, which are based on averaging several different methods. 
A detailed analysis by Lee et al. (2007a,b) demonstrates that systematic metallicity
differences between the methods used in averaging do not exceed $\sim$0.1 dex. 
A comparison with Galactic open and globular clusters indicates that the adopted 
metallicity scale systematically overestimates metallicity by $\sim$0.15 dex for 
$[Fe/H] < -2$ and  underestimates metallicity by up to $\sim$0.3 dex for stars 
near Solar metallicity (the metallicity offsets have been improved recently, and
are now essentially nil, but for the purpose of this paper, we have made use of
a previous version of the SSPP, hence the sytematics remain present). 

Only a few percent of stars in SDSS spectroscopic sample are giants.
For this reason, we consider only the main sequence stars, using the selection
criteria described below. Although we address photometric estimates of 
effective temperature, the main goal of this section is to derive
a robust and accurate photometric metallicity estimator.

\subsubsection{ Sample Selection }
\label{sample}

\begin{figure}
\plotone{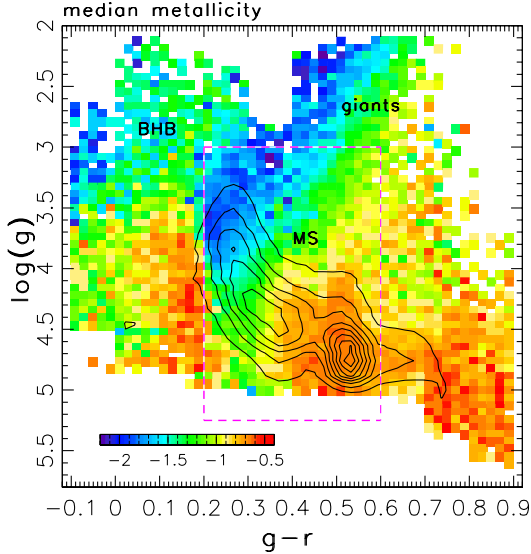}
\caption{
The linearly-spaced contours show the distribution of $\sim$110,000 stars with
$g<19.5$ and $-0.1 < g-r < 0.9$ (corresponding to effective temperatures in the
range 4500~K to 8200~K)  from the SDSS DR6 spectroscopic sample in the log(g) vs.
$g-r$ plane. The multi-modal distribution is a result of SDSS target selection
algorithm. The color scheme shows the median metallicity in all 0.02 mag by 0.06
dex large pixels that contain at least 10 stars. The fraction of stars with
log(g)$<$3 (giants) is 4\%, and they are mostly found in two color regions:
$-0.1 < g-r < 0.2$ (BHB stars) and $0.4 < g-r < 0.65$ (red giants). They are
dominated by low-metallicity stars ($[Fe/H]<-1$). The dashed lines outline the
main-sequence (MS) region selected for deriving photometric estimates for effective 
temperature and metallicity.
\vspace{1em}}
\label{logggr}
\end{figure}

We begin by selecting bright stars from the main stellar locus (Lenz et al. 1998; 
Fan 1999;, Finlator et al. 2000; Smol\v{c}i\'{c} et al. 2004), with colors 
located in the proper range for the application of photometric metallicity 
method (roughly\footnote{At the $\sim$0.05 mag accuracy level, 
$B-V$=0.949$(g-r)$+0.20; for more accurate ($<$0.01 mag) transformations 
see Ivezi\'{c} et al. (2006a).} $0.4 < B-V < 0.8$, Carney 1979), and
from sky regions with modest interstellar dust extinction (SDSS utilizes the 
Schlegel, Finkbeiner \& Davis 1998 maps). The specific criteria applied
to 130,620 entries from the so-called {\it sppParams} table\footnote{Available
from http://www.sdss.org/dr6/products/spectra/spectroparameters.html} 
that have $log(g)>0$ are 
\begin{enumerate}
\item The interstellar extinction in the $r$ band below 0.3: [106,816]
\item $14 < g <19.5$: [104,844]
\item $0.2 < (g-r) < 0.6$: [75,928] 
\item $0.7 < (u-g) < 2.0$ and $-0.25 < (g-r) - 0.5(u-g) < 0.05$: [68,306]  
\item $-0.2 < 0.35(g-r) - (r-i) < 0.10$: [66,496]
\end{enumerate}
The numbers in brackets indicates the number of stars left after each selection
step.

Using a photometric parallax relation based on observations of globular
clusters (see Appendix A for a detailed discussion), 
\begin{equation}
\label{mainPP}
              M_r(g-i,[Fe/H]) = M_r^0(g-i) + \Delta M_r([Fe/H]),
\end{equation}
where $\Delta M_r([Fe/H])$ and $M_r^0(g-i)$ are given by eqs.~\ref{GCppFeH} 
and \ref{GCppFinal}, respectively, we further limit the sample to 61,861 stars 
in the 1--10 kpc distance range. Due to 
the small $r-i$ color range spanned by F/G stars, when comparing results to
J08 it is better to estimate the $r-i$ color from the better measured $g-i$ 
color using a stellar locus relation\footnote{J08 uses a maximum likelihood 
projection on the mean stellar locus, which avoids this problem. At the bright 
end that is relevant here, the two methods produce essentially the same results;
we opted for the simpler one.}   
\begin{eqnarray}
\label{locus}
	g-r = 1.39 (1-\exp[-4.9(r-i)^3 \\ \nonumber
 	   - 2.45(r-i)^2 - 1.68(r-i) - 0.050] ).  
\end{eqnarray}

The selected stars span the 5000--7000 K temperature range (with a 
median of 5,900 K) and 99.4\% have metallicity in the range $-3$ to 0 (with 
a median of $-1.0$). While the sample is dominated by main sequence stars 
(the median log(g) is 4.1, with an rms scatter of 0.44 dex), a small fraction 
of stars have gravity estimates consistent with giants (see
Figure~\ref{logggr}). We exclude $\sim$3\% of stars with log(g)$<$3 (which typically 
have lower metallicity than dwarfs, with a median $[Fe/H] = -1.5$;
see Figure~\ref{logggr}), resulting in a final calibration sample of 59,789 stars.
This fraction of giants is
relatively high because the $g-r\sim0.5$ color range, where the fraction of
giants is the highest, was deliberately targeted for SDSS spectroscopy; about 
7\% of stars in the subsample with $0.4 < g-r < 0.6$ have log(g)$<$3.

\subsubsection{ Effective Temperature}

\begin{figure*}
\epsscale{0.8}
\plotone{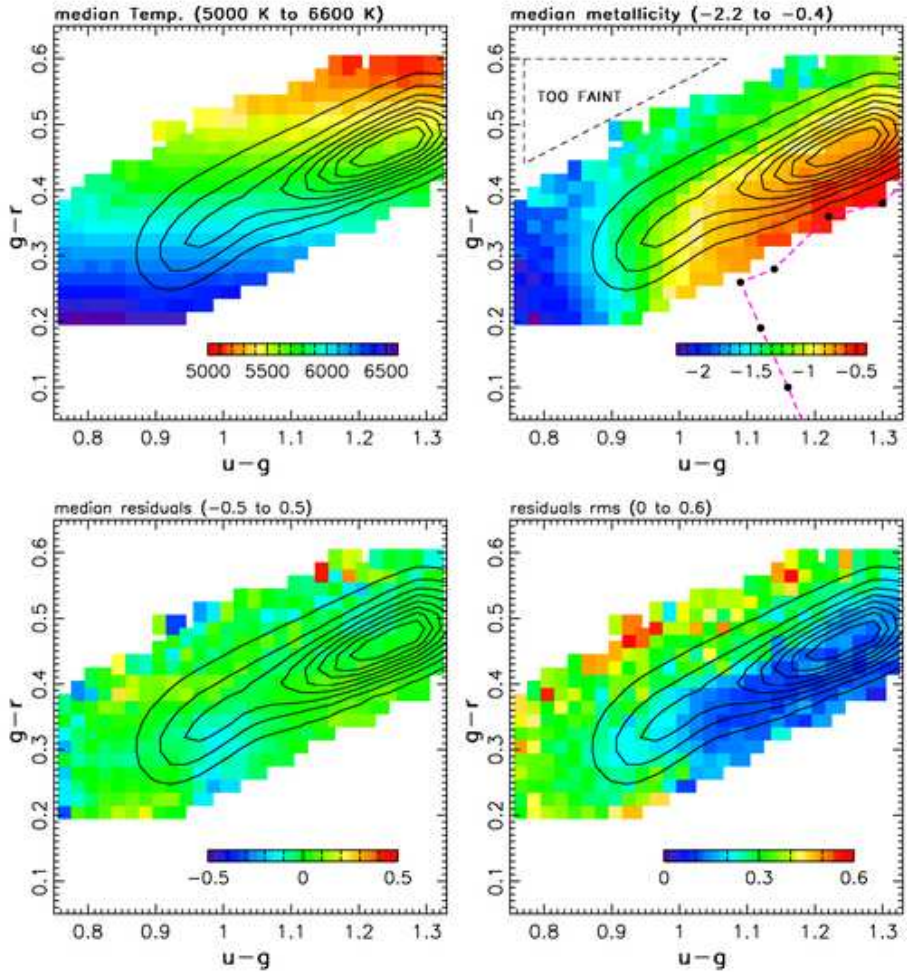}
\caption{
The correlation of spectroscopic effective temperature (top left) and
metallicity (top right) with the position of a star in the $g-r$ vs. $u-g$
color-color diagram. The color scheme shows the median values in all 0.02 mag by
0.02 mag large pixels that contain at least 10 stars. The distribution of stars
in an {\it imaging} sample with $g<19.5$ is shown by linearly-spaced contours.
The dots show the synthetic colors for the Pickles (1998) solar metallicity
standards (F0, F2, F5, F6, F8, and G0, from bottom to top), taken from Covey et
al. (2007). The triangular region marked ``TOO FAINT'' in the top right panel
contains no stars, due to the $g<19.5$ flux limit and the fact that low-metallicity
stars are generally more distant and fainter than high-metallicity stars. The
bottom left panel shows the median residuals between the spectroscopic metallicity
and photometric estimates based on eq.~\ref{Zphotom}. Their root-mean-square
scatter (over all pixels) is 0.06 dex. The bottom right panel shows a map of the
root-mean scatter of metallicity for individual stars in each pixel. Its median
value is 0.21 dex. The scatter is larger for weak-lined low-metallicity stars
($\sim$0.3 dex) than for high-metallicity stars ($\sim$0.15 dex), as expected.}
\label{panels1}
\end{figure*}

The dependence of the median effective temperature and metallicity on the position
in the $g-r$ vs. $u-g$ color-color diagram  for the final sample of 59,789
stars is shown in Figure~\ref{panels1}. The top left panel demonstrates that 
the effective temperature, $T_{\rm eff}$,
can be determined from the $g-r$ color alone, with a negligible dependence on the 
$u-g$ color (the gradient of $\log(T_{\rm eff})$ with respect to the $g-r$ color 
is at least $\sim60$ times as large as the gradient with respect to the 
$u-g$ color). This difference in gradients is due to a general insensitivity 
to metallicity of relationships between effective temperature and colors at 
wavelengths longer than 0.4 $\mu m$ (Sandage \& Smith 1963; Mannery 
\& Wallerstein 1971). The best-fit expression,
\begin{equation}
\label{logT}
          \log(T_{\rm eff} / {\rm K}) = 3.872 - 0.264\,(g-r)
\end{equation}
reproduces SDSS spectroscopic temperature for 59,789 main sequence stars
selected from the $0.2 < g-r < 0.6$ color range with an rms scatter of 0.007
dex (corresponding 
to $\sim$100 K). When residuals are binned in 0.01 mag wide $g-r$ bins, 
the largest median residual is 0.003 dex ($\sim$40 K at the median temperature
of 5900 K), demonstrating that a linear fit is sufficient. When residuals are 
binned in 0.1 dex wide metallicity bins, the largest median residual is 
also 0.003 dex. There is no discernible dependence of residuals on metallicity 
for stars with $[Fe/H] < -1$, while for stars with $-1 < [Fe/H] < -0.5$ a
gradient of $log(T)$ is 0.008 per decade of metallicity (dex per dex) is present. 

This behavior is consistent with 
results based on temperatures derived with the infrared flux method 
(Ram\'{i}rez \& Mel\'{e}ndez 2005; Casagrande, Portinari \& Flynn 2006, 
hereafter CPF). For example, the expression for effective temperature as a 
function of $B-V$ color and metallicity from CPF predicts an effective temperature 
of 5700 K for $B-V=0.6$ ($g-r$=0.425) and metallicity of $-1.0$, with the latter 
corresponding to the median metallicity of stars in the SDSS spectroscopic sample. 
The effective temperature predicted by eq.~\ref{logT} is 5750 K (a discrepancy 
of 0.004 dex), and the median spectroscopic temperature for stars with 
$0.42<g-r<0.43$ is 5730 K. We note that both the CPF relation and 
Ram\'{i}rez \& Mel\'{e}ndez (see Figs. 1 and 10) predict a steeper dependence
of effective temperature on metallicity: at  $B-V=0.6$ the predicted
effective temperature increases by 180 K as metallicity increases from 
$-1.5$ to $-0.5$, while in the SDSS spectroscopic sample the corresponding 
temperature increase is 50 K. Discrepancies with the expression proposed
by Sekiguchi \& Fukugita (2000) are somewhat larger. Their effective 
temperature scale is cooler by $\sim$130 K than the SDSS scale, and 
$\log(T_{\rm eff})$ residuals are correlated with metallicity and log(g) 
with gradients of about 0.01 dex per dex. Further details about the behavior of
photometric temperature estimator are discussed in Appendix B.

\subsubsection{                           Metallicity                        }
\label{metalSec}
As first suggested by Schwarzschild, Searle, \& Howard (1955), the depletion of 
metals in a stellar atmosphere has a detectable effect on the 
emergent flux, in particular in the blue region where the density of metallicity 
absorption lines is highest (Beers \& Christlieb 2005, and references therein).
The median metallicity of stars selected from the SDSS spectroscopic sample as a 
function of the $u-g$ and $g-r$ colors shows a complex behavior that is consistent 
with expectations: the detailed dependence of the UV excess (i.e., the $u-g$ color) 
on metallicity varies with effective temperature (i.e., the $g-r$ color). Even 
when the $g-r$ vs. $u-g$ plane is separated by $g-r=0.4$ into two regions suggested 
by the metallicity map, at least second-order polynomials, or several piecewise 
linear fits, are required to avoid systematic errors larger than 0.1 dex. 
In order to do so for the entire map with a single function, we find 
that third-order terms are required, and model the map as:
\begin{eqnarray}
\label{Zphotom}
  [Fe/H]_{ph} & = & A + Bx + Cy + Dxy + Ex^2 + Fy^2 \\ \nonumber
	& & + Gx^2 y + Hxy^2 + Ix^3 + Jy^3,
\end{eqnarray}
where
\begin{itemize}
\item $x=(u-g)$ for $(g-r)\le0.4$, and $x=(u-g)-2(g-r)+0.8$ for $(g-r)>0.4$
(this dual definition is required to describe the map with a single set
of coefficients, $A$--$J$)
\item $y=(g-r)$, and 
\item
($A$--$J$) = ($-$4.37, $-$8.56, 15.5, $-$39.0, 23.5,  20.5, 12.1, 7.33, $-$10.1, $-$21.4). 
\end{itemize}
The above expression describes the median metallicity map shown in the
top right panel in Figure~\ref{panels1} with a root-mean-square (rms) 
scatter of 0.09 dex. This level of systematic calibration errors is negligible 
compared to random errors per star ($\sim0.2$ dex, due to photometric errors), 
discussed below, and is comparable to systematic errors in the SDSS spectroscopic
metallicity estimates\footnote{The systematic errors are much larger for 
stars with log(g)$<$3: for example, for stars with $0.4<g-r<0.6$ and
log(g)=2.5, the photometric metallicity estimate is 0.5 dex larger than the spectroscopic
metallicity (when the systematic error vs. log(g) trend of about 0.35 dex per dex 
is corrected for, the rms scatter of the metallicity residuals for
log(g)$<$3.5 is $\sim0.3$ dex).}. 
A map of the median residuals, when fitting the median metallicity map
using eq.~\ref{Zphotom}, in the $g-r$ vs. $u-g$ plane is shown in the bottom 
left panel in Figure~\ref{panels1}. It illustrates that there is no strong 
correlation between systematic errors in photometric metallicity 
computed with eq.~\ref{Zphotom} ($\la$0.1 dex) and the $u-g$ and $g-r$ colors. 

We compute photometric metallicity estimates for all 59,789 stars in the sample
using eq.~\ref{Zphotom}, and compare these to the spectroscopic metallicity
determinations.
The rms scatter of metallicity residuals is 0.24 dex (determined from 
the interquartile range), and the distribution of residuals is only slightly 
non-Gaussian (97\% of the sample is contained within a $\pm3\sigma$ range). 

The rms scatter of the metallicity residuals depends on both the apparent magnitude
and color of the star under consideration. The bottom right panel in Figure~\ref{panels1} illustrates 
the color dependence: for low-metallicity stars the rms increases
to $\la$0.3 dex, while it is about 0.15 dex or less for high-metallicity
stars. This is expected, due to the weaker spectral lines in low-metallicity 
stars (e.g., Du et al. 2004; Keller et al. 2007).

The rms scatter of metallicity residuals increases with the $g$-band
magnitude from 0.18 dex for $g<$17, to 0.25 dex at $g=18$, and 0.45 dex
at $g=19.5$. The random metallicity errors are dominated by the errors in the 
$u$-band magnitudes. 
The ratio of this scatter to the scatter expected due to
photometric errors (which is readily computed from eq.~\ref{Zphotom}) 
is 1.7, and is nearly independent of magnitude. The implied {\it random} errors 
in spectroscopic and photometric metallicity estimates are thus comparable, and have 
similar signal-to-noise properties. In particular, we estimate that 
random errors in spectroscopic metallicity estimates increase from 0.15 dex
for $g<17$ to 0.36 dex at $g=19.5$ (for comparison, the corresponding
values for photometric metallicity estimates are 0.10 dex and 0.30 dex,
respecitvely). 
This seemingly surprising result, that the estimated errors for photometric 
metallicity are {\it smaller} than those obtained for spectroscopic metallicity
estimates, despite
the former being calibrated off the latter, is due to the averaging of 
{\it many} spectroscopic estimates in a given small color-color bin
when calibrating photometric metallicity, and the fact that the signal for
photometric metallicity estimates predominantly comes from wavelengths shorter
than 0.4 $\mu$m, while for spectroscopic metallicity estimates are obtained from longer 
wavelengths.

{\it This error behavior limits the application of photometric metallicity
estimates, based on SDSS data, to about $g<19.5$}. This limit is essentially 
set by the precision of the $u$-band photometry ($u<20.5$). Somewhat 
coincidentally\footnote{The similar depths are not entirely independent,
as they both reflect the atmospheric and sky properties, and various
scientific tradeoffs, but this discussion is beyond the scope of this paper.
}, this is about the same limiting depth as for the SDSS 
spectroscopic sample (the spectroscopic targeting limit for the SEGUE
survey is $g<20$). Despite this limitation, the photometric metallicity 
estimator given by eq.~\ref{Zphotom}
is a valuable tool, because it allows metallicity to be determined 
for {\it all} main sequence SDSS stars in the $0.2 < g-r < 0.6$ color range. For example, 
in SDSS DR6, out of $\sim$5.7 million point sources from this color 
range that are brighter than $g=19.5$, SDSS spectra classified as stars 
are available only for $\sim$94,000 objects. This implies a sample size increase
by about factor of 60 when using photometric metallicity estimates. Furthermore, when 
deeper data are available, the photometric metallicity  estimator can be 
used to study the metallicity distribution in the Galaxy to distances beyond the reach of 
main sequence stars in the spectroscopic sample (a small number of giants in 
the spectroscopic sample, which reach to distances $\sim$100 kpc, cannot be easily recognized 
using photometry alone; however, see Helmi et al. 2003). Further details 
about the behavior of the photometric metallicity estimator are discussed in 
Appendix C.

\section{ Analysis of the Stellar Photometric Metallicity Estimates }

\begin{deluxetable}{cccrcc}
\tablenum{1} \tablecolumns{6}
\tablecaption{Sample Distance Limits}
\tablehead{$(g-r)$ &  $(r-i)^a$  & $M_g^b$ & $D^c$ & $(B-V)^d$ & $M_V^e$}   
\startdata
0.2 & 0.03 & 3.25 & 17.8 & 0.35 & 3.11 \\
0.3 & 0.08 & 4.55 &  9.8 & 0.46 & 4.36 \\
0.4 & 0.12 & 5.48 &  6.4 & 0.57 & 5.25 \\
0.5 & 0.16 & 6.20 &  4.6 & 0.68 & 5.92 \\
0.6 & 0.20 & 6.77 &  3.5 & 0.77 & 6.43 \\
0.7 & 0.24 & 7.24 &  2.8 & 0.86 & 6.84 \\
0.8 & 0.28 & 7.64 &  2.4 & 0.95 & 7.19 \\
0.9 & 0.32 & 8.02 &  2.0 & 1.04 & 7.50 \\
1.0 & 0.37 & 8.40 &  1.7 & 1.12 & 7.83 \\
1.1 & 0.42 & 8.81 &  1.4 & 1.21 & 8.18 \\
1.2 & 0.48 & 9.27 &  1.1 & 1.29 & 8.58 \\
\enddata
\tablenotetext{a}{The mean $r-i$ color on the main stellar locus for 
the $g-r$ color listed in first column, evaluated using eq.~\ref{locus}.}
\tablenotetext{b}{The absolute magnitude in the $g$ band, evaluated for a fiducial
                   $[Fe/H]=-1.0$ using eq.~\ref{mainPP}.} 
\tablenotetext{c}{The distance for a star with $g=19.5$ (kpc).} 
\tablenotetext{d}{The Johnson $B-V$ color, computed for convenience from SDSS 
                  photometry using transformations from Ivezi\'{c} et al. (2006a).} 
\tablenotetext{e}{The absolute magnitude in the Johnson $V$ band, computed from $M_g$.} 
\end{deluxetable}

We now use the photometric metallicity estimator developed above to study the
stellar metallicity distribution as a function of position in the Galaxy and
stellar kinematics. We consider stars in a restricted color range, $0.2 < g-r <
0.4$, because the redder stars ($0.4 < g-r < 0.6$) do not extend as far into the
halo (due to their smaller luminosities; Table 1). The small color range also
minimizes various selection effects that could be larger for a wider
color/luminosity range (such as uncertainties in the photometric parallax
relation and contamination by giants). As an additional motivation, in this
color range metallicity is nearly a function of the $u-g$ color alone
(eq.~\ref{ZphotomHighT} in Appendix C), which allows a simple assessment of the
impact of photometric errors in the $u$ band on derived metallicity. The adopted
$0.2 < g-r < 0.4$ color range spans about 10 MK spectral subtypes (from $\sim$F5
to $\sim$G5; Bailer-Jones et al. 1997, 1998). The median absolute magnitude in
this color range is $M_g=4.6$, with an rms scatter of 0.3 mag and a difference
of $\sim$2.2 mag in $M_g$ between the blue and red ends (for a fiducial
$[Fe/H]=-1$, see Table 1). 

We consider two photometric catalogs constructed using SDSS data. A catalog of
coadded repeated observations (10 on average; Ivezi\'{c} et al. 2007), known as
the SDSS Stripe 82 catalog, provides improved photometric accuracy to a fainter
flux limit in $\sim$300 deg$^2$ of sky. For example, while single-epoch SDSS
data deliver a median $u-g$ error of 0.06 mag at $g=19.5$ (for point sources
with $0.2 < g-r < 0.4$), the same level of accuracy is extended to beyond
$g=20.5$ in the coadded catalog. This allows us to study the metallicity
distribution all the way to the disk-halo interface, at several kpc from the
Galactic plane, with small metallicity errors. At the bright end, the random
errors in the $u-g$ error are 0.01 mag in the coadded catalog, and 0.025 mag in
single-epoch data (an error in the $u-g$ color of 0.02 mag induces a metallicity
error in $[Fe/H]$ that varies from 0.02 dex at $[Fe/H]=-0.5$ to 0.11 dex at
$[Fe/H]=-1.5$). This improvement in photometric metallicity accuracy by more
than a factor of two enables robust estimates of the metallicity distribution
width for disk stars. However, an important disadvantage of using the coadded
catalog is its very small sky coverage. Hence, we extend our study to a
significant fraction of the Galaxy by using a wide-area catalog based on
SDSS Data Release 6 (DR6 catalog, hereafter). This catalog covers an area $\sim$30
times larger than the deep coadded catalog, at the expense of a $\sim$1 mag
shallower sample. 

We begin our analysis with a discussion of the stellar distribution in the $g$
vs. $u-g$ color-magnitude diagram, which reveals several features that are
central to the conclusions of this paper. While this diagram maps well to a
distance vs. metallicity plane, as discussed in \S~\ref{metalDistrib} further
below, we choose to first describe these features using directly observed
quantities. When discussing positions of stars in the Milky Way, we use the
usual cylindrical coordinate system ($R, \phi, Z$) aligned with the Galactic
center (assumed to be at a distance of 8.0 kpc) and with the $Z$ axis towards the
north Galactic pole. For projections parallel to the Galactic plane, we follow
J08 and use right-handed $X$ and $Y$ coordinates, with the Sun at $X=8.0$ kpc
and the positive $Y$ axis pointing towards $l=270^\circ$.

\subsection{   The Bimodal $u-g$ Distribution of F/G Stars     }

\begin{figure*}
\plotone{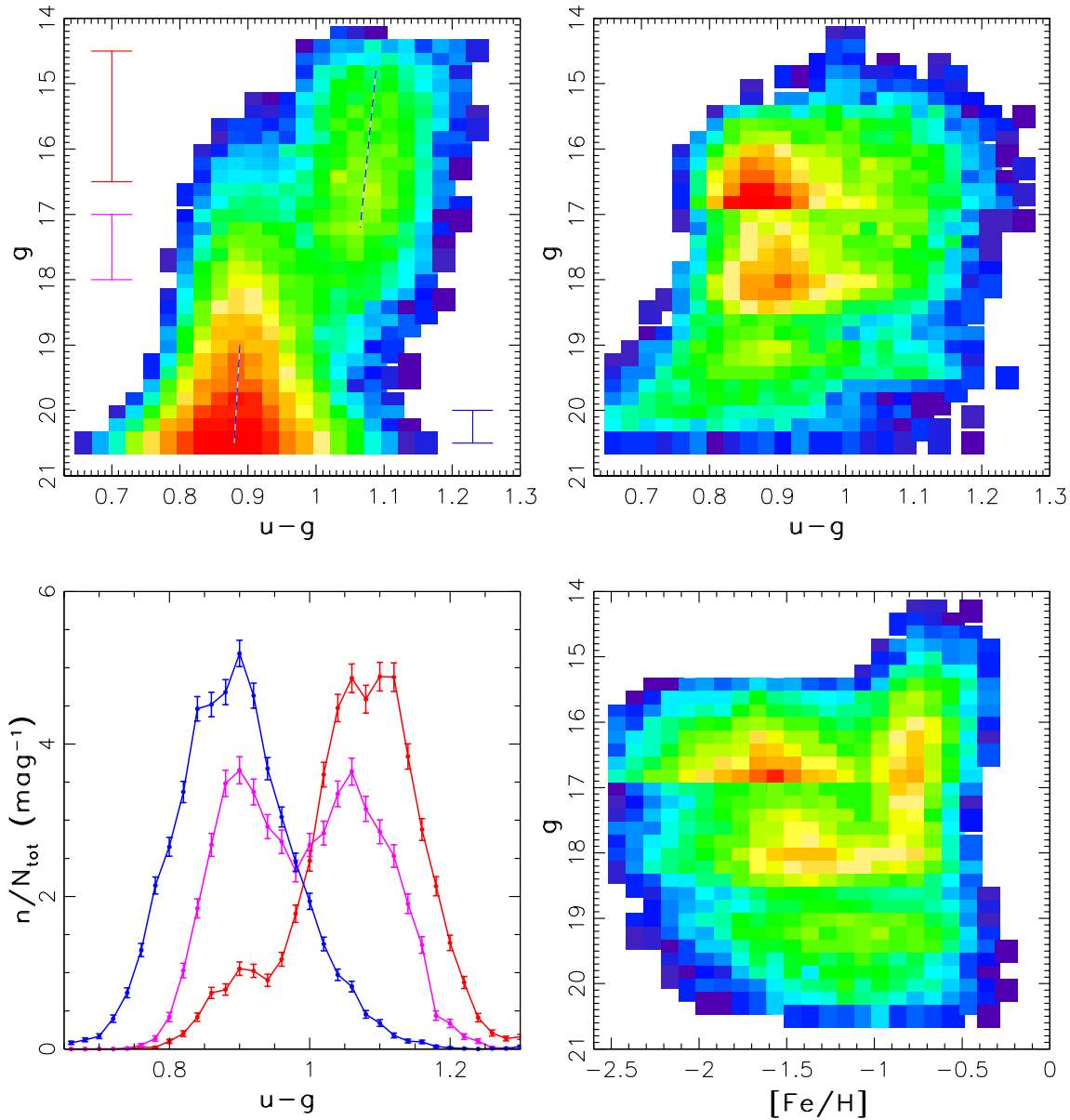}
\caption{
The top left panel shows the distribution of stars (logarithm of counts in each
bin; low to high is blue to green to red) from the SDSS Stripe 82 catalog with 
$0.2 < g-r <0.4$ in the $g$ vs. $u-g$
color-magnitude diagram. In this $g-r$ range, the $u-g$ color is a proxy for
metallicity (see eq.~\ref{ZphotomHighT} in Appendix C). The two concentrations
of stars correspond to disk ($u-g\sim1.1$) and halo ($u-g\sim0.9$) stars, with
the dashed lines indicating the change of the median $u-g$ color with magnitude
for each concentration. The $u-g$ color distributions in three magnitude slices,
marked by vertical bars in the top left panel, are shown in the bottom left
panel. All three histograms can be approximately described by a sum of two
$\sim$0.07 mag wide Gaussians centered on $u-g$=0.90 and 1.08, with the number
ratio of blue to red component increasing with magnitude from 1:7 to 20:1. For
detailed fits to the $u-g$ color distribution as a function of magnitude, see
Figure~\ref{panels3x}. The top right panel is analogous to the top left panel,
except that a complete imaging sample of stars is replaced by stars from the
SDSS spectroscopic survey. The spectroscopic sample is highly incomplete, as
evident from the patchy distribution. The bottom right panel shows the same
sample of stars from the spectroscopic survey, with the $u-g$ color replaced by
spectroscopic metallicity.}
\label{panels2}
\end{figure*}

\begin{figure}
\plotone{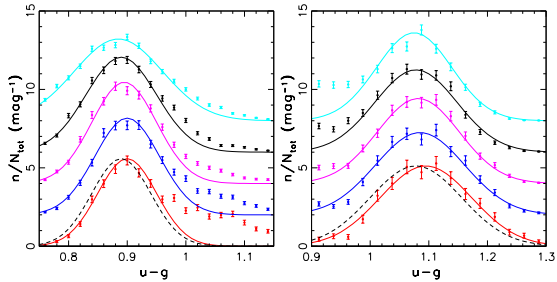}
\caption{
The symbols with error bars show the $u-g$ color distributions in 0.5 mag wide
magnitude slices (see the top left panel in Figure~\ref{panels2}), in the range
$g=18-20.5$ (left) and $g=14.5-17.0$ (right). The bottom histograms correspond
to the brightest bin, and each histogram is successively offset by 2 for
clarity. The solid lines show the best-fit Gaussians, fit to data with
$u-g<0.95$ (left) and $u-g>1.0$ (right), to minimize contamination by disk and
halo stars, respectively. The best-fit parameters are listed in Table 2. The
dashed black lines are the same as the solid black lines, and are added to
illustrate the shift of histograms towards bluer colors for the faint bins. The
gradients of the $u-g$ color with respect to the $g$ magnitude are
$-0.006\pm0.002$ mag/mag (left) and $-0.012\pm0.002$ mag/mag (right). }
\label{panels3x}
\end{figure}

We selected 110,363 sources from $\sim$1.01 million entries in the Stripe 82
coadded catalog\footnote{This catalog is publicly available from http:
//www.sdss.org/dr6/products/value\_added/index.html.} by requiring at least 4
detections in the $u$ band, $0.20 < g-r < 0.40$ and $g<20.5$. These sources have
$|\delta_{J2000}|<$ 1.266$^\circ$ and right ascension in the range 20h 34m to 4h
00m. For reference, Galactic coordinates, ($l$,$b$), are (46,$-$24), (96,$-$60)
and (190,$-$37) for $\alpha_{J2000}$=$-$50$^\circ$, 0$^\circ$ and 60$^\circ$ (at
$\delta_{J2000}$=0$^\circ$).

The distribution of these stars in the $g$ vs. $u-g$ color-magnitude diagram is
shown in the top left panel of Figure~\ref{panels2}. Bright red ($g<18$,
$u-g\sim1.1$) and faint blue ($g>18$, $u-g\sim0.9$) features are clearly
discernible, and are roughly separated by the $u-g=1$ line (corresponding to
$[Fe/H] \sim -1.0$). The marginal $u-g$ distributions for three $g$ slices are
shown in the bottom left panel. They can be approximately described by a sum of
two $\sim$0.07 mag wide Gaussians centered on $u-g$=0.90 and 1.07, with the
number ratio of the blue to red component increasing with magnitude from 1:7 in
the blue bin to 20:1 in the red bin. The blue and red components correspond to
distant metal-poor halo stars, and more metal-rich and closer disk stars,
respectively, as discussed further below. The width of 0.07 mag is sufficiently
larger than the median error in the $u-g$ color (0.05 mag at $g$=20.5, 0.04 at
$g$=20, 0.02 at $g$=19 and 0.01 at $g$=17.5) to provide a robust measure of the
intrinsic width of the $u-g$ distribution. 

\begin{deluxetable}{ccccc}
\tablenum{2} \tablecolumns{5}
\tablecaption{Best-fit Parameters for the Gaussians Shown in Fig.~\ref{panels3x}}
\tablehead{$g$ range$^a$ &  N$^b$  & $\mu^c$ & $\sigma^d$ & $\mu$ error$^e$}   
\startdata
 15.0--15.5 & 1,087 & 1.095 & 0.078 & 2.4 \\
 15.5--16.0 & 1,605 & 1.085 & 0.076 & 1.9 \\
 16.0--16.5 & 1,911 & 1.082 & 0.070 & 1.6 \\
 16.5--17.0 & 2,328 & 1.078 & 0.070 & 1.5 \\
 17.0--17.5 & 2,590 & 1.075 & 0.064 & 1.3 \\
 18.0--18.5 & 3,348 & 0.899 & 0.050 & 0.9 \\
 18.5--19.0 & 3,745 & 0.901 & 0.055 & 0.9 \\
 19.0--19.5 & 4,504 & 0.895 & 0.055 & 0.8 \\
 19.5--20.0 & 5,893 & 0.891 & 0.060 & 0.8 \\
 20.0--20.5 & 8,712 & 0.886 & 0.075 & 0.8 \\
\enddata
\tablenotetext{a}{The $g$ magnitude range.}
\tablenotetext{b}{The number of stars in the bin.} 
\tablenotetext{c}{The best-fit mean $u-g$ color (only 
data with $u-g>1.0$ are fit in the five brightest bins,
and data with $u-g<0.95$ in the five faintest bins (see 
Figure~\ref{panels3x}).} 
\tablenotetext{d}{The best-fit distribution width.} 
\tablenotetext{e}{The statistical error in the mean (millimag).} 
\end{deluxetable}
 
In addition to an overall blueing of the median $u-g$ color towards the faint
end induced by the varying number ratio of the two components, {\it the median
$u-g$ color for each component also becomes bluer}, as illustrated in
Figure~\ref{panels3x} and summarized in Table 2. We measure gradients of
$-$0.012 mag/mag for the disk component\footnote{This gradient was accounted for
in the definition of the $s$ color by Ivezi\'{c} et al. (2004), but its meaning
was not understood at that time.} ($14.5 < g < 17$) and $-$0.006 mag/mag for the
halo component ($18 < g < 20.5$), with statistical errors of $\sim$0.002
mag/mag. Using an approximate mapping from magnitude to distance, these color
differences could be produced by a gradient of roughly 0.02 mag/kpc between $|Z|$
= 1 kpc and 2.5 kpc for disk stars, and 0.003 mag/kpc between $|Z|$ = 4 kpc and 10
kpc for halo stars. Hence, the color gradient {\it per kpc} is about 7 times
larger for disk stars. 

The detected color gradient cannot be caused by potential errors in the applied
corrections for interstellar extinction (Schlegel, Finkbeiner
\& Davis, 1998). The median value of the $u-g$ reddening correction
is only 0.05 mag, and even the closest stars (at 500 pc) are well beyond the
$\sim$100 pc thick dust layer (J08). Such a gradient ($\sim$0.05 mag between
$u=14$ and $u=19$) could be caused by a non-linearity in the $u$-band
measurements (based on a comparison with independently measured Stetson
standard stars, this effect is ruled out for the $g$-band measurements; Ivezi\'c
et al. 2007). However, a $u$-band non-linearity at the 0.05 mag level is
excluded by {\it in situ} measurements of the hardware response curve, and a
comparison of reductions of SDSS data using several different pipelines
(SExtractor, DAOPhot, and DoPhot; Becker et al. 2007) excludes such a large
software error in the SDSS photometric pipeline. We proceed with an assumption that
this gradient is not a problem in the SDSS data. 

For selected stars with $0.2 < g-r < 0.4$, the $u-g$ color measures metallicity
(see Appendix C), and the observed color scatter and color gradients correspond to 
the metallicity distribution width and metallicity gradients. Because the 
imaging sample is defined by a simple flux limit, these measurements
are relatively easy to interpret. On the contrary, the SDSS spectroscopic 
sample has an extremely biased distribution (by design) in the $g$ vs. $u-g$ 
color-magnitude diagram, as illustrated in the two right panels of
Figure~\ref{panels2}; it would not be easy to derive a robust 
selection function. Using eqs.~\ref{mainPP} and \ref{Zphotom}, we 
find that the color gradients measured for the imaging sample roughly correspond 
to a $\sim$0.06 dex/kpc metallicity gradient for disk stars at $|Z| \sim$1.5 kpc, 
and a $\sim$0.01 dex/kpc for halo stars in the |Z|=4-10 kpc range (given the
distance and sky coordinates, the three-dimensional position in the Galaxy can
be trivially computed). In the remainder of this section we remap the $g$ vs.
$u-g$ color-magnitude diagram to a distance-metallicity diagram, discuss the
metallicity distribution as a function of the position in the Galaxy, develop a
model that captures the data behavior, and correlate the metallicity with the
observed stelar kinematics.

\subsection{ The Bimodal Metallicity Distribution of Thick Disk and Halo Stars  }
\label{metalDistrib}

\begin{figure*}
\plotone{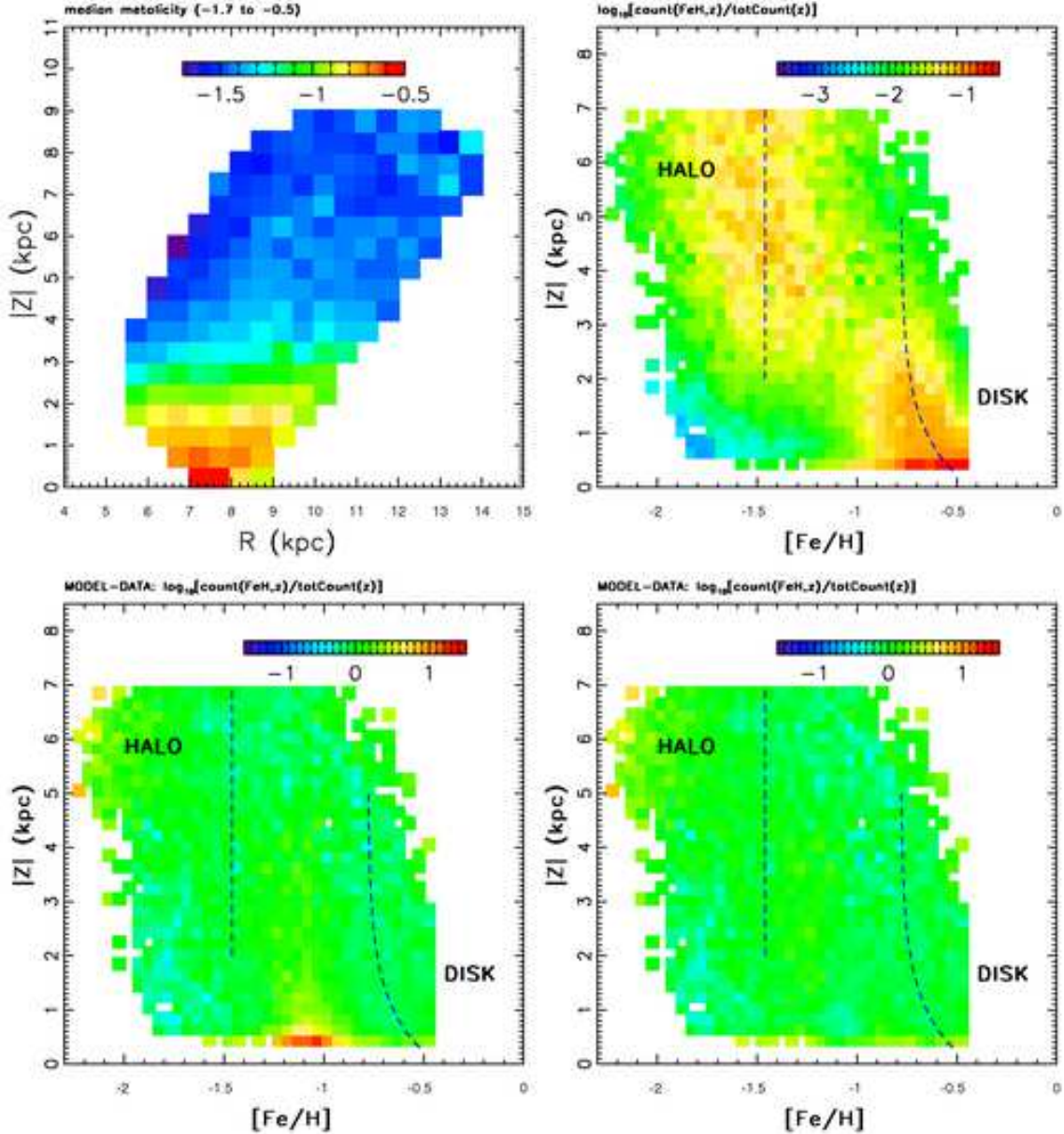}
\caption{
The top left panel illustrates the dependence of the median photometric
metallicity for $\sim$76,000 stars from the SDSS stripe 82 coadded photometric
catalog with $14 < g < 20.5$, $0.2 < g-r < 0.4$ and photometric distance in the
0.5--10 kpc range, on cylindrical Galactic coordinates $R$ and $Z$. Note that
the $Z$ gradient is much larger than the $R$ gradient ($\sim$0.1 dex/kpc vs.
$<$0.01 dex/kpc). The top right panel shows the conditional metallicity
probability distribution at a given distance from the Galactic plane for
$\sim$34,000 stars with $0.5 < Z/{\rm kpc} < 7$ and $7 < R/{\rm kpc} < 9$ (the
probability density is shown on a logarithmic scale, with its integral
normalized to 1). The two concentrations of stars correspond to disk ($[Fe/H]
\sim -0.7$) and halo ($[Fe/H] \sim -1.5$) stars. The difference between this map
and a two-component Gaussian model described in 
\S~\ref{metalDistribModel} is shown in the bottom left panel with the
same dynamic range for color coding as used in the top right panel.
The residual feature visible around $[Fe/H] \sim -1.1$ and $Z\sim1$ kpc
can be modeled either as a third Gaussian component, or by adopting
a non-Gaussian metallicity distribution for the disk component.
The residuals map for the former is shown in the bottom right panel.
}
\label{panels3}
\end{figure*}

Despite its small area, the Stripe 82 catalog covers a substantial range of $R$
and $|Z|$, as shown in the top left panel in Figure~\ref{panels3}. As evident from
the dependence of the median metallicity on $R$ and $|Z|$, the $|Z|$ gradient is
much larger than the radial gradient (by about a factor of 10). Given this large
difference in metallicity gradients, we proceed by making the assumption that
the metallicity distribution is a function of the $Z$ coordinate only (in
\S~\ref{subStruct} below, we critically examine and justify this assumption
using the DR6 catalog). To further minimize the effect of any radial gradient,
we constrain the sample to $\sim$34,000 stars with $7 < R/{\rm kpc} < 9$. 

The top right panel of Figure~\ref{panels3} shows the resulting conditional
metallicity probability distribution for a given $Z$, $p([Fe/H]|Z)$. This
distribution is computed as metallicity histograms in narrow $Z$ slices, and
normalized by the total number of stars in a given slice. Apart from
renormalization and the applied $7 < R/{\rm kpc} < 9$ selection, this is
essentially an upside-down warped version of the $g$ vs. $u-g$ color-magnitude
diagram shown in the top left panel of Figure~\ref{panels2}. The bright red and
faint blue components from Figure~\ref{panels2} are now readily identifiable as
the relatively close metal-rich disk component, and the more distant metal-poor halo
component, respectively.

\subsection{ A Simple Model for the  Conditional Metallicity Probability Distribution}
\label{metalDistribModel}

\begin{figure}
\plotone{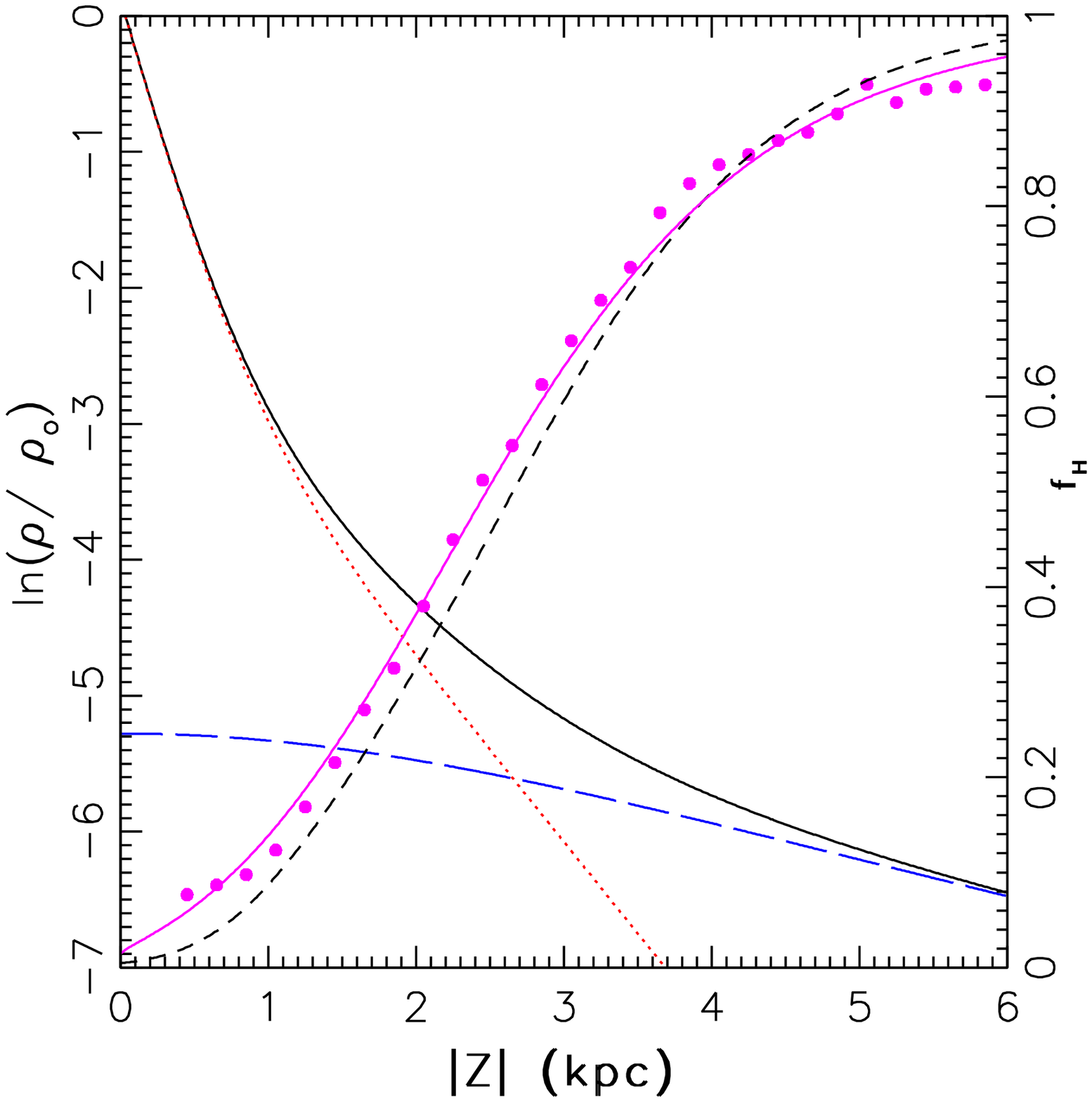}
\caption{
The symbols show the number ratio of stars with $[Fe/H]<-1$ relative to those with
$[Fe/H]>-1$, for stars with 7 kpc $<R<$ 9 kpc (with the ratio labeled on the
right y axis). Beyond $Z\sim2.5$ kpc, metal-poor stars dominate the counts. The
solid line passing through the symbols is a three-parameter best fit used in
modeling the conditional metallicity probability distribution (equal to
$1/(1+70\eta)$, with $\eta=\exp(-|Z/240 {\rm pc}|^{0.62})$). The short-dashed
line is a prediction for the halo-to-disk counts ratio based on a best-fit
Galaxy model to stellar counts from Juri\'{c} et al. (2008). The model includes
an oblate power-law halo, and exponential thin and thick disks (see
\S~\ref{J08}). The disk contribution to the counts is shown by the dotted line (with
the ln(counts) labeled on the left y axis), and the long-dashed line shows the
halo contribution. The sum of disk and halo contributions is shown by the solid
line.}
\label{Zcounts}
\end{figure}

As is evident from the $p([Fe/H]|Z)$ map shown in the top right panel of Figure~\ref{panels3}, 
the $Z$ gradient of the median metallicity map shown in the top left panel of 
Figure~\ref{panels3} is due to the varying contributions of a metal-rich disk 
component and a metal-poor halo component. We first attempt to model 
the $p([Fe/H]|Z)$ map using two Gaussian components with a $Z$-dependent
ratio of their area (components' number ratio)
\begin{eqnarray}
\label{metModel}
p(x=[Fe/H]|Z) = (1-f_H) & & G(x|\mu_D, \sigma_D) \\ \nonumber 
	+ f_H & & G(x|\mu_H, \sigma_H),
\end{eqnarray}
where
\begin{equation}
   G(x|\mu,\sigma) = {1 \over \sqrt{2 \pi} \sigma} {\rm e}^{- (x-\mu)^2 \over 2 \sigma^2}.
\end{equation}
The distribution width for both components can be modeled as spatially
invariant, $\sigma_D=0.16$ dex, and $\sigma_H=0.30$ dex, as is the case for 
the median halo metallicity, $\mu_H=-1.46$. The median and dispersion for 
metallicity distribution of halo stars is in good agreement with previous 
work (e.g., Ryan \& Norris 1991). Due to the small errors in the $u-g$ color 
for the coadded data, the contribution of measurement errors to $\sigma_D$ and 
$\sigma_H$ is very small: the implied intrinsic widths are 0.16 dex and 
0.29 dex, respectively. 

Inspection of the $p([Fe/H]|Z)$ map suggests that the variation of the median 
metallicity for the disk component\footnote{An obvious question is whether the 
observed variation of the median metallicity for the disk component simply reflects the varying 
contributions of thin- and thick-disk stars. This question in addressed in 
detail in \S~\ref{TvsT}.} in the $0.5 < |Z|/{\rm kpc} < 5$ range can be
described as 
\begin{equation}
\label{muD}
    \mu_D(Z) = \mu_\infty + \Delta_\mu \,\exp(-|Z|/H_\mu) \,\, {\rm dex},
\end{equation}
with the best-fit values $H_\mu=1.0$ kpc, $\mu_\infty=-0.78$ and
$\Delta_\mu=0.35$. The best fit is shown by the curved dashed line in the top 
right panel in Figure~\ref{panels3}. The exponential ``height'', $H_\mu$, 
is constrained to only about 50\% due to covariances with $\mu_\infty$ and
$\Delta_\mu$ (which are constrained to about 0.05 dex). The implied median 
metallicity values agree well with a value of $-0.7$ obtained by Gilmore 
\& Wyse (1985) (they did not detect a metallicity gradient). 

The best-fit $\mu_D(Z)$ given by eq.~\ref{muD} is valid only for $|Z|>500$ pc 
because of the sample bright limit. Close to the plane, the mean and 
rms scatter of the metallicity distribution are $-0.14$ and 0.19 for F/G- 
type dwarfs (Nordstr\"{o}m et al. 2004; Allende Prieto et al. 2004), and 
$-0.12$ and 0.18 for
K-type giants (Girardi \& Salaris 2001), respectively.  Hence, the vertical metallicity 
gradient close to the plane must be larger than $\sim0.35$ dex/kpc implied
by the {\it extrapolation} of eq.~\ref{muD}  (because stars on average become 
more metal poor by about 0.5 dex between $Z=0$ and $|Z|=1$ kpc).

To set the relative normalization of the two Gaussians, $f_H(Z)$, we 
approximately separate halo and disk components by isolating stars
with $[Fe/H] < -1.1$ and $[Fe/H] > -0.9$, respectively. A good description of 
the data, shown by symbols in Figure~\ref{Zcounts}, is provided by a best-fit 
function with three free parameters
\begin{equation}
\label{fH}
          f_H(Z) = {1 \over 1 + a \exp[-(|Z|/b)^c]},
\end{equation}
with $a=70$, $b=240$ pc, and $c=0.62$. We discuss this function further
in \S~\ref{J08}.

\subsubsection{The ``metal-weak thick disk'': a third Gaussian component or 
                                         a non-Gaussian Distribution? }
\label{thirdGauss}

\begin{figure*}
\plotone{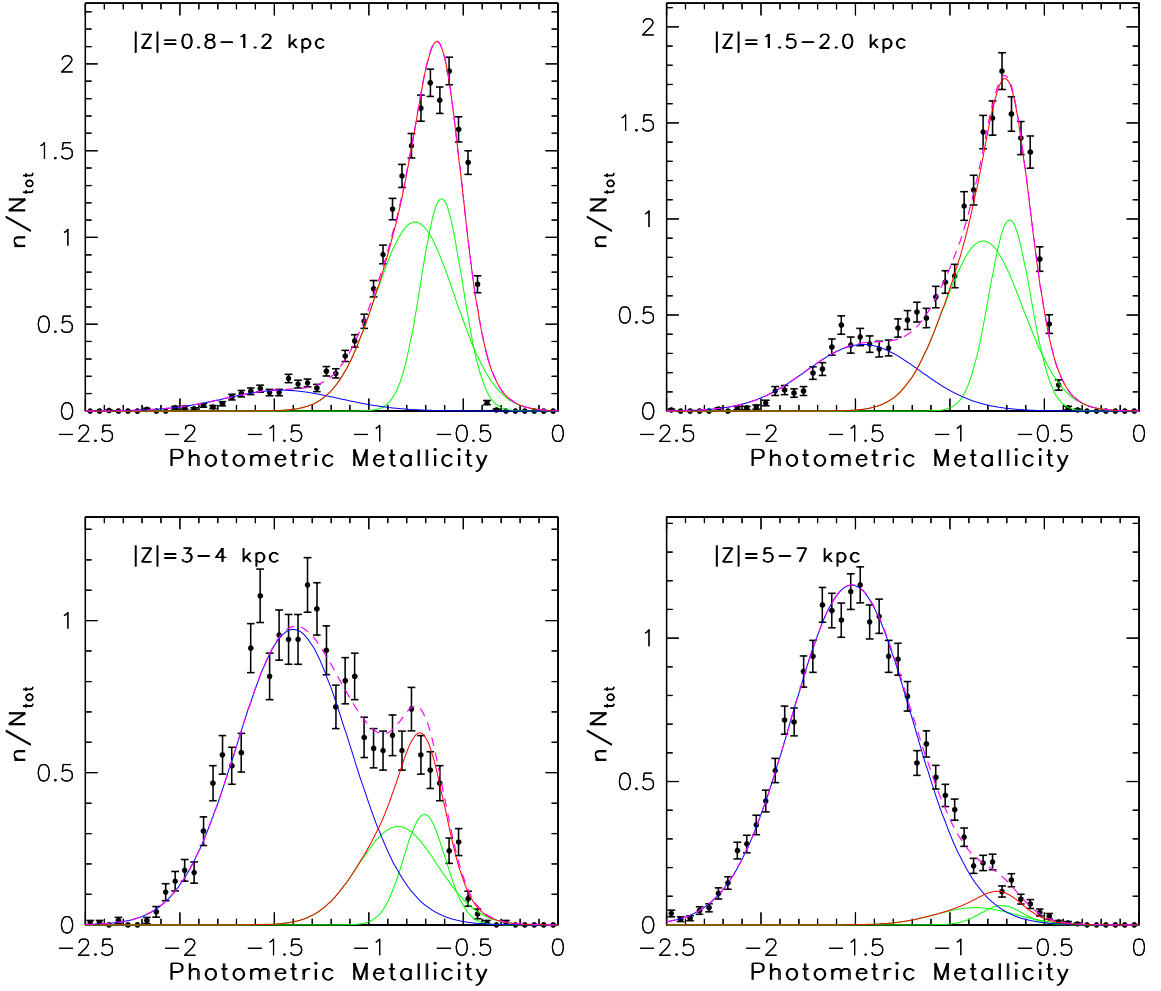}
\caption{
The symbols with error bars show the measured photometric metallicity distribution 
for stars with $0.2 < g-r < 0.4$, 7 kpc $<R<$ 9 kpc, and distance from 
the Galactic plane in the range 0.8--1.2 kpc (top left, $\sim$6,200 stars), 
1.5--2.0 kpc (top right, $\sim$3,800 stars), 3.0--4.0 kpc (bottom left,
$\sim$2,800 stars) and 5.0--7.0 kpc (bottom right, $\sim$6,000 stars).
The histograms are essentially horizontal (parallel to $x$ axis) slices 
at corresponding $|Z|$ intervals through the map shown in the top right panel 
of Figure~\ref{panels3}. The dashed magenta lines show a best-fit 
two-component, halo plus disk, model. The blue lines show the halo
contribution, modeled as a single Gaussian, and the red lines show the 
contribution of a non-Gaussian disk model, modeled as a sum of two 
Gaussians shown by the green lines. See \S~\ref{thirdGauss} and Table 3 
for the best-fit parameters.}
\label{Zhist}
\end{figure*}

The difference between the data and a two-Gaussian model described above is
shown in the bottom left panel of Figure~\ref{panels3}. As is evident, the
overall behavior of the two dominant components is captured, but the residual
map reveals a feature that contains intermediate-metallicity stars
($[Fe/H]\sim-1.0$) within $\sim$2-3 kpc from the plane. This feature includes
about 5\% of stars in the sample and is reminiscent of the so-called
``metal-weak thick disk'' (Morrison, Flynn \& Freeman 1990; Chiba \& Beers 2000;
Beers et al. 2002). Indeed, it can be satisfactory modeled as a third Gaussian
component with $\mu=-1.0$, $\sigma=0.10$ dex and a strength of 20\% relative to
the metal-rich component, as illustrated by the smooth residual map shown in the
bottom right panel of Figure~\ref{panels3}. 

An alternative to postulating a third Gaussian component for eq.~\ref{metModel}
is to adopt a skewed metallicity distribution for the disk component whose shape
need not vary with the distance from the plane (i.e., replacing $G(x|\mu_D,
\sigma_D)$ with a non-Gaussian distribution). A skewed shape for the metallicity
distribution of local F/G dwarfs was also measured by Gilmore, Wyse \& Jones
(1995), but with an overall offset of $\sim$0.5 dex towards higher metallicity,
as would be expected compared to our data at $|Z|=1$ kpc. 

\begin{deluxetable*}{cccccccc}
\tablenum{3} \tablecolumns{8} \tablewidth{5in}
\tablecaption{Best-fit Parameters for $p([Fe/H]|Z)$, Shown in Fig.~\ref{Zhist}}
\tablehead{$|Z|$ range$^a$ & $\langle |Z| \rangle^b$ &  N$^c$ & $f_H^d$ & $f_H^e$ & $f_H^f$  & $\mu_H^g$ & $\sigma_H^h$}   
\startdata
   0.8--1.2   &  0.98 & 6,187  &  0.08  &  0.14  &  0.09  & -1.46 &  0.30 \\
   1.5--2.0   &  1.72 & 3,842  &  0.24  &  0.30  &  0.26  & -1.46 &  0.30 \\
   3.0--4.0   &  3.47 & 2,792  &  0.71  &  0.73  &  0.73  & -1.40 &  0.30 \\
   5.0--7.0   &  5.79 & 6,025  &  0.97  &  0.95  &  0.95  & -1.52 &  0.32 \\
\enddata
\tablenotetext{a}{The $|Z|$ range for each bin (kpc).}
\tablenotetext{b}{The median $|Z|$ in each bin (kpc).}
\tablenotetext{c}{The number of stars in the corresponding bin.} 
\tablenotetext{d}{The halo-to-disk number ratio predicted by the J08 best-fit model (see \S~\ref{J08}).} 
\tablenotetext{e}{The halo-to-disk number ratio predicted by eq.~\ref{fH}.} 
\tablenotetext{f}{The best-fit halo-to-disk number ratio. The halo parameters 
                  are only weakly constrained in the first bin.} 
\tablenotetext{g}{The best-fit mean metallicity for the halo component.} 
\tablenotetext{h}{The best-fit distribution width for the halo component.} 
\end{deluxetable*}

A quantitative validation of such a universal shape for the disk metallicity
distribution is shown in Figure~\ref{Zhist}. In each $Z$ slice that shows
evidence for the disk component, the shape of its skewed metallicity distribution
can be modeled\footnote{We attempted to fit this skewed distribution using a
log-normal distribution, but the detailed shape could not be reproduced.} as a
sum of two Gaussians with fixed positions relative to each other (offset of 0.14
dex), fixed widths (0.21 dex for the more metal-poor component, and 0.11 dex for
the other), and fixed relative normalization (1.7-to-1 number ratio in favor of
the more metal-poor component). The values of these four parameters were
obtained by {\it simultaneously} fitting the four histograms shown in
Figure~\ref{Zhist}, with the position of the disk distribution fixed at values
computed from eq.~\ref{muD} (the difference between the median given by
eq.~\ref{muD} and the mean for the more metal-poor disk Gaussian is 0.067 dex).
The halo parameter $f_H$ was initially estimated using eq.~\ref{fH}, while
$\mu_H=-1.46$ and $\sigma_H=0.30$ were kept constant. A few minor adjustments to
these three parameters, listed in Table 3, improved the fits. The most notable
change is a shift of the halo metallicity distribution by $\pm$0.06 dex for
$Z>3$ kpc.

The best-fit values of $f_H$ from Table 3 are consistent with eq.~\ref{fH}. By
retaining that function, and adopting the above $Z$-independent two-Gaussian
shape description for the disk component, we obtain a residual map that is
indistinguishable from that obtained using a third Gaussian component. Hence, we
conclude that either hypothesis can explain the data. While both are formally
based on three Gaussian components, the ``universal shape'' hypothesis
demonstrates that the data do not require the second disk Gaussian to be
independent of the first.

It is tempting to identify the two Gaussians that describe the disk component
with the thin- and thick-disk contributions. However, the fits presented above
are inconsistent with this interpretation. The double-exponential best-fit to
observed spatial profile (with scale heights of 245 pc and 743 pc, and a
relative normalization of 0.13, see \S~\ref{J08} and Figure~\ref{Zcounts}) imply
that the ratio of thick-to-thin disk stars should strongly vary from 1.9 in the
$Z=0.8-1.2$ kpc bin to 14 in the $Z=1.5-2.0$ kpc bin, and $>1000$ in the
$Z=3.0-4.0$ kpc bin. Yet, the metallicity distributions admit a fit based on a
{\it constant normalization ratio} for the two Gaussian components. Of course,
this fitting success alone does not necessarily prove that traditional
decomposition into two fixed distributions with a varying normalization ratio is
inconsistent with the data. We return to the problem of distinguishing these two
hypotheses, which have very different implications for galaxy formation and
evolution theories, when discussing correlations with kinematics further below
(\S~\ref{TvsT}).

\subsubsection{The Effects of Systematic Errors on the Photometric Parallax Relation} 
\label{sysErr}

Various systematic errors in the metallicity and distance estimates affect the
best-fit models for the metallicity distribution described above. The main
sources of systematic errors in the adopted photometric parallax relation are
binarity effects (age effects can contribute at most 0.2 mag systematic
uncertainty in absolute magnitude at the median color of the sample, $g-r=0.3$,
and $\sim$0.1 mag at $g-r=0.4$; see Appendix A). As discussed in detail by J08,
binarity effects are expected to decrease the inferred distances and disk
exponential scale height by about 15\%. The impact of binarity on the metallicity
determination is hard to estimate without detailed knowledge of both the incidence
of binaries and their mass ratio distributions. By performing simulations
similar to those described by J08, we find that the worst-case scenario is a
system consisting of a luminous low-metallicity primary with $u-g=0.8$ and
$g-r=0.2$, and a secondary with $u-g \sim 1.0$ and $g-r=0.5$ (the redder
secondaries are too faint to have a larger impact). When such a system is (mis)
interpreted as a single star, the distance is underestimated by 15\%, the
effective temperature is underestimated by $\sim$240 K, and the metallicity is
overestimated by $\sim$0.2 dex. For components that have mass ratios closer
to unity (as suggested by, e.g., Reid et al. 2002), the bias in metallicity will
be much smaller, while the bias in distance estimates can increase up to
$\sim$40\%.

\subsubsection{The Edge of the Thick Disk at $|Z|\sim5$ kpc? }

Using photographic data for 250 stars, Majewski (1992) advocated an ``edge'' of
the Galactic thick disk at about 5.5 kpc above the Galactic plane. Indeed, the
map of conditional metallicity distribution shown in the top right panel of
Figure~\ref{panels3} suggests a tantalizing possibility that the metal-rich
component does not extend beyond 4--5 kpc from the plane. This visual impression
is addressed quantitatively by the histogram shown in the bottom right panel of
Figure~\ref{Zhist}. The extrapolation of the disk component to the 5--7 kpc bin
predicts that the disk should be detectable as a factor of $\sim$2 excess around
$[Fe/H]\sim-0.8$, on top of the underlying halo distribution. Such an excess
seems consistent with the data, and argues against a cutoff in the distribution
of disk stars. Due to the small sample size ($\sim$4,800 stars), the noise is
large and the significance of this excess is not overwhelming. Even when the $7
< R/{\rm kpc} < 9$ constraint is removed, the $5 < |Z|/{\rm kpc} < 7$ subsample
is still too small ($\sim$8,600 stars) to significantly overcome counting noise.

We conclude that Stripe 82 catalog is not sufficiently large to convincingly
demonstrate the lack of an edge in the distribution of disk stars. We improve
the statistical power of this analysis by using a $\sim$30 times larger sky
coverage provided by the DR6 catalog, as discussed below.

\subsubsection{ A comparison with results from Juri\'{c} et al. (2008) }
\label{J08} 

The expression given by eq.~\ref{fH} is only a convenient fit that involves a
small number of free parameters. The halo-to-disk number ratio, $f_H$, depends
on the disk and halo spatial profiles, both of which are usually modeled using
numerous free parameters. As discussed by J08, a unique solution is not possible
without a large sky coverage, and even in such a case it is difficult to obtain.
Nevertheless, we can test whether the data for $f_H$ discussed here are {\it
consistent} with the best-fit spatial profiles obtained by J08, {\it which do not
incorporate metallicity information}. 

J08 show that the stellar number density distribution, $\rho(R,Z,\phi)$ 
can be well described (apart from local overdensities) as a sum of two 
cylindrically symmetric components
\begin{equation} 
      \rho(R,Z,\phi) = \rho_D(R,Z) + \rho_H(R,Z).
\end{equation}
The disk component can be modeled as a sum of two exponential disks
\begin{eqnarray} 
\rho_D(R,Z)&=&\rho_D(R_\odot,0) [{\rm e}^{-|Z+Z_\odot|/H_1-(R-R_\odot)/L_1} \\ \nonumber
                    & &   + \epsilon_D {\rm e}^{-|Z+Z_\odot|/H_2-(R-R_\odot)/L_2} ],
\end{eqnarray}
while the halo component requires an oblate power-law model
\begin{equation} 
 \rho_H(R,Z)= \rho_D(R_\odot,0)\,\epsilon_H\, \left({R_\odot^2 \over R^2 + (Z/q_H)^2}\right)^{n_H/2}.
\end{equation} 
The best-fit parameters are discussed in detail by J08. We have adopted
the following values for parameters relevant in this work (second column in 
Table 10 from J08): $Z_\odot$=25 pc, $H_1=245$ pc, $H_2=743$ pc, 
$\epsilon_D=0.13$, $\epsilon_H=0.0051$, $q_H=0.64$, and $n_H=2.77$.

The fraction of halo stars implied by this model, 
\begin{equation} 
      f_H(R,Z) = {\rho_H(R,Z) \over  \rho_D(R,Z) + \rho_H(R,Z)},
\end{equation} 
agrees reasonably well with $f_H$ determined here (see Figure~\ref{Zcounts})
{\it without any modification}. Given that J08 determined their 
best-fit parameters using counts at $Z\ga10$ kpc and that the adopted photometric parallax
relations are somewhat different, this agreement is remarkable. The
short-dashed line representing the J08 model in
Figure~\ref{Zcounts} can be brought into essentially perfect agreement 
with the data points by decreasing $H_1$ and $H_2$ by 4\% of
their values (i.e., by much less than their uncertainty, 20\%, quoted by J08). 
Alternatively, data points can be moved closer to the J08 model by increasing
halo normalization by 24\%, to $\epsilon_H=0.0063$ ($\sim 1\sigma$ change). 
Hence, the results presented here validate the best-fit model from J08.

The best-fit model for stellar counts provides the useful guidance that the counts
of thin- and thick-disk stars become equal around $|Z|\sim1$ kpc, and that the
counts of disk and halo stars become equal around $|Z|\sim2.5$ kpc. At the
bright end of the sample (500 pc), thin-disk stars contribute $\sim$70\% of
stars to the disk component (with a halo contribution of $\sim$1\%); at the
faint end (5 kpc), halo stars contribute $\sim$90\% of the sample. Hence, the
SDSS imaging data are well suited for studying the metallicity distribution all the
way from thin disk to halo, through the thick-disk transition, using an essentially
complete flux-limited sample of numerous main-sequence stars.

\subsection{    Analysis of the Large Area SDSS Data Release 6 Sample  }

\begin{figure*}
\plotone{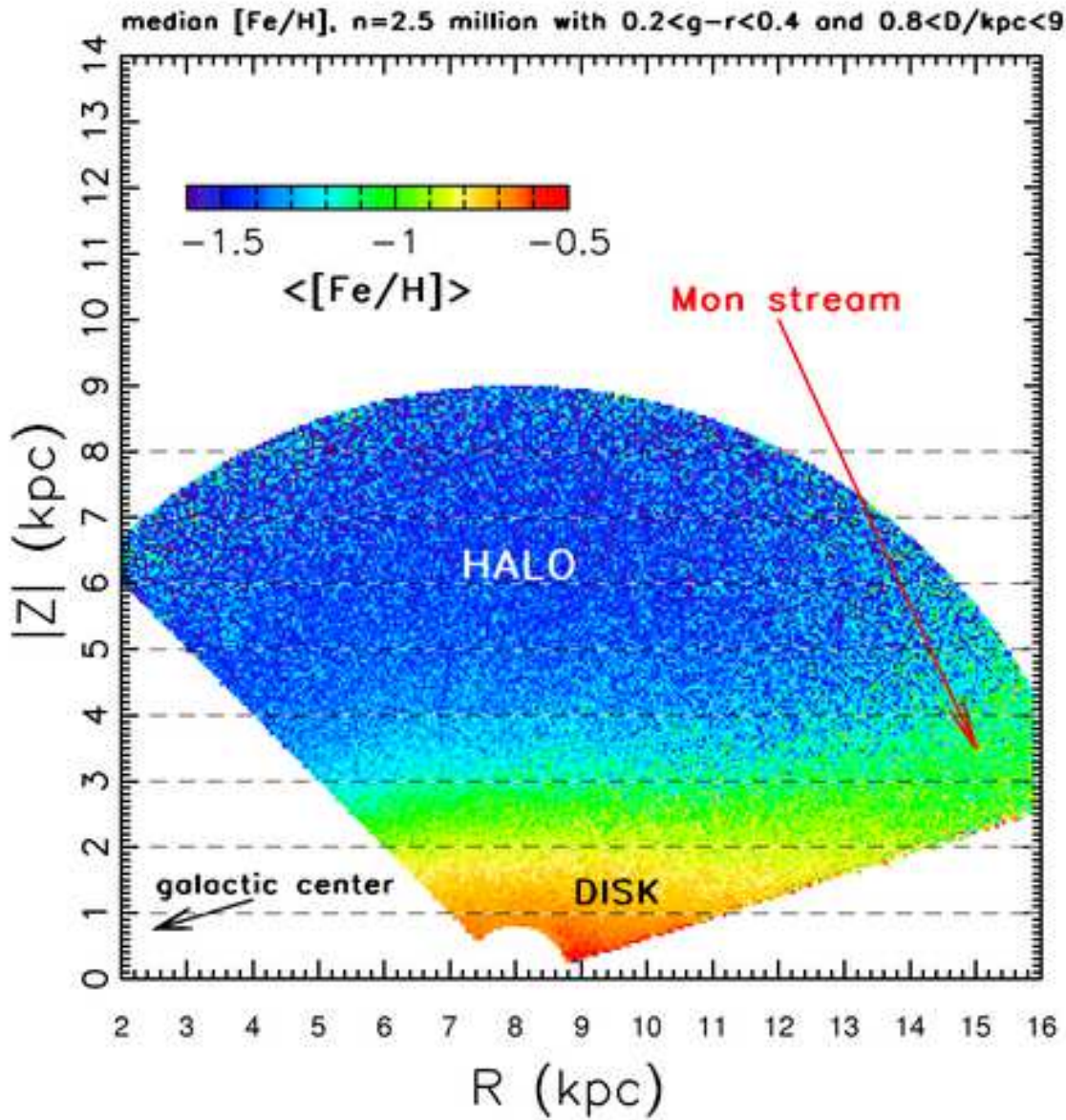}
\caption{The dependence of the median photometric metallicity for $\sim$2.5
million stars from SDSS Data Release 6 with $14.5 < r < 20$, $0.2 < g-r < 0.4$,
and photometric distance in the 0.8--9 kpc range, in cylindrical
Galactic coordinates $R$ and $|Z|$ (compare to the top left panel of
Figure~\ref{panels3}). There are $\sim$40,000 pixels (50 pc $\times$ 50 pc)
contained in this map, with a minimum of 5 stars per pixel and a median of 33
stars. Note that the gradient of the median metallicity is essentially parallel
to the $|Z|$ axis, except in the Monoceros stream region, as marked. }
\label{pretty}
\end{figure*}

\begin{figure*}
\plotone{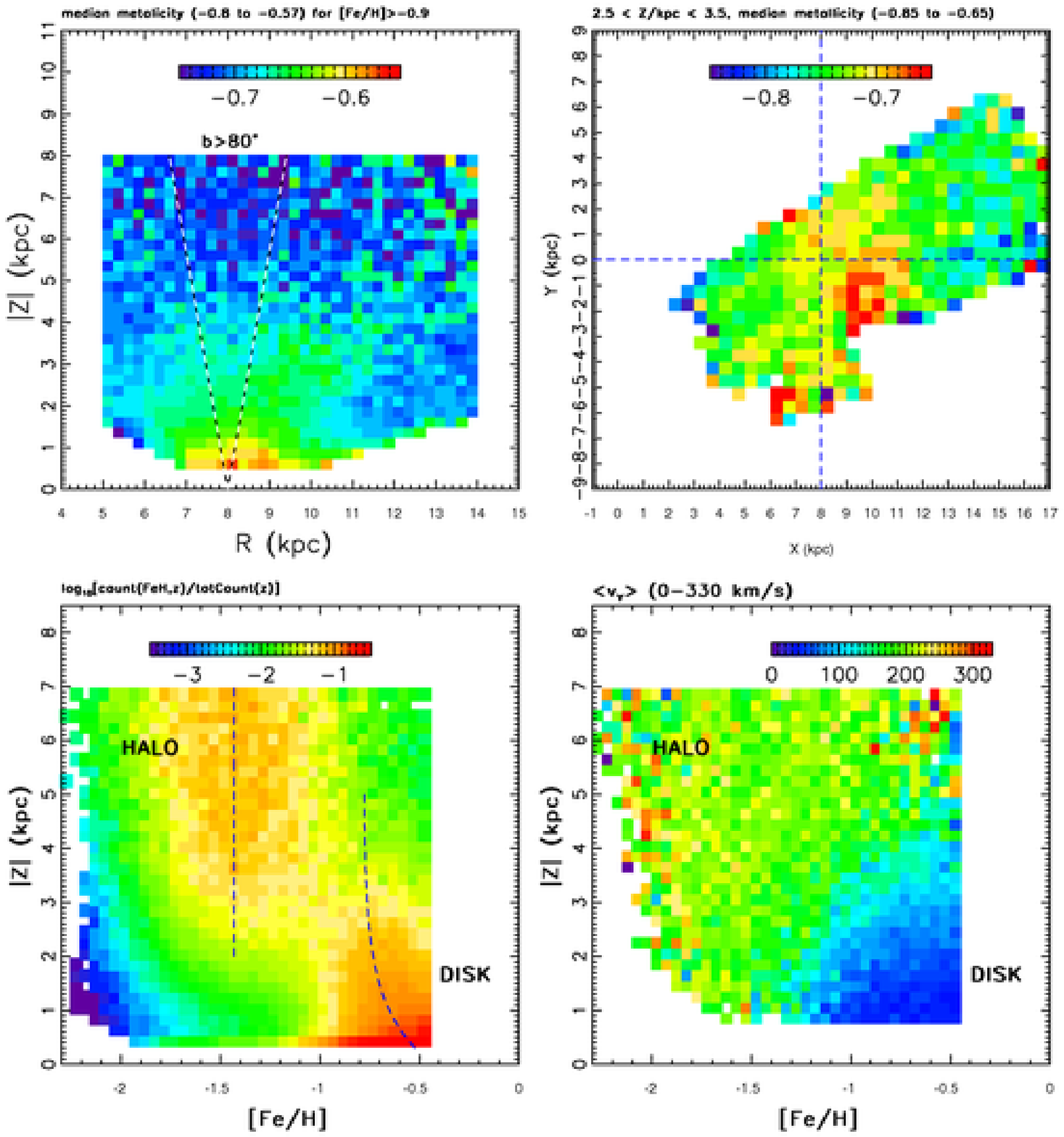}
\caption{
The top left panel is analogous to Figure~\ref{pretty}, except that only 1.1
million stars with $[Fe/H]>-0.9$ are used to compute the median $[Fe/H]$ (and
the display stretch is harder, as indicated in the panel). Note a coherent feature
at $R\sim10$ kpc and $Z\sim3$ kpc. Its extent parallel to the Galactic plane is
shown in the top right panel, which shows the median metallicity for stars with
$2.5 < Z/{\rm kpc} < 3.5$ (note the shifted color map). The bottom left panel
shows the conditional metallicity probability distribution for $\sim$300,000
stars from a cylinder perpendicular to the Galactic plane, centered on the Sun,
and with a radius of 1 kpc. The dashed lines are the same fiducials as shown in
the top right panel of Figure~\ref{panels3}. The bottom right panel shows the
median $v_Y$ velocity component (heliocentric; the value of $\sim$220 km/s
corresponds to no rotation) as a function of metallicity and distance from the
Galactic plane for $\sim$41,000 stars with $Z<7$ kpc and $b>80^\circ$ (see the
dashed lines in the top left panel). Note the coherent feature with slightly
larger $\langle v_Y \rangle$ (by $\sim$ 20-30 km/s) at $[Fe/H]\sim -0.6$ and
$Z\sim 6.5$ kpc.}
\label{panels6}
\end{figure*}

SDSS Data Release 6 provides photometry for $\sim$300 million objects detected
over $\sim$10,000 sq.deg. of sky. Using selection criteria listed in
\S~\ref{sample}, and additionally requiring $0.2<g-r<0.4$ and distances in the
0.5-12 kpc range, we selected 2.53 million stars (the extended $0.2<g-r<0.6$
color range includes 5.7 million stars). The significantly increased sky
coverage comes at the expense of photometric metallicity precision: for
metal-rich disk stars random errors increase from 0.05 dex at a distance of 1
kpc to 0.12 dex at 4 kpc, and for metal-poor halo stars they increase from 0.20
dex at 4 kpc to 0.36 dex at 7 kpc (about 3 times as large as for the coadded data
discussed above).

The median metallicity map shown in Figure~\ref{pretty} is analogous to that
shown in Figure~\ref{panels3} except for the significantly larger sky coverage.
The conclusion about the vertical metallicity gradient being much stronger than
the radial gradient remains valid. The strong $Z$ gradient is a result of the
low-metallicity halo component becoming dominant at $Z$ beyond $\sim$3 kpc. The
only deviation from this overall trend is seen in the region associated with the
Monoceros Stream (see \S~\ref{metalMon} for a detailed discussion). 

An analogous map constructed using only stars with $[Fe/H]>-0.9$ is shown in 
the top left panel of Figure~\ref{panels6}. While it also displays a much
stronger gradient in $Z$ direction, a local maximum with an amplitude of
$\sim$0.1 dex is discernible at $R\sim10$ kpc and at $\sim$2.5-3.5 kpc above the
plane. The $X-Y$ median metallicity map for this $Z$ slice, shown in the top
right panel of Figure~\ref{panels6}, places this maximum at $X\sim10$ kpc and
$Y\sim-2$ kpc. We address such localized metallicity inhomogeneities further
below (\S~\ref{subStruct}). 

The conditional metallicity distribution for DR6 stars within from a solar cylinder 
having a radius of 1 kpc is shown in the bottom left panel of Figure~\ref{panels6}.  
As evident, there is a close resemblance between the conditional 
metallicity distribution map constructed using the Stripe 82 catalog
(Figure~\ref{panels3}) and the map based on the DR6 catalog. The best-fit 
parameters listed in Table 3 that describe the histograms shown in 
Figure~\ref{Zhist} are consistent with the DR6 map, when the increased 
measurement errors are taken into account (disk and halo components are
convolved with 0.10 dex and 0.20 dex wide Gaussians, respectively).

\subsubsection{The Edge of the Thick Disk Revisited}
\label{diskEdge}

\begin{figure}
\plotone{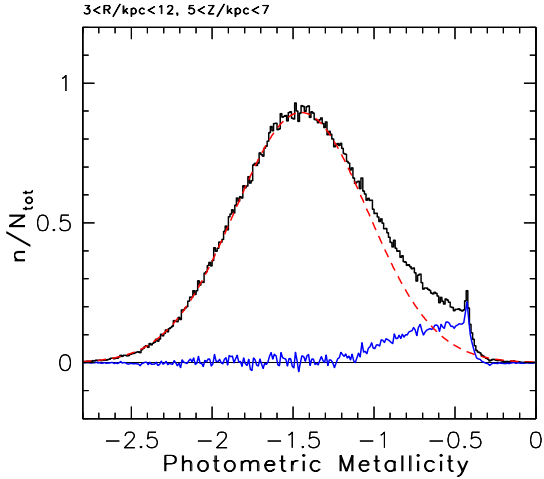}
\caption{The black histogram shows the photometric metallicity
distribution for $\sim$295,000 stars with $3 < R/{\rm kpc} < 12$ and $5 < |Z|/{\rm
kpc} < 7$ (compare to the bottom right panel in Figure~\ref{Zhist}). The dashed
red line is a Gaussian with a width of 0.41 dex (implying an intrinsic width of
0.32 dex), centered on $[Fe/H]=-1.45$, and an area of 0.92. The difference
between the observed distribution and this Gaussian is shown by the blue line,
and presumably corresponds to disk stars. The small peak around $[Fe/H]=-0.5$ is
likely an artifact of the photometric metallicity estimation method. }
\label{diskCutoff}
\end{figure}

The large number of stars in the DR6 sample enables a statistically robust analysis
of the suspected cutoff in the distribution of thick-disk stars around
$|Z|\sim5$ kpc. Figure~\ref{diskCutoff} shows the metallicity distribution of
$\sim$295,000 stars with $3 < R/{\rm kpc} < 12$ and $5 < Z/{\rm kpc} < 7$. When
a best-fit Gaussian is subtracted from the observed distribution, a highly
significant residual at $[Fe/H]>-1$ remains. This residual feature contains
about 8\% of the stars ($\sim$24,000, so counting noise is not an issue) and
suggests that, even so far away from the plane, stars with $[Fe/H]>-0.7$ are
dominated by disk stars (the disk-to-halo number ratio is about 2:1). The value
of 8\% is in excellent agreement with the model discussed in \S~\ref{J08} (8.8\%
at $Z=6$ kpc). An alternative interpretation is that halo metallicity
distribution is not Gaussian, but skewed towards high $[Fe/H]>-1$ values, with a
disk cutoff at $|Z|\la5$ kpc. Given the remarkable agreement with a Gaussian for
$[Fe/H] < -1.1$, it seems more likely that disk is indeed traceable to beyond 5
kpc from the plane.

The small peak in the observed metallicity distribution visible around
$[Fe/H]=-0.5$ is probably an artifact of the photometric metallicity estimator.
As discussed in \S~\ref{sample} and \S~\ref{metalSec}, stars with $[Fe/H] >
-0.5$ may be biased towards somewhat lower metallicity values, which may explain the
observed (minor) effect.

\subsubsection{The Metallicity--Kinematics Maps for Stars around the North Galactic Pole} 

\begin{figure*}
\plotone{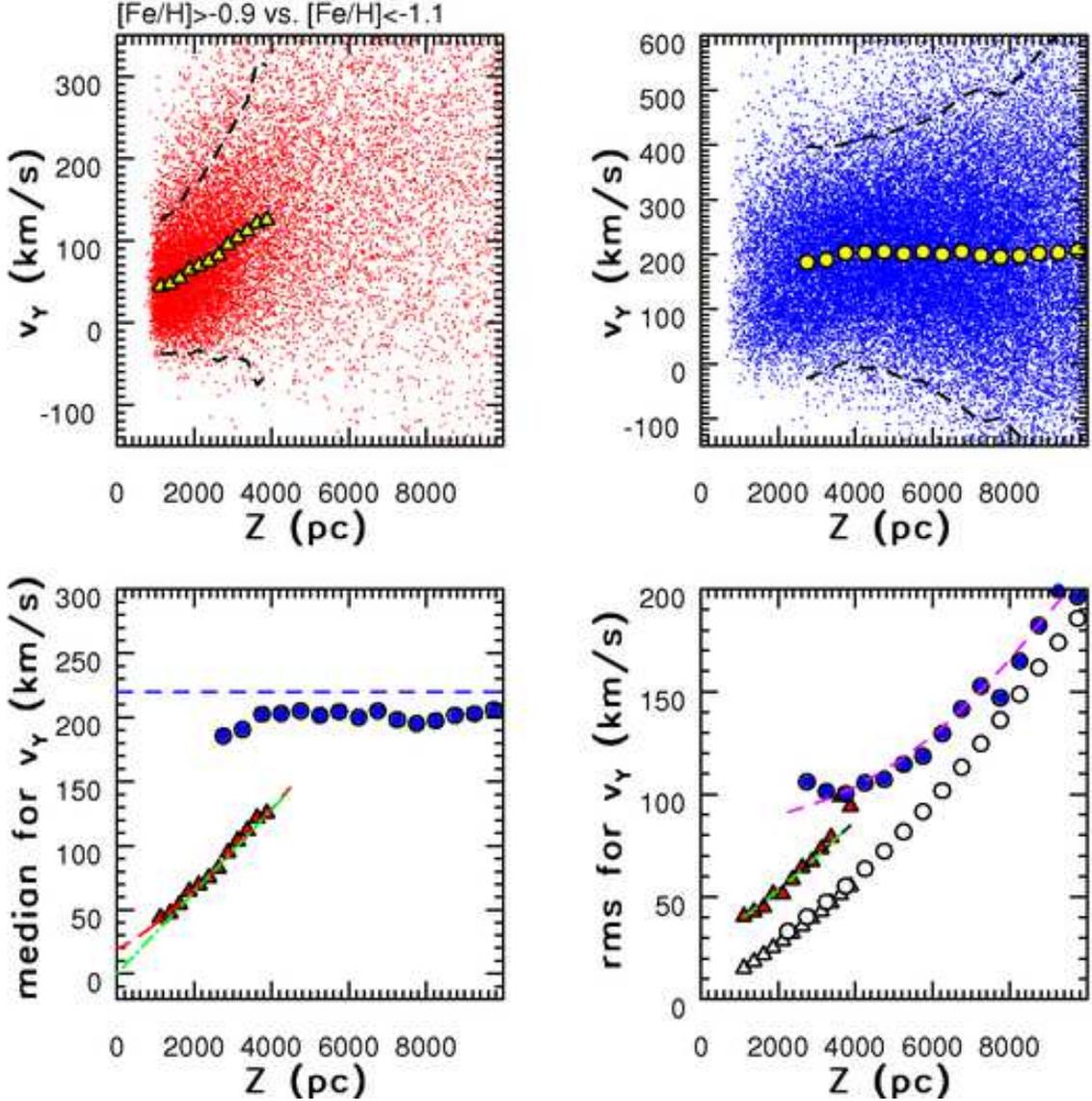}
\caption{
The dependence of $v_Y$ (heliocentric rotational velocity component) on distance from 
the plane, $|Z|$, for $\sim18,000$ high-metallicity ($[Fe/H]>-0.9$, top left) and 
$\sim40,000$ low-metallicity ($[Fe/H]<-1.1$, top right) stars with $b>80^\circ$. 
In the top two panels individual stars are shown by small dots, and the medians 
in bins of $Z$ are shown by large symbols. The 2$\sigma$ envelope around the 
medians is shown by dashed lines. The bottom two panels compare the medians 
(left) and dispersion (right) for the two subsamples shown in the top panels
(dots for low-metallicity and triangles for high-metallicity subsamples). 
The horizontal dashed line at $v_Y$=220 km/s in the bottom left panel is
added to guide the eye. The other two dashed lines in the bottom left panel 
are best fits to the observed $v_Y(|Z|)$ for high-metallicity stars (see text).
The open symbols in the bottom right panel show the median random velocity
measurement errors (circles for low-metallicity and triangles for
high-metallicity subsamples),
and the two dashed lines are best fits to the observed increase of velocity 
dispersion with $|Z|$. They assume that the intrinsic halo velocity dispersion is 
constant, $\sigma_Y$=90 km/s, and that the intrinsic disk velocity dispersion 
increases as $\sigma_Y=26+10\,|Z|/{\rm kpc}$ (km/s).}
\label{FigVvsZ}
\end{figure*}

Kinematic measurements can offer additional insight into Galactic components
revealed by the metallicity distribution. The sky coverage of the DR6 catalog
includes the anticenter ($l\sim180^\circ$, AC hereafter) and north Galactic pole
(NGP hereafter) regions, where proper motion measurements provide a robust
constraint on the rotational velocity component even without knowledge of
radial velocity. We take proper motion measurements from the Munn et al. (2004)
catalog based on astrometric measurements from SDSS and a collection of Schmidt
photographic surveys. {\it The proper motion measurements publicly available as
a part of SDSS DR6 are known to have significant systematic errors (Munn et al.,
in prep.). Here we use a revised set of proper motion measurements which will
become publicly available as a part of SDSS DR7.}

Despite the sizable random and systematic astrometric errors in the Schmidt
surveys, the combination of a long baseline ($\sim$$50$~years for POSS-I
survey), and a recalibration of the photographic data using positions of SDSS
galaxies, results in median random errors for proper motions of only
$\sim3$~mas~yr$^{-1}$ (per component) for $g<19.5$. Systematic errors are
typically an order of magnitude smaller, as robustly determined by B08 using
$\sim$80,000 spectroscopically confirmed SDSS quasars from Schneider et al.
(2007). At a distance of 1 kpc, a random error of 3~mas~yr$^{-1}$ corresponds to
a velocity error of $\sim$15 km/s, which is comparable to the radial velocity
accuracy delivered by the SDSS stellar spectroscopic survey. At a distance of 7
kpc, a random error of 3~mas~yr$^{-1}$ corresponds to a velocity error of 100
km/s, which still represents a usable measurement for large samples, given that
the systematic errors are much smaller ($\sim$10 km/s at a distance of 7 kpc). The
faint limit of this catalog ($g\sim20$) is well matched to the depth of the SDSS
photometric metallicity sample, and thus proper motion measurements are
available for more than 90\% of the $0.2 < g-r< 0.4$ sample. The kinematics of
the SDSS stellar sample, including the mutual consistency of kinematics based on
radial velocity and proper motion measurements, are discussed in detail by B08.
Here we briefly present a few results that are directly related to the
conclusions of this paper. 

For stars towards the Galactic poles, the proper motion measures the radial and
rotational velocity components. We select 55,429 unresolved sources closer than
10 kpc with $0.2 < g-r< 0.4$ and $b>80^\circ$ from SDSS DR6. We use the $v_X$ and
$v_Y$ velocity notation to emphasize the spatially constrained nature of this
sample, and to make a distinction from velocity components computed using a
measured radial velocity; in these directions, $v_X \sim v_R$ and $v_Y \sim
v_\Phi$. We do not correct velocities for the solar motion relative to the Local
Standard of Rest (LSR; $v_X^\odot = -10.0\pm0.4$ km/s, $v_Y^\odot = -5.3\pm0.6$ km/s;
and $v_Z^\odot = 7.2\pm0.4$ km/s; Dehnen \& Binney 1998). Therefore, the mean
value of $v_Y$ for a non-rotating population is $v_Y = -v_Y^\odot + v_Y^{LSR}
\sim 225$ km/s, where $v_Y^{LSR} \sim 220$ km/s is the rotational velocity of
the local standard of rest (Gunn, Knapp \& Tremaine 1979). 

The median heliocentric rotational velocity component as a function of
metallicity and distance from the Galactic plane in shown in the bottom right
panel of Figure~\ref{panels6}. The detailed kinematic behavior shows the same
two halo and disk components as implied by the metallicity distribution. The
high-metallicity disk component at $[Fe/H]>-1$ and $Z<3$ kpc lags the local
rotation by up to $\la$100 km/s, while the low-metallicity halo component at
$[Fe/H]<-1$ and $Z>3$ kpc has $\langle v_Y \rangle \sim 200$ km/s. This strong
metallicity-kinematic correlation is qualitatively the same as discussed in the
seminal paper by Eggen, Lynden-Bell \& Sandage (1962), except that here it is
reproduced {\it in situ} with a $\sim$100 times larger, nearly-complete sample,
extending it beyond the solar neighborhood. 

The variation of the median $v_Y$ as a function of distance from the Galactic
plane is shown separately for high-metallicity ($[Fe/H]>-0.9$) and
low-metallicity ($[Fe/H]<-1.1$) subsamples in Figure~\ref{FigVvsZ}. The median
$v_Y$ for $\sim$40,000 low-metallicity stars is 205 km/s. The systematic
velocity uncertainty, set by the systematic distance and proper motion errors,
is about 10-20 km/s. Therefore, this measurement implies that the halo does not
rotate at that level of accuracy, in contradiction with a result based on
similar type of analysis (proper motions and photometric parallax) by Majewski
(1992), that the halo counter-rotates with a speed of $-55 \pm 16$ km/s relative
to the LSR (based on a sample of a few hundred stars towards the north Galactic
pole). In order to make our $v_Y$ for halo stars agree with Majewski's result,
the distance scale given by photometric parallax relation needs to be increased
by 25\% ($\sim$0.5 mag offset in absolute magnitude scale), which is unlikely
(see Appendix A). An alternative explanation that proper motion measurements
used here are systematically overestimated by $\sim$2 mas/yr is ruled out by
independent data (B08). In addition, the radial velocity measurements from SDSS
spectroscopic survey also indicate no significant halo rotation at the level of
$\sim$10 km/s (Allende Prieto et al. 2006; Ivezi\'c et al. 2006; B08). Recall,
as mentioned previously, that our data do not extend very far into the region
that Carollo et al. (2007) have argued is dominated by a proposed outer,
counter-rotating, halo component.

The median $v_Y$ for $\sim$18,000 high-metallicity stars increases with $Z$. 
We obtained a best fit
\begin{equation}
\label{vMedianZ}
     \langle v_Y \rangle = 20.1 + 19.2 \,|Z/{\rm kpc}|^{1.25} \,\,\, {\rm km/s}.
\end{equation}
An alternative linear fit
\begin{equation}
\label{vMedianZ2}
        \langle v_Y \rangle = 32.2 \,|Z/{\rm kpc}|  \,\,\, {\rm km/s},
\end{equation}
also provides a good description for $\langle v_Y \rangle$ as a function of $Z$
for stars with $0.2<g-r<0.4$ and $1 < Z / {\rm kpc} < 4$. However, B08 show that
the detailed kinematics of red stars ($g-r>0.6$), which sample the smaller $Z$ range 
($<$1 kpc) where the two fits differ by $\sim$20 km/s, prefer the non-linear 
form given by eq.~\ref{vMedianZ}. 

The observed velocity dispersion for both halo and disk subsamples increases
with $Z$ (see the bottom right panel of Figure~\ref{FigVvsZ}). This
increase is dominated by increasing measurement errors (mostly due to 
the fact that even for constant proper motion errors, the velocity error 
increases proportionally to distance). We find that the observed velocity
dispersion for halo stars can be modeled with a constant intrinsic dispersion of 
$\sigma_Y^H$ = 90 km/s. For disk stars, the intrinsic dispersion increases
with $Z$ as
\begin{equation}
\label{sigYD}
          \sigma_Y^D = 25.8 + 10.1 \,|Z/{\rm kpc}| \,\,\, {\rm km/s},
\end{equation}
from $\sim$36 km/s at $Z=1$ kpc to $\sim$66 km/s at $Z=4$ kpc. 

The decrease of rotational velocity with $Z$ (sometimes called velocity lag)
was first convincingly detected by Majewski
(1992), using photographic data for 250 stars. At $Z\sim1.5-2.0$ kpc, he 
found a lag of 50$\pm$10 km/s, in good agreement with the value of 
59 km/s given by eq.~\ref{vMedianZ}. Chiba \& Beers (2000) measured 
a somewhat steeper gradient, of 30$\pm$3 km/s/kpc. Using proper motion data for 
about a million M dwarfs, Ivezi\'{c}  et al. (2005) reported essentially the same 
behavior of rotational velocity at $Z<1$ kpc, with a median value of 34 km/s at 
$Z=1$ kpc, which agrees well with the lag of 39 km/s obtained 
here using F dwarfs. Using SDSS radial velocity data,
Allende Prieto et al. (2006) measured a vertical rotational velocity gradient 
of 16 km/s, which is similar to the results based on proper motion measurements.
Most recently, Girard et al. (2006) 
used data for 1200 red giants and detected ``velocity shear'' towards the south 
Galactic pole. The velocity gradient of 29 km/s/kpc given by 
eq.~\ref{vMedianZ} at $Z=2$ kpc is consistent with their value of 30$\pm$3 km/s/kpc. 
They also detected a vertical gradient in the rotational velocity 
dispersion of (10$\pm$3) km/s/kpc, in excellent agreement with the value obtained
here. Given the different tracers, analysis methods, and the possibility for north-south 
asymmetry, the results presented here and those from Girard et al. (2006) are 
remarkably consistent.

\subsubsection{A Model for the Rotational Velocity Distribution}
\label{vPhiDist}

\begin{figure*}
\plotone{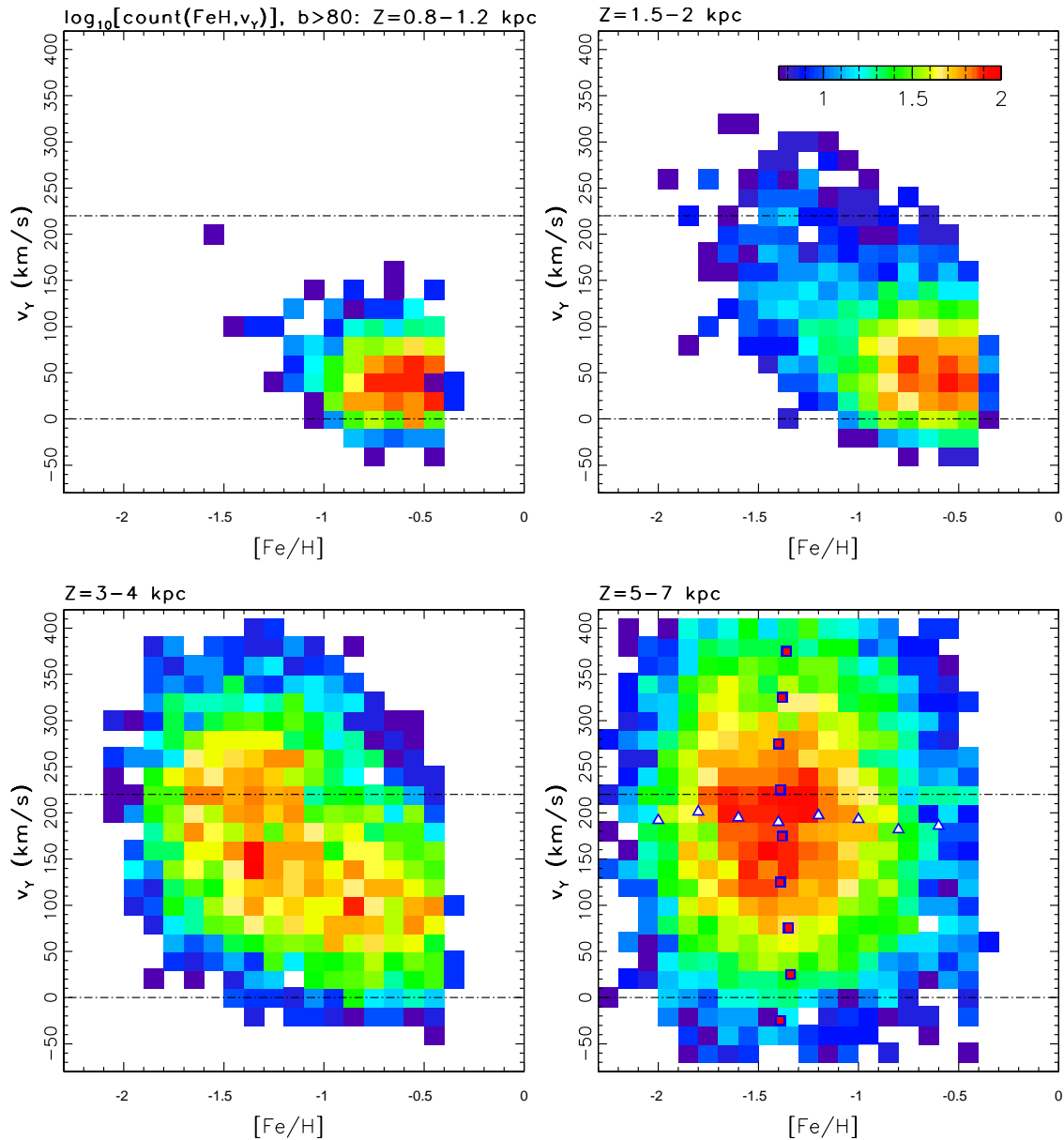}
\caption{
The distribution of stars with $0.2 < g-r < 0.4$ observed towards the north 
Galactic pole ($|b|>80^\circ$) in the velocity-metallicity diagrams, and as 
a function of distance from the plane in the range 0.8--1.2 kpc (top left, 
$\sim$1,500 stars), 1.5--2.0 kpc (top right, $\sim$4,100 stars), 3.0--4.0 kpc 
(bottom left, $\sim$6,400 stars) and 5.0--7.0 kpc (bottom right, $\sim$12,500 
stars). The maps show the logarithm of counts in each pixel, according to the 
legend shown in the top right panel. Towards the north Galactic pole, the 
plotted heliocentric velocity component ($v_Y$) corresponds to the rotational 
component. Its median value for subsamples selected by $[Fe/H]>-1$ in the top 
right panel is 59 km/s and 117 km/s in the bottom left panel. The median value
for subsamples selected by $[Fe/H]<-1$ is 192 km/s and 203 km/s in the
bottom two panels, respectively. For marginal $v_Y$ distributions, see 
Figure~\ref{Vphihist}. The horizontal lines at  $v_Y=0$ and  $v_Y=220$ km/s
are added to guide the eye. The symbols in the bottom right panel show
the median values of metallicity (squares) and $v_Y$ (triangles) in narrow
bins of the other coordinate.}
\label{panels9}
\end{figure*}

\begin{figure*}
\plotone{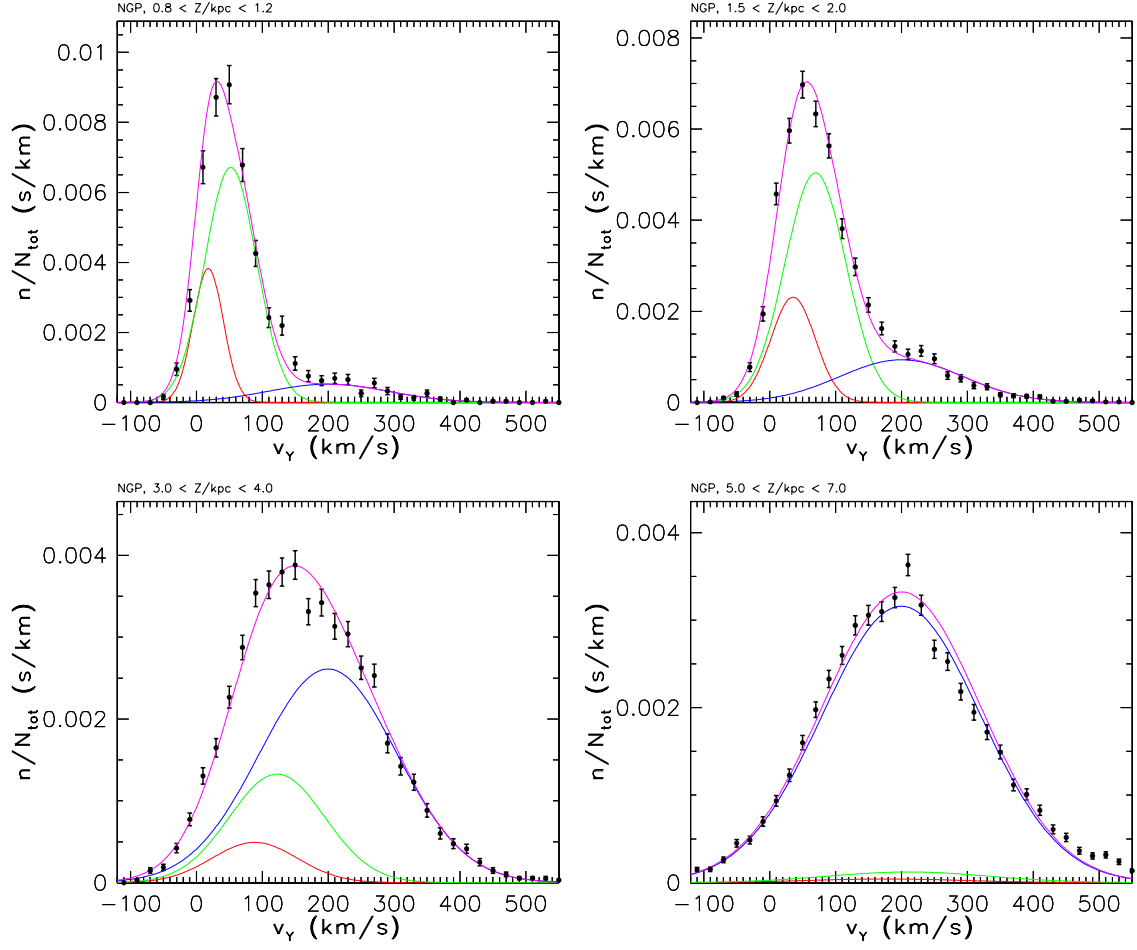}
\caption{
The symbols with error bars show the measured rotational velocity distribution 
for stars with $0.2 < g-r < 0.4$, $b>80^\circ$, and the distance from 
the Galactic plane in the range in the range 0.8--1.2 kpc (top left, $\sim$1,500
stars), 1.5--2.0 kpc (top right, $\sim$4,100 stars), 3.0--4.0 kpc (bottom left,
$\sim$6,400 stars) and 5.0--7.0 kpc (bottom right, $\sim$12,500 stars).
These histograms are the marginal $v_Y$ distributions for the maps shown in
Figure~\ref{panels9}. The red and green curves show the contribution of a 
non-Gaussian disk model (a sum of two Gaussians with fixed, 1:3, relative
normalization), the blue curves show halo contribution (a Gaussian), and the 
magenta curves are their sum (see \S~\ref{vPhiDist} for details and Table 4 
for best-fit parameters). }
\label{Vphihist}
\end{figure*}

\begin{figure*}
\plotone{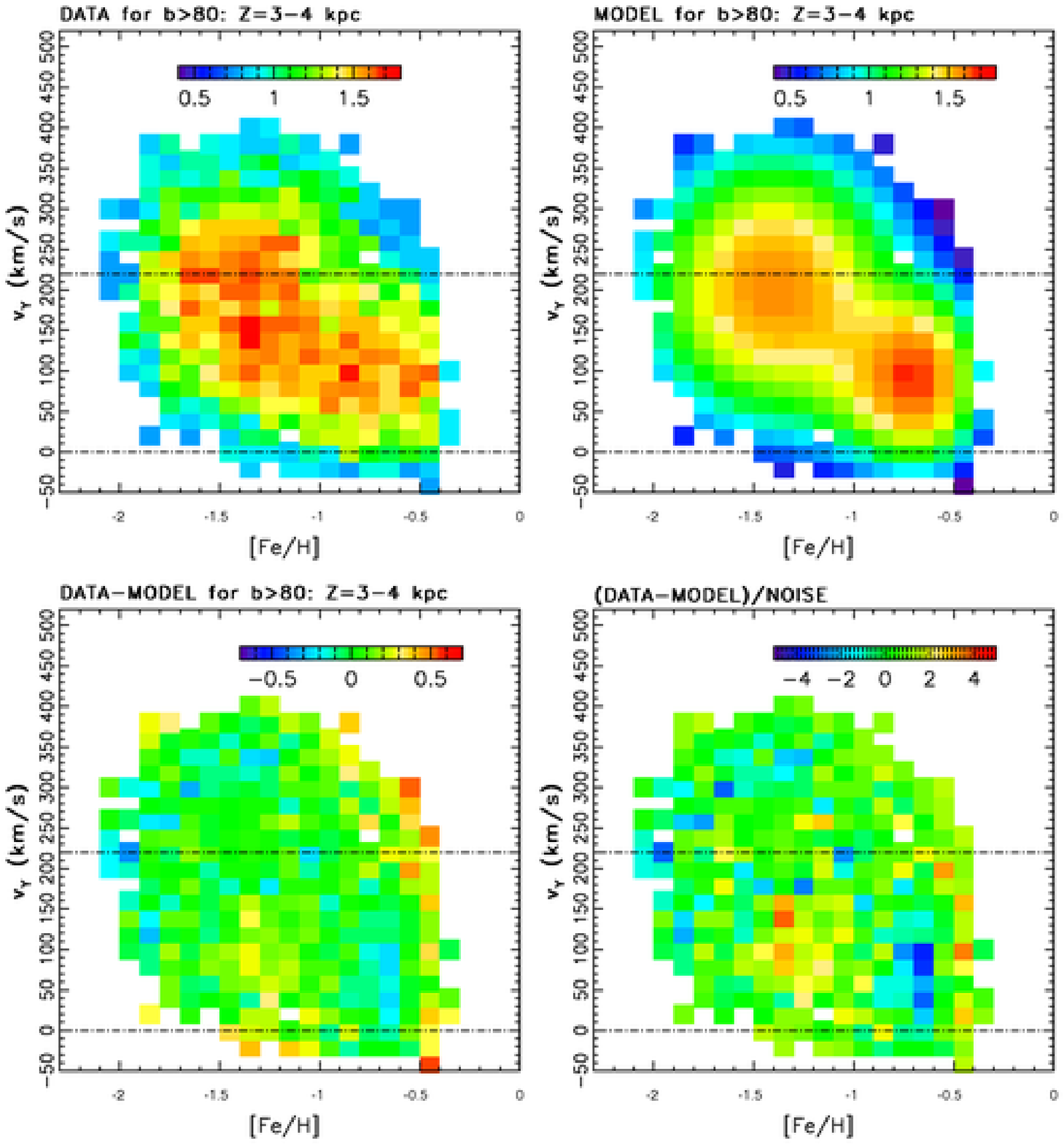}
\caption{
The top left panel shows distribution of stars (logarithm of counts in each bin)
observed towards the north Galactic pole ($|b|>80^\circ$), and with $3.0 <
|Z|/{\rm kpc} < 4$ kpc, in the velocity-metallicity diagram (same map as in the
bottom left panel in Figure~\ref{panels9}). The top right panel shows a best-fit
model assuming that velocity and metallicity distributions are uncorrelated,
when disk and halo components are treated separately, and are given by the best
fits shown in the bottom left panel of Figure~\ref{Zhist} for the metallicity
distribution, and in the bottom left panel of Figure~\ref{Vphihist} for the velocity
distribution. The bottom left panel shows the logarithm of the data/model ratio,
displayed with the same dynamic range as the top two panels. The observed counts
are predicted with an rms scatter of 33\%. This scatter is consistent with
the expected statistical noise. The bottom right panel shows the data-model
difference map normalized by the expected noise. The rms width of the
distribution is 1.09. }
\label{panels14}
\end{figure*}

The full metallicity/velocity distribution in four $Z$ slices is shown in 
Figure~\ref{panels9}. In agreement with the median behavior shown in 
Figure~\ref{FigVvsZ}, close to the Galactic plane the sample is dominated by 
high-metallicity stars with a small velocity lag, and gradually becomes
dominated by low-metallicity stars with no net rotation at large $Z$. 
We find no correlation between velocity and metallicity for distant
low-metallicity stars (see the bottom right panel in Figure~\ref{panels9}).

The marginal rotational velocity distributions for maps from Figure~\ref{panels9} 
are shown in Figure~\ref{Vphihist}. Analogously to modeling the metallicity
distributions in \S~\ref{thirdGauss}, we first attempt to fit the 
dependence of rotational velocity distribution on distance from the Galactic 
plane, $Z$, using a Gaussian for the halo component and a universal shape, modeled 
as two coupled Gaussians, for the disk component (an alternative approach 
based on thin/thick disk decomposition is described below).

We heuristically model the {\it shape} of the disk velocity distribution as a
sum of two Gaussians with fixed positions relative to each other (an offset of 
34 km/s), fixed widths (12 km/s and 34 km/s), and fixed relative normalization 
(3-to-1 number ratio in favor of the more metal-poor component). Motivated
by the behavior of $\langle v_Y \rangle$ for the $[Fe/H] > -0.9$ subsample
(eq.~\ref{vMedianZ}), we describe the central velocity of the narrower
Gaussian as
\begin{equation}
\label{vMeanZ}
                v_n(Z) = -3.0 + 19.2 \,|Z/{\rm kpc}|^{1.25} \,\,\, {\rm km/s},
\end{equation}
where the offset of 23 km/s is due to the difference between the median
of the adopted skewed velocity distribution and the mean for the 
narrower Gaussian component. The values of these four free parameters were 
obtained by {\it simultaneously} fitting the four histograms shown in 
Figure~\ref{Vphihist}. The halo velocity distribution was kept fixed as a 
Gaussian centered on 200 km/s and with a width of 90 km/s. 

\begin{deluxetable*}{ccrrrrccccrr}
\tablenum{4} \tablecolumns{12}
\tablecaption{Best-fit Parameters for $p(v_Y)$, Shown in Fig.~\ref{Vphihist}}
\tablehead{$|Z|$ range$^a$ & $\langle |Z| \rangle^b$ & N$^c$ & $\langle g \rangle^d$ & $v_{d1}^e$ 
& $v_{d2}^f$ & $\sigma_{d1}^g$  & $\sigma_{d2}^h$ & $f_H^i$ & $\sigma_{H}^j$ & $\sigma_v^k$ & $\sigma_v^l$ }
\startdata
 0.8--1.2 & 1.08 &  1,526 & 15.2 &  18 &  57 &  23 &  39 & 0.12 &  92 & 14 & 17 \\
 1.5--2.0 & 1.75 &  4,076 & 16.2 &  36 &  70 &  34 &  46 & 0.22 &  94 & 23 & 26 \\
 3.0--4.0 & 3.49 &  6,445 & 17.8 &  88 & 122 &  65 &  72 & 0.68 & 104 & 48 & 52 \\
 5.0--7.0 & 5.99 & 12,452 & 18.9 & 176 & 211 & 118 & 122 & 0.95 & 123 & 93 & 84 \\
\enddata
\tablenotetext{a}{The $|Z|$ range for the bin (kpc).}
\tablenotetext{b}{The median $|Z|$ in the bin (kpc).}
\tablenotetext{c}{The number of stars in the bin.} 
\tablenotetext{d}{The median $g$-band magnitude in the bin.} 
\tablenotetext{e}{The mean velocity for the first Gaussian disk component, in km/s,
                  computed using eq.~\ref{vMeanZ}.} 
\tablenotetext{f}{The mean velocity for the second Gaussian disk component, $v_{d2}=v_{d1}+34$, (km/s).} 
\tablenotetext{g}{The velocity dispersion for the first Gaussian disk component (12 km/s) 
                  convolved with measurement errors (km/s).} 
\tablenotetext{h}{The velocity dispersion for the second Gaussian disk component (34 km/s) 
                  convolved with measurement errors (km/s).} 
\tablenotetext{i}{The best-fit halo-to-disk number ratio.} 
\tablenotetext{j}{The velocity dispersion for the halo component (90 km/s)
                  convolved with measurement errors (km/s).} 
\tablenotetext{k}{The median velocity measurement error for stars with $[Fe/H]>-0.9$ (km/s).}
\tablenotetext{l}{The median velocity measurement error for stars with $[Fe/H]<-1.1$ (km/s).}
\end{deluxetable*}

In each $Z$ bin, the expected velocity measurement error (determined from
expected random errors in distance and proper motion) was added in quadrature 
to the widths of all three Gaussians. For the disk component, we also
add in quadrature $10.1 \,|Z/{\rm kpc}|$ km/s based on eq.~\ref{sigYD}. 
The best-fit parameters are listed in Table 4 and the best fits are shown in 
Figure~\ref{Vphihist}. As is evident, this simple model of a skewed distribution
that slides linearly with $Z$ provides a satisfactory description of the data. 

We use the best fits for the marginal metallicity distribution described in
\S~\ref{thirdGauss} (see Figure~\ref{Zhist}), and the best fits for marginal 
rotational velocity distribution discussed above to model the observed joint 
distributions in the rotational velocity vs. metallicity plane, shown in 
Figure~\ref{panels9}. In each $Z$ bin, we simply multiply the best-fit marginal 
distributions and subtract their product from the observed map. There are two 
important assumptions underlying this approach: (1) The disk and halo components
used to model the two marginal distributions map well onto each other, and (2) 
the velocity and metallicity distributions of each individual component 
are uncorrelated. 

The above assumptions are borne out by the data. As an illustration of the residual 
and $\chi^2$ maps, we show the $Z=3-4$ kpc bin, which contains similar
fractions of disk and halo stars (Figure~\ref{panels14}). The observed distribution is 
successfully modeled to within expected statistical noise ($\sim$30\% 
for counts per pixel, on average). The observed lack of a correlation between 
velocity and metallicity distributions for disk component is at odds with the
traditional thin/thick disk decomposition, which we address next.

\subsubsection{ Difficulties with the Thin--Thick Disk Separation? }
\label{TvsT}

\begin{figure*}
\plotone{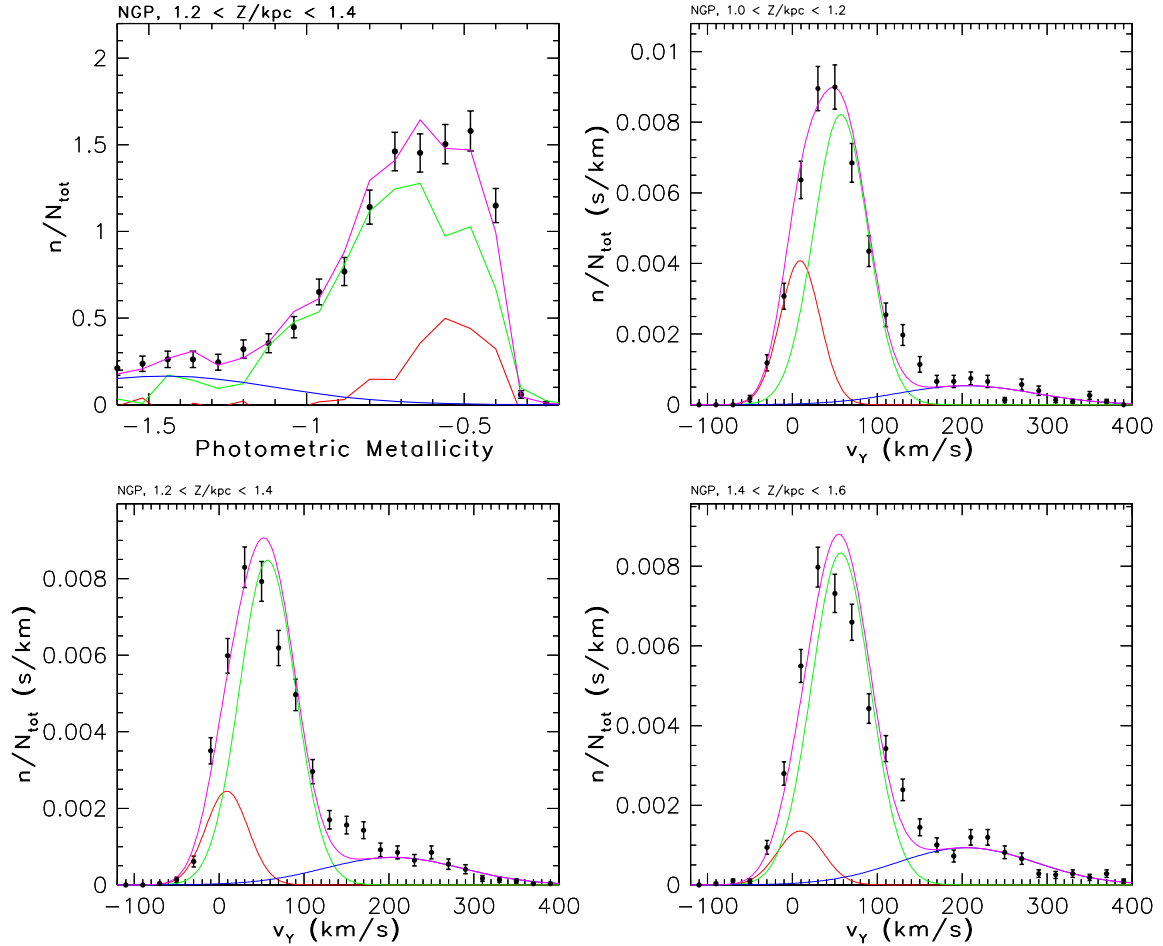}
\caption{The top left panel illustrates a test of the hypothesis that
the observed variation of metallicity distribution with $|Z|$ is due to a
varying mixing ratio of the independent thin and thick disk metallicity
distributions. The latter are determined using metallicity distributions in the
$|Z|=1.0-1.2$ kpc and $|Z|=2.0-2.5$ kpc bins, and are shown as red (thin disk)
and green (thick disk) lines. They are scaled using the 4.6:1 thick-to-thin disk
number ratio expected for the displayed $|Z|=1.2-1.4$ kpc bin from the counts
profile. The data for this $|Z|$ bin are shown by symbols with error bars. The
blue line shows halo contribution (15\%), and the magenta line is the sum of all
three components. The remaining three panels show measured rotational velocity
distributions in three $Z$ slices (symbols with error bars; 1.0-1.2 kpc: top
left; 1.2-1.4 kpc: bottom right; 1.4-1.6 kpc: bottom right). The blue curves are
Gaussians corresponding to halo stars, with the parameters listed in Table 5.
The red lines are Gaussians centered on 9 km/s, and the green lines are
Gaussians centered on 57 km/s, with widths listed in Table 5. The increase of
their widths with $|Z|$ is consistent with estimated measurement errors; the
implied intrinsic widths are 18 km/s for narrow Gaussians, and 28 km/s for wide
Gaussians. These curves are normalized according to thick-to-thin number ratio
and halo contribution listed in Table 5. }
\label{singleGauss}
\end{figure*}

\begin{figure*}
\plotone{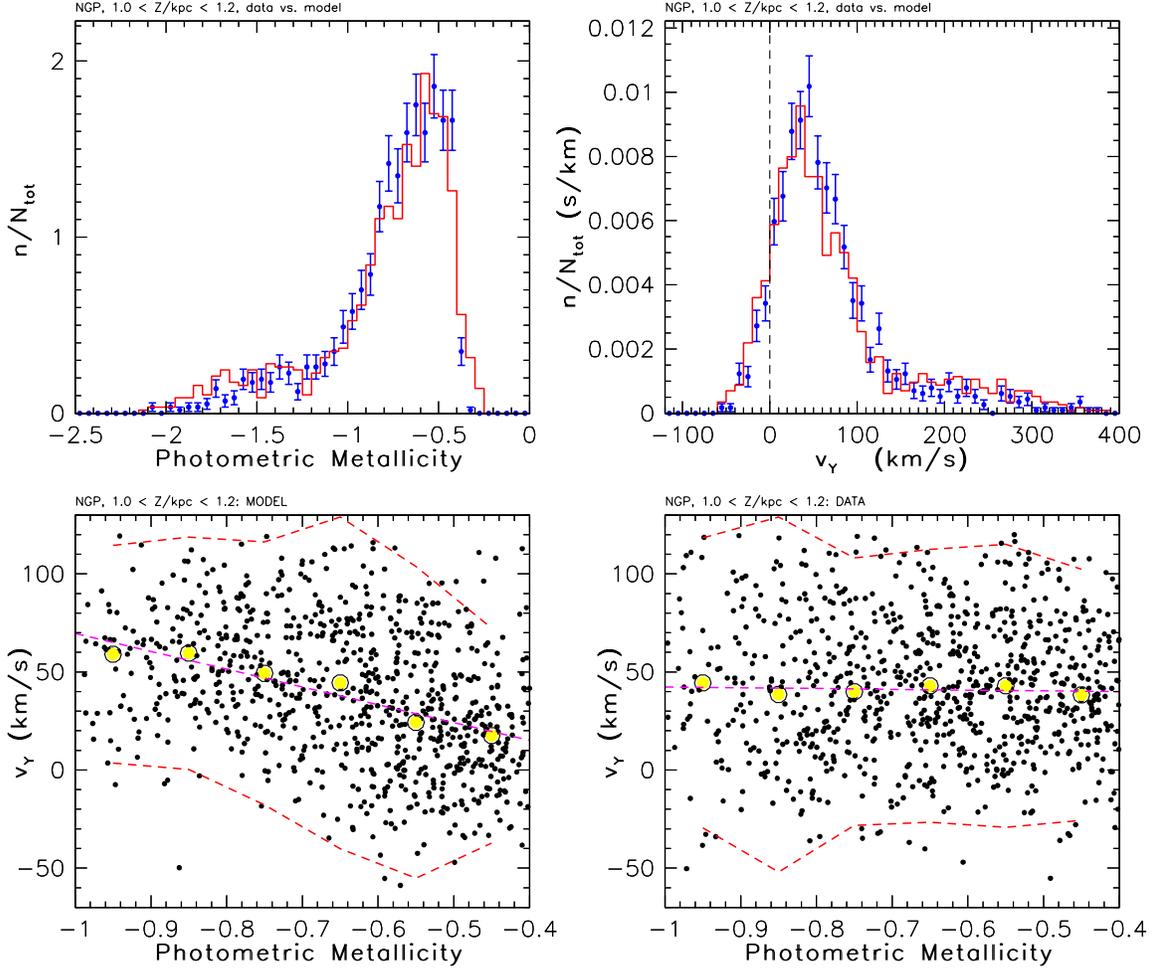}
\caption{A comparison of data for 1,142 stars with $b>80$ and
$1.0 < |Z|/{\rm kpc} < 1.2$ kpc, and a model based on a traditional disk
decomposition into thin and thick components. The model assumes three Gaussians
with contributions of thin disk, thick disk and halo stars equal to 28\%, 52\%,
and 20\%, respectively. The Gaussians describing the metallicity distribution
(top left panel, symbols represent data and histogram represents model) are
centered on $[Fe/H] = -0.50, -0.72$ and $-1.37$, and have widths of 0.04, 0.15,
and 0.32, dex, respectively. The Gaussians describing the velocity distribution
(top right panel) are centered on 9, 57 and 205 km/s, and have widths of 25, 33,
and 80 km/s, respectively. The two bottom panels show the two-dimensional
distributions in the velocity/metallicity diagram for the model (left) and the
data (right), with individual stars shown as small dots. The large dots show the
median velocity in 0.1 dex wide metallicity bins, for stars with velocity in the
$-60$ km/s to 120 km/s range. The 2-$\sigma$ envelope around these medians is
shown by the two outer dashed lines. The dashed lines in the middle show the
best linear fit to the median velocity, with slopes of $-91$ km/s/dex (model)
and $-4.1$ km/s/dex (data).}
\label{corrComp}
\end{figure*}

\begin{figure*}
\plotone{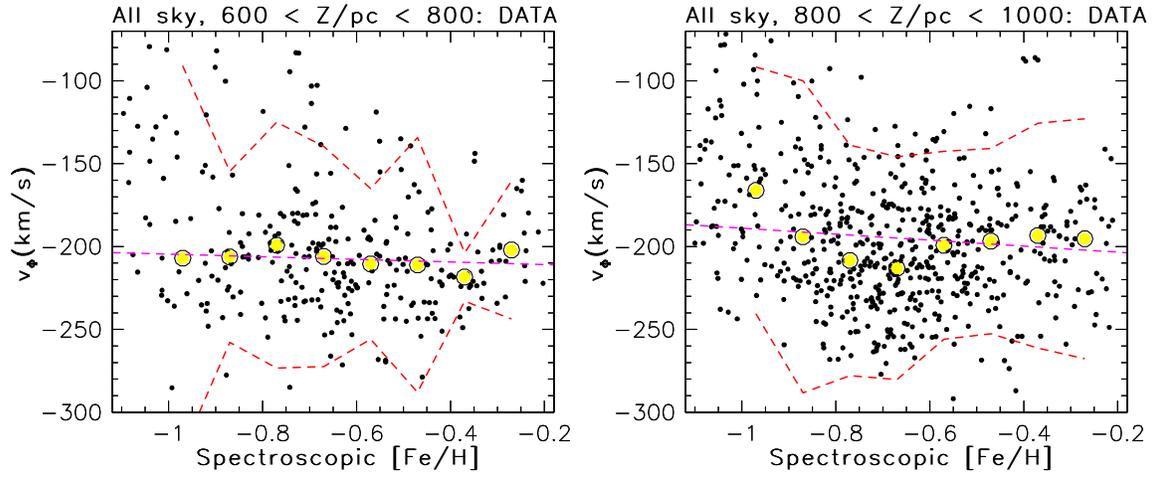}
\caption{
Analogous to the bottom right panel of Fig.\ref{corrComp}, except that SDSS
spectroscopic sample of stars is used ($v_Y$ from Fig.\ref{corrComp} corresponds
to $v_\Phi$+225 km/s). Its all-sky distribution and radial velocity measurement
enable the sampling of regions closer to the Galactic plane than with the
$b>80^\circ$ sample (left: 285 stars with $|Z|$=600-800 pc; right: 583 stars with
$|Z|$=800-1000 pc; both with $0.2<g-r<0.4$). Although the metallicity distribution
of spectroscopic sample is highly biased (see the top right panel in
Fig.~\ref{panels2}), the dependence of the median rotational velocity for narrow
metallicity bins, as a function of metallicity, is not strongly affected. Note
the absence of a strong $v_\Phi$ vs. $|Z|$ gradient ($\la$20 km/s/dex). }
\label{f17}
\end{figure*}

Our model fits above show that the data for both the metallicity and rotational
velocity distributions of disk stars can be fit with shapes that do not vary
with $Z$. While formally these skewed distributions are modeled as sums of two
Gaussians, these Gaussians cannot be readily identified with the classical thin
and thick disks. First, their normalization ratio is constant, while
double-exponential best-fits to the observed spatial profile predict that the
fraction of stars in each component should strongly vary with $Z$. Secondly, the
normalization ratios are {\it different}, 1.7 for the metallicity distributions and
3.0 for the rotational velocity distributions, respectively, which implies that the two components
do not perfectly map onto each other.

Despite these difficulties, it is instructive to attempt a traditional thin- plus
thick-disk decomposition using the data presented here. We seek metallicity and
velocity distributions whose linear combination, with weights given by the
observed spatial profiles, reproduces the variation of the observed distributions
with $Z$. This can be done with only a minimal reliance on models, because the
data span a wide range of $Z$, {\it and because there are clearly detected $Z$
gradients in both the metallicity and velocity distributions, which cannot be
attributed to halo stars}. The disk spatial profile (with the J08 best-fit
parameters) implies that the $|Z|=2.0-2.5$ kpc bin is dominated by thick-disk stars
(only 2\% of disk stars from this bin are expected to belong to the thin-disk
component, with a 38\% contribution from halo stars), while they contribute, for
example, $\sim$67\% of all disk stars in $|Z|=0.8-1.2$ kpc bin. Therefore, one
can subtract the $\sim$38\% halo contribution from the metallicity distribution
in $|Z|=2.0-2.5$ kpc bin, then re-normalize the difference, and treat it as the
thick-disk metallicity distribution. It can then be subtracted, after renormalization, from
the disk-dominated metallicity distribution in the $|Z|=0.8-1.2$ kpc bin. The
renormalized difference can be treated as a pure thin-disk distribution. The two
distributions can then be linearly combined and compared to distributions
observed in the intermediate bins.  

The top left panel of Figure~\ref{singleGauss} shows the result of this
comparison for the $|Z|=1.2-1.4$ kpc bin. The metallicity distributions for both
components appear non-Gaussian, with a difference of their means of about 0.2
dex. Their expected linear combination reveals discrepancies with the data, but
they are not overwhelming, and could be due to uncertainties in the decomposition
procedure itself. 

A similar linear decomposition method cannot be applied to the kinematic data
analyzed here because the velocity measurement errors increase too much over the
relevant range of $Z$. Instead, we choose to model the observed velocity
distributions using two Gaussians, with fixed mean velocities and dispersions.
The widths are convolved with known, $Z$-dependent, velocity measurement errors
when fitting the four free parameters. We normalize the two Gaussians assuming
their relative contributions (and halo normalization) predicted by the density
(counts) profiles shown in Figure~\ref{Zcounts}. 

\begin{deluxetable*}{ccrrrrrrr}
\tablenum{5} \tablecolumns{10} \tablewidth{5in}
\tablecaption{Best-fit Parameters for $p(v_Y)$, Shown in Fig.~\ref{singleGauss}}
\tablehead{$|Z|$ range$^a$ & $\langle |Z| \rangle^b$ & N$^c$ & $\langle g \rangle^d$ 
& $\sigma_{d1}^e$  & $\sigma_{d2}^f$ & $\sigma_{H}^g$ & $f_H^h$ & $r_{DD}^i$ }
\startdata
 1.0--1.2 &  1.11 &  1,142 & 15.3 &  23 &  32  & 80 & 11 & 2.7   \\     
 1.2--1.4 &  1.39 &  1,480 & 15.6 &  25 &  33  & 81 & 15 & 4.6   \\ 
 1.4--1.6 &  1.50 &  1,603 & 15.9 &  27 &  34  & 81 & 19 & 7.9   \\ 
\enddata
\tablenotetext{a}{The $|Z|$ range for the bin (kpc).}
\tablenotetext{b}{The median $|Z|$ in the bin (kpc).}
\tablenotetext{c}{The number of stars in the bin.} 
\tablenotetext{d}{The median $g$-band magnitude in the bin.} 
\tablenotetext{e}{The velocity dispersion for the first Gaussian disk component 
                  convolved with measurement errors (km/s).} 
\tablenotetext{f}{The velocity dispersion for the second Gaussian disk component 
                  convolved with measurement errors (km/s).} 
\tablenotetext{g}{The velocity dispersion for the halo component 
                  convolved with measurement errors(km/s).} 
\tablenotetext{h}{The fraction of halo stars from J08 model (\%).}
\tablenotetext{i}{The thick-to-thin disk star number ratio from J08 model (see \S~\ref{J08}).}
\end{deluxetable*}

The best fits are shown in Figure~\ref{singleGauss} for three representative $Z$
bins, with the best-fit parameters listed in Table 5. The presumed thick-disk
Gaussian has a velocity lag of 48 km/s, relative to the first Gaussian (centered
at 57 km/s and 9 km/s, respectively). The implied intrinsic velocity dispersions
are 18 km/s and 28 km/s. The fits are not as good as those shown in
Figure~\ref{Vphihist}, but the discrepancies are not too large. They are,
however, formally highly statistically significant, due to the large number of
stars, but there is always a possibility for hidden systematic errors. If the
thick-disk Gaussian were replaced by a non-Gaussian velocity distribution, it is
likely that the most egregious discrepancies around $v_Y \sim 150$ km/s could be
resolved. 

Therefore, the marginal metallicity and velocity distributions do {\it not}
strongly rule out the thin/thick disk linear combination hypothesis, but {\it
only if} both disks have non-Gaussian metallicity and velocity distributions.
The choice for both metallicity and rotational velocity distributions is then
between a linear combination of two fixed non-Gaussian distributions, or a
single non-Gaussian distribution that slides with $Z$. A difficulty with the
former hypothesis is that the implied thin-disk metallicity distribution has a
mean of about $-0.55$, which is too metal poor by about 0.4 dex, compared to
local measurements (Girardi \& Salaris 2001; Nordstr\"{o}m et al. 2004). On the
other hand, there is a worrisome possibility of a metallicity ``compression''
discussed in Appendix C, which could bias thin disk measurements to lower
values. 

While fitting the marginal metallicity and kinematic distributions separately
does not strongly discriminate among models, the thin/thick disk linear
combination hypothesis makes a very strong prediction about the joint
two-dimensional distribution. Because the individual components are offset by
$\sim$0.2 dex in metallicity, and by $\sim$50 km/s in rotational velocity, this
hypothesis predicts a correlation between metallicity and rotational velocity
for samples selected from narrow $Z$ slices in the $Z$ range where neither
component dominates. To compute the strength of this correlation, we generated a
Monte Carlo sample for $Z=1.0-1.2$ kpc bin, where the correlation should be
strong, with the same size as the observed sample (the expected thick-to-thin
disk star number ratio is 2.7). The distributions of the data and model stars in
the velocity vs. metallicity plane are shown in Figure~\ref{corrComp}.  

Even without any computation, it is evident from the lower left panel of
Figure~\ref{corrComp} that the model distributions are strongly correlated: the
28\% contribution of thin-disk stars, which have smaller mean velocity and higher metallicity
than thick-disk stars, induce a net metallicity-velocity correlation ($\sim-90$
km/s/dex). The value of Kendall's rank correlation coefficient (e.g., see Lupton
1993) for the model distribution is $-0.30\pm0.04$, which is significantly
inconsistent with the observed value of $0.017\pm0.018$ (we limit velocities to
120 km/s, and metallicity to $[Fe/H] > -1$, to exclude halo stars). 

In order to test whether this observed lack of correlation is localized to the
north Galactic pole, we have also evaluated the Kendall's rank correlation
coefficient for stars with $Z=1.0-1.2$ kpc in three 10 degree wide patches along
the $l=180^\circ$ line and with $b$=50$^\circ$, 30$^\circ$, and $-$50$^\circ$.
None of these patches exhibits a statistically significant correlation between
velocity and metallicity, with the value of the correlation coefficient towards the
north Galactic pole remaining the most negative. In order to test whether the
measured correlation coefficient depends on the selection of the $Z$ bin, we
evaluated it for seven 200 pc thick bins in the range $Z$=0.5--2.1 kpc, and
found values statistically consistent with the value for the $Z=1.0-1.25$ kpc bin.

For another test, one that is less sensitive to errors in the adopted
metallicity scale, we compared velocity histograms for stars with $-1 < [Fe/H] <
-0.8$ and $-0.6 < [Fe/H] < -0.4$. The observed median rotation velocities differ
by 1 km/s, while the model values differ by 25 km/s, or eight times more than
the expected statistical noise. To explain the observed flat median rotation
velocity vs. metallicity behavior as due to errors in adopted absolute
magnitudes, adopted $M_r$ for stars with $-1 < [Fe/H] < -0.8$ would have to be
too faint by $\sim1$ mag, and $\sim1.5$ mag too bright for stars with $-0.6 <
[Fe/H] < -0.4$. Such large errors are not consistent with the constraints on
photometric parallax relation discussed in Appendix A, nor with plausible age
effects. 

These tests would be statistically even stronger if the samples extended to at
least $Z=0.75$ kpc, where the expected fractions of thin and thick disk stars
are equal. This is not possible with the $b>80^\circ$ sample due to image
saturation (at $r\sim14$), while for stars observed at lower Galactic latitudes,
the radial velocity measurement is required to compute the rotational velocity
component. Hence, we use the SDSS spectroscopic survey to extend the sample to
$Z<1$ kpc. In addition, this sample also critically tests a possibility that
these thin/thick disk separation difficulties are caused by photometric
metallicity problems. 

Figure~\ref{f17} shows the $v_\Phi$ vs. $[Fe/H]$ velocity diagrams for two
samples of stars with SDSS spectra that have $Z$=600-800 pc and $Z$=800-1000 pc.
Here, $v_\Phi$ is the rotational velocity component corrected for the LSR and
peculiar solar motion (towards the NGP, $v_\Phi \sim v_Y$-225 km/s). Although the
metallicity distribution of the SDSS spectroscopic sample is highly biased (see the
top right panel in Figure~\ref{panels2}), the dependence of the median
rotational velocity for narrow metallicity bins, as a function of metallicity,
is not strongly affected. These two subsamples are also inconsistent with the
strong velocity--metallicity correlation expected from traditional thin-thick
decomposition.   

We conclude that {\it the absence of a correlation between the observed velocity
and metallicity distributions for disk stars represents a major problem for the
interpretation of vertical velocity and metallicity gradients as being due to a
varying linear combination of two fixed distributions}. 

Our conclusion does not contradict previous work on this subject, because the
older samples did not include {\it simultaneous} distance, velocity, and
metallicity measurements for a sufficient number of stars with the appropriate
distance distribution. A test based on a randomly selected ten-times-smaller
subsample, with photometric errors and proper motion errors increased by a
factor of two, resulted in an inconclusive difference in Kendall's correlation
coefficients between the ``data'' and a thin/thick disk model. 

It is noteworthy that Norris (1987) proposed a Galaxy model which does not
assume that thin and thick disks are discrete components, but instead form a
kinematical and chemical continuum. Stars traditionally associated with the
thick disk belong to an ``extended'' disk (in terms of spatial distribution) in
Norris' terminology, and represent an extreme tail of metallicity and kinematic
distributions. Our results appear roughly consistent with Norris' proposal, and
provide a quantitative support for such ``continuous'' view of the disk over the
entire relevant range of distances from the Galactic plane.

\subsubsection{The ``Metal-weak Thick Disk'' Revisited}

In order to test whether the third Gaussian component discussed in
\S~\ref{thirdGauss} has the same kinematics as the rest of disk stars, we
compare the $v_Y$ histograms for $-1.1 < [Fe/H] < -0.8$ and $-0.6 < [Fe/H] <
-0.5$ subsamples in two $Z$ bins: 0.5--1 kpc and 1.5--2.0 kpc. We find no
statistically significant differences, with an upper limit on the relative
velocity offset of $\sim$15 km/s. This kinematic similarity supports the view
that stars with $-1.1 < [Fe/H] < -0.8$ reflect a non-Gaussian metallicity
distribution of disk stars, rather than a separate entity (Beers et al. 2002). 
Further insight will be obtained from the Rockosi et al. (in prep.) discussion
of the properties of the metal-weak thick disk as revealed by the spectroscopic
SEGUE sample.

\subsection{Spatially  Localized Deviations from the Mean Metallicity Distribution  }
\label{subStruct}

\begin{figure*}
\plotone{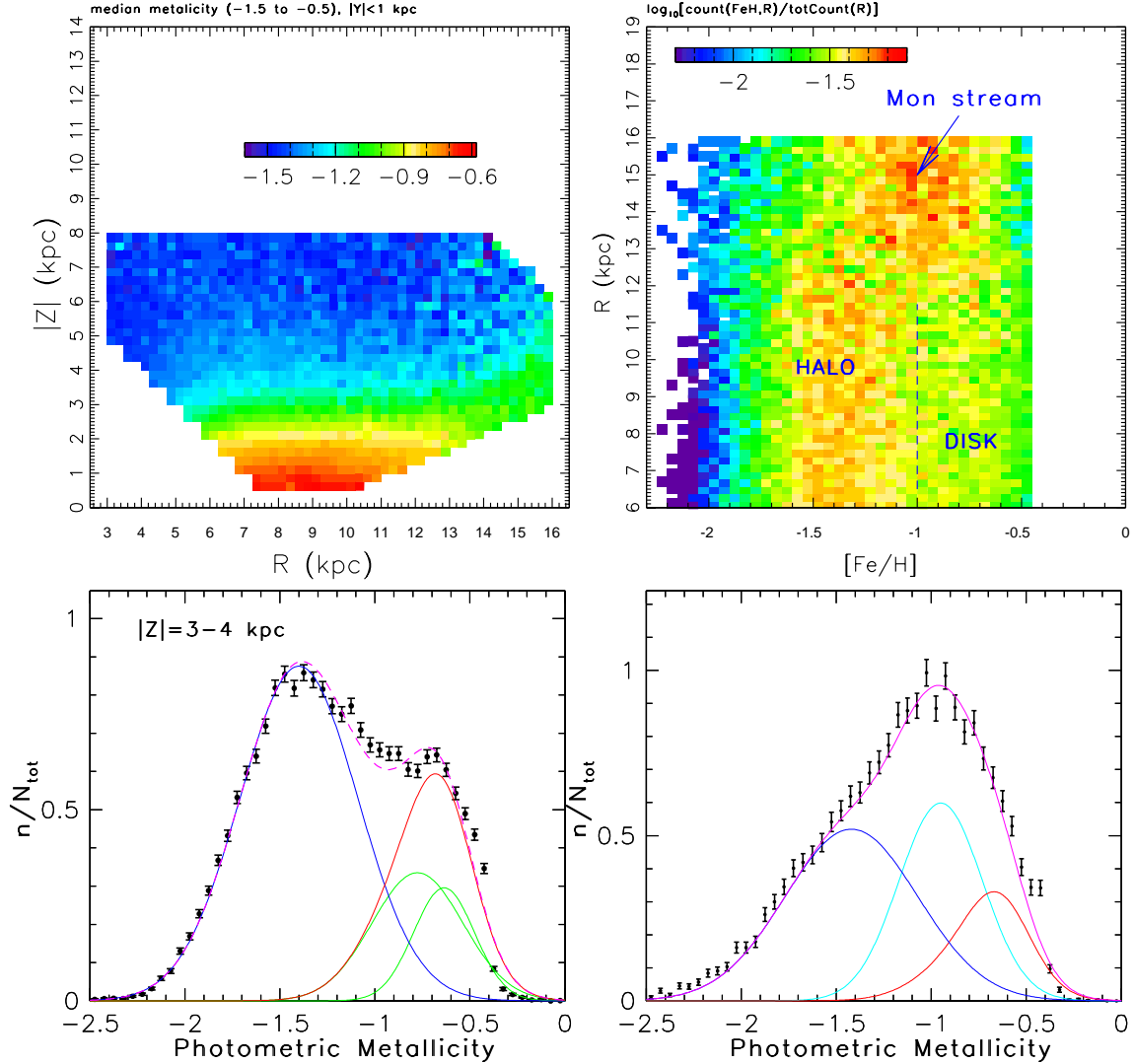}
\caption{
The top left panel illustrates the dependence of the median photometric
metallicity for $\sim$1.04 million stars from SDSS Data Release 6 with $14.5 < r
< 20$, $0.2 < g-r < 0.4$ and $|Y|<1$ kpc, in cylindrical Galactic coordinates
$R$ and $|Z|$. This $Y$ range is selected to include the Monoceros stream, which
represents an overdensity by a factor of $\sim$1.5-2 in a region around
$R\sim15$ kpc and $|Z|\sim$3--4 kpc. As discernible from the map, this region has
a larger median metallicity than expected for this $|Z|$ range from smaller $R$.
The top right panel shows the conditional metallicity probability distribution
for a subsample of $\sim$111,000 stars with $3 < |Z|/{\rm kpc} < 4$. The strong
overdensity at $R>12$ kpc is the Monoceros stream. The bottom panels show the
metallicity distribution (symbols with error bars) for a subsample of
$\sim$40,000 stars with $6 < R/{\rm kpc} < 9$ (left) and for $\sim$12,000 stars
with $13 < R/{\rm kpc} < 16$. The curves have the same meaning as in
Figure~\ref{Zhist}, with the addition of the cyan curve in the right panel. This
curve is a 0.22 dex wide Gaussian centered on $[Fe/H]=-0.95$, and accounts for
33\% stars in the sample that presumably belong to the Monoceros stream (see 
\S~\ref{metalMon} for details). }
\label{panels7}
\end{figure*}

The rich metallicity and kinematic data presented here enable more powerful
searching for Milky Way substructure than possible using the stellar number
density maps alone (e.g., see J08). We utilize the large sky coverage of the DR6
catalog to quantify spatial deviations from the conditional metallicity
distribution discussed in \S~\ref{metalDistrib} and \ref{metalDistribModel}. We
first constructed 51 maps, such as that shown in the top right panel of
Figure~\ref{panels3}, for regions defined by a 1 kpc by 1 kpc square footprint
in the $X-Y$ plane. In each map, we compute the median metallicity in two $Z$
slices that are dominated by disk (1--2 kpc) and halo (5--7 kpc) stars. The
range of cylindrical radius, $R$, is 6--14 kpc for the disk slice, and 5--15 kpc
for the halo slice.

In the range $6 < R/{\rm kpc} < 10$, the median disk metallicity is
$[Fe/H]=-0.72$, with an rms scatter of 0.05 dex, and a median distribution
width of 0.27 dex. There is no discernible radial metallicity gradient, with an
upper limit of 0.01 dex/kpc. For $R>10$ kpc, the median metallicity is
$[Fe/H]=-0.80$, with the shift of 0.08 dex probably due to stars from the
Monoceros stream, as discussed below. The median halo metallicity is
$[Fe/H]=-1.40$, with an rms scatter of 0.03 dex, and the median distribution
width of 0.41 dex. There is no discernible radial metallicity gradient, with an
upper limit of 0.005 dex/kpc. 

The mean expected statistical scatter in the medians is 0.005 dex for the disk and
0.011 for the halo (with a wider distribution and fewer stars for the latter), suggesting
that the variation of the median halo metallicity is probably insignificant,
while the rms variation of the median disk metallicity of 0.05 dex appears real.
The photometric calibration errors in individual SDSS runs, which could produce
a metallicity scatter of a similar magnitude, are averaged out because many runs
contribute to each map. Furthermore, such an instrumental effect is ruled out by
the fact that the $u$-band calibration errors would have to have an rms of 0.02
mag to produce the disk median metallicity rms of 0.05 dex, and only 0.006 mag
to produce the halo rms of 0.03 dex. Hence, were the disk rms due to calibration
errors, the halo rms would have been 0.1 dex, and not 0.03 dex. 

As an additional method to search for localized substructure, we subtracted a
best-fit model from each map (such as those described in
\S~\ref{metalDistribModel}), and visually inspected residual maps. The only
strong feature found in residual maps was localized at $Y\sim$0, $Z\sim3-4$ kpc
and $R\sim15$ kpc, and represents an excess of $[Fe/H]\sim-1$ stars. It is
clearly visible in the median metallicity $RZ$ map and, as an especially
striking feature, in a conditional metallicity distribution map shown in
Figure~\ref{panels7}. Using its spatial position, we identify this feature as
the Monoceros stream\footnote{Immediately following its discovery, it was not
clear whether the Monoceros stellar overdensity was a ring, stream or due to
disk flaring. Subsequent work has demonstrated its stream-like profile, see,
e.g., maps in J08.}, discovered in SDSS data using stellar counts by Newberg et
al. (2002).

\subsubsection{The Metallicity Distribution  for the Monoceros Stream }
\label{metalMon}
The conditional metallicity map from Figure~\ref{panels7} demonstrates 
that regions with $R<12$ are not strongly affected by the Monoceros
stream. We compare the metallicity distributions for stars with 
$6 < R/{\rm kpc} < 9$ (control sample), and for stars with 
$13 < R/{\rm kpc} < 16$, in the two bottom panels in Figure~\ref{panels7}. 
The metallicity distribution of the control sample is consistent with the 
halo and disk metallicity distributions described in \S~\ref{thirdGauss}, 
with a few minor adjustments: the disk distribution is shifted by 0.07 dex,
and the halo distribution by 0.02 dex, towards higher metallicity,
the fraction of halo stars is changed from 61\% to 55\%, and 0.16 dex 
is added in quadrature to the widths of the three Gaussians to account 
for the increased metallicity errors in single-epoch DR6 data. 

The subsample containing the Monoceros stream can be described using 
the same function as for the control sample, and an additional 0.22 dex 
wide Gaussian component centered on $[Fe/H]=-0.95$, and with a relative 
normalization of 33\%. When corrected for measurement errors, the implied 
width of the metallicity distribution for the Monoceros stream is 0.15 dex. 
The best-fit normalization is in good agreement with spatial profiles 
from J08, which suggests that the Monoceros stream is about a factor 
of two overdensity over the local background counts (i.e., a relative
normalization of 50\%).

\subsubsection{The Kinematics of the Monoceros Stream }

\begin{figure*}
\plotone{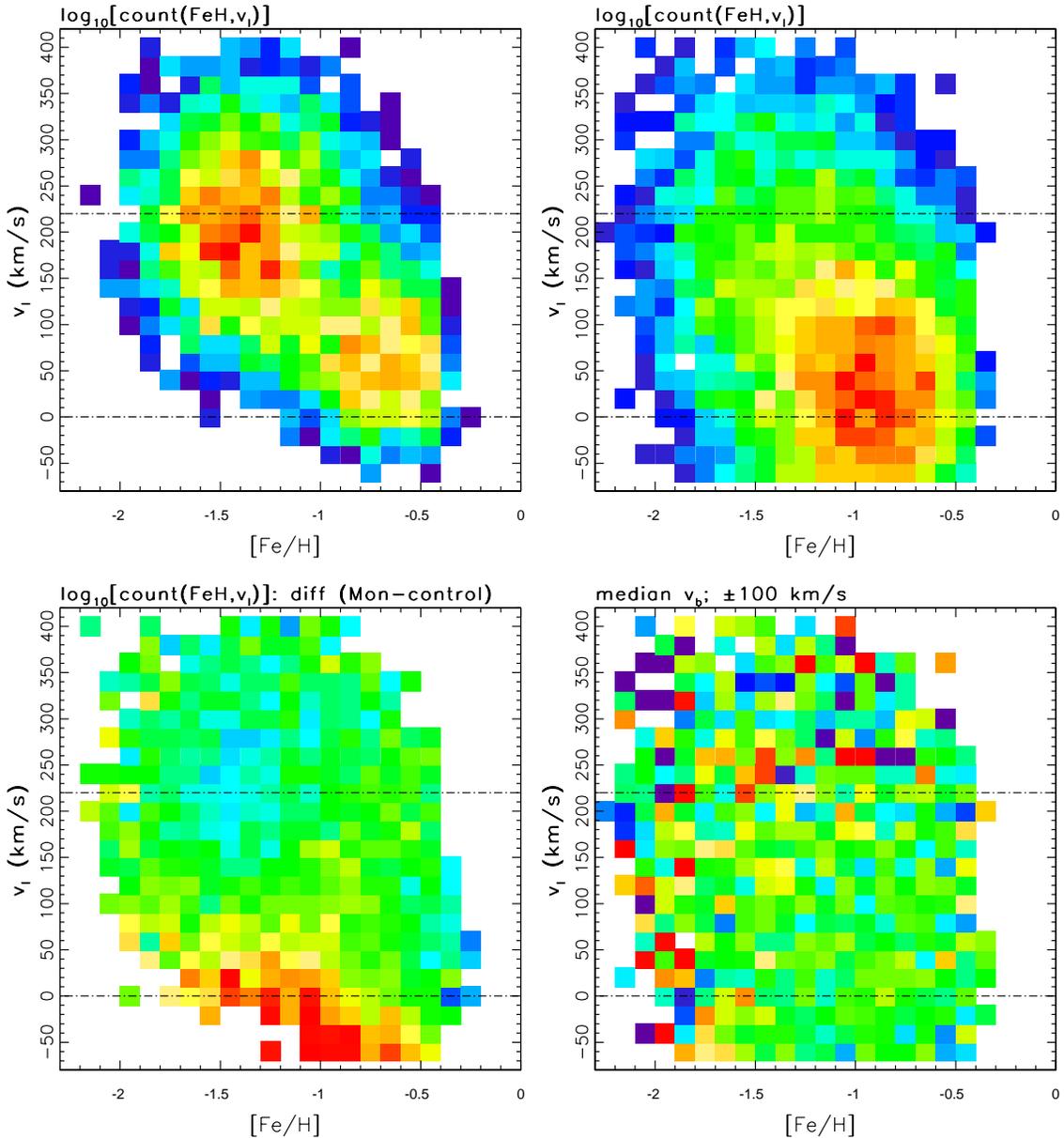}
\caption{
The top left panel shows the distribution of $\sim$7,200 stars (logarithm of
counts in each bin) with $|Y|<1$ kpc, $3 < |Z|/{\rm kpc} < 4$, $9 < R/{\rm kpc} <
12$ and $170^\circ < l < 190^\circ$, in the longitudinal velocity vs.
photometric metallicity plane (a slice through the map shown in the top right
panel of Figure~\ref{panels7}). The longitudinal velocity, $v_l$, is based on
proper motion measurements, and at selected $l\sim180^\circ$ corresponds to
the rotational component (heliocentric). The top right panel is analogous, except
that $\sim$12,000 stars with $170^\circ < l < 190^\circ$ are selected from the
$13 < R/{\rm kpc} < 16$ radial range, which maximizes the fraction of stars from
the Monoceros stream (clustered around $[Fe/H]\sim -1$ and $v_l \sim 25$ km/s.
The difference of these two maps is shown in the bottom left panel. The bottom
right panel shows the median latitudinal velocity ($v_b$) for stars in the
latter (Monoceros) subsample ($\pm$100 km/s stretch, green corresponds to 0
km/s). The analogous map for the $9 < R/{\rm kpc} < 12$ subsample appears
similar.}
\label{panels8}
\end{figure*}

We select a subsample of $\sim$11,000 stars that maximizes the fraction of stars
from the Monoceros stream, and allows an estimate of rotational velocity using
only longitudinal proper motion, by requiring $|Y|<1$ kpc, $3 < Z/{\rm kpc} <
4$, $13 < R/{\rm kpc} < 16$ and $170^\circ < l < 190^\circ$. The distribution of
these stars in the rotational velocity vs. metallicity diagram is shown in the
top right panel of Figure~\ref{panels8}. It is discernible that the Monoceros
stream has kinematics more similar to disk stars than to halo stars. We obtain a
more accurate assessment with the aid of an analogous map for a control sample
selected using similar criteria, except that $9 < R/{\rm kpc} < 12$ (top left
panel). The difference of these two maps is shown in the bottom left panel. 

The excess stars are centered on $[Fe/H]\sim -1$, as expected from the best-fit
described above. Their velocity distribution shows a strong peak at $v_l \sim
v_Y \sim -50$ km/s, with a long tail towards more positive velocities. This
residual map indicates that most of the Monoceros stream stars move in the
direction of LSR rotation with velocities of up to $\sim$270 km/s. This result
is qualitatively in agreement with Penarrubia et al. (2005), who ruled out
retrograde motions using models and proper motion measurements.

The latitudinal velocity, $v_b$, based on the latitudinal proper motion component,
is a linear combination of the radial and vertical velocity components. The median
latitudinal velocity of stars from the Monoceros stream region in the rotational
velocity vs. metallicity plane is shown in the bottom right panel of
Figure~\ref{panels8}. There is no significant offset from 0 ($<20$ km/s) in
parts of the diagram where the excess of the Monoceros stream stars is the
highest. This presumably indicates that contributions from the radial and
vertical motion for the Monoceros stream stars cancel (for disk stars, the
medians for both components should be 0). Together with the radial velocity
measurements obtained by Conn et al. (2005) and Martin et al. (2006), this
information can be used to further refine models, such as those described by
Penarrubia et al. (2005).

\section{           Discussion and Conclusions                     }

The spectroscopic stellar parameters for over 60,000 F and G dwarfs, 
computed by the SEGUE Stellar Parameters Pipeline (Beers et al. 2006;
Allende Prieto et al. 2006, 2007; Lee et al. 2007ab) using the SDSS spectroscopic 
database, allowed us to derive photometric estimators for effective temperature 
and metallicity in the SDSS photometric system. The availability of the SDSS 
imaging survey, with its accurate $ugr$ photometry, then enabled an unbiased 
volume-limited study of the stellar metallicity distribution to a distance 
limit of $\sim$10 kpc.

\subsection{Photometric Estimates for Effective Temperature and Metallicity}

The photometric metallicity estimator based on the SDSS $u-g$ and $g-r$ 
colors is reminiscent of the traditional $\delta(U-B)_{0.6}$ method based on the 
$UBV$ photometry. It reflects the same physics, and can be applied to F 
and G type main sequence stars ($0.2<g-r<0.6$). For the SDSS single-epoch 
main survey data, it provides metallicity accurate to 0.2 dex or better
for stars brighter than about $g=19.5$.  In this magnitude and color 
range, the photometric effective temperature estimator reproduces 
spectroscopic temperature with a root-mean-square scatter of only $\sim$100 K.
The accuracy\footnote{Here, precision may be a more appropriate terminology
than accuracy because the photometric estimates are tied to SDSS spectroscopic
values rather than to an absolute metallicity scale. That is, all systematic
errors in SDSS spectroscopic parameters are inherited by photometric estimators.} 
of 100 K for effective temperature, and 0.2 dex or better for 
metallicity, are comparable to parameter accuracy achieved using artificial neural 
networks with spectroscopic observations (Snider et al. 2001), and the 
estimated accuracy of parameters determined from SDSS spectra (Beers 
et al. 2007). It is plausible that increased photometric accuracy 
would further improve these values. For example, photometry accurate to
1\% may enable MK spectral type determination with errors smaller than
one subtype. Derived mapping from color space to metallicity implies that,
at least formally, an error in the $u-g$ color of 0.01 mag induces a
metallicity error that varies from 0.01 dex at $[Fe/H]=-0.5$ to 0.06 dex 
at $[Fe/H]=-1.5$. This is as good a performance as obtained by the
Str\"{o}mgren $uvby\beta$ narrow-band photometric system which was 
optimized for this purpose (e.g., Str\"{o}mgren 1966; Nordstr\"{o}m et
al. 2004). In other words, the increase of metallicity errors due to 
increased bandpass width can be compensated for by an improved photometric 
accuracy.

We apply these methods to a photometric catalog of coadded SDSS observations 
from the so-called Stripe 82 (Ivezi\'{c} et al. 2007). These deeper and 
photometrically exceedingly precise coadded data allowed us to measure an unbiased 
metallicity distribution for a {\it volume limited sample} of $\sim$200,000 F 
and G type stars in the 0.5--8 kpc distance range. We also study the
metallicity distribution using a shallower, but much larger, sample of several 
million stars in 8500 sq. deg. of sky provided by SDSS Data Release 6.  
The large sky coverage tests the conclusions derived using the relatively 
small Stripe 82 sample and enables detection of coherent substructures in the 
kinematics--metallicity space, such as the Monoceros stream.

\subsection{The Milky Way Structure and  Multi-dimensional Stellar Counts}

From an observer's point of view, the ultimate goal of Milky Way 
studies is to measure and describe the distribution (counts) of stars in the space 
spanned by apparent brightness, colors in multiple bandpasses\footnote{We 
limit this analysis to two colors, $u-g$ and $g-r$, which provide good 
estimates of effective temperature and metallicity for F and G main-sequence
stars. However, an alternative is to consider a full spectral shape,
instead of colors, which carries all the relevant information about
the three main stellar parameters ($T_{eff}$, $log(g)$), $[FE/H]$.}, 
two proper motion components ($\mu_l$, $\mu_b$), radial velocity ($v_{rad}$), 
and position on the sky. Specialized to SDSS data, 
we seek to understand the 8-dimensional probability density function, 
$p(g,u-g,g-r, \mu_l, \mu_b, v_{rad}, l, b)$. This function could be
described empirically, without any reference to stellar and Galactic
structure, but in practice measurements are used to place constraints
on the latter. From a theorist's point of view, the problem of interpreting 
data can be rephrased as: given a small control volume centered on ($R,Z,\phi$), 
\begin{itemize}
\item What is the distribution of stars as a function of luminosity\footnote{At 
least in principle, it would be desirable to measure the stellar luminosity 
function, and other quantities, as a function of stellar age, but estimating
age for individual stars is very difficult (e.g., Nordstr\"{o}m et al. 2004;
Jorgensen \& Lindegren 2005), and beyond the scope of this paper.}, $L$,
\item 
What is the metallicity distribution for a given $L$, and
\item
What are the distributions of three velocity components, $v_\phi$, $v_R$ and $v_Z$,
for a given $L$ and metallicity?
\end{itemize}
Guided by these questions, we can write 
\begin{eqnarray}
\label{pFunc}
p(g,u-g,g-r, \mu_l, \mu_b, v_{rad}, l, b) = \\ \nonumber
   \,\, p_1(g,g-r|l,b) \\ \nonumber
   \,\, \times \, p_2(u-g|g,g-r,l,b) \\ \nonumber
   \,\, \times \, p_3(\mu_l, \mu_b, v_{rad}|u-g,g,g-r,l,b),
\end{eqnarray}
where the three functions on the right side of eq.~\ref{pFunc} are discussed 
in this and two companion papers: 
\begin{enumerate}
\item 
The function $p_1(g,g-r|l,b)$ describes the behavior of the $g$ vs. $g-r$ color-magnitude 
(Hess) diagram as a function of position on the sky. This behavior is a 
reflection of stellar luminosity function, $\Phi(L)$, and the dependence 
of stellar number density on spatial coordinates, $\rho(R,Z,\phi)$. A best-fit
model for $\rho(R,Z,\phi)$ developed by J08 is discussed in \S~\ref{J08}.
\item
The function $p_2(u-g|g,g-r,l,b)$ describes the $u-g$ color distribution for
a given bin in the $g$ vs. $g-r$ color-magnitude diagram, and as a function of 
position on the sky. The $u-g$ color distribution reflects the metallicity 
distribution, $p([Fe/H]|R,Z,\phi)$, and $\rho(R,Z,\phi)$. In this contribution, 
we show that, similarly to $\rho(R,Z,\phi)$, $p([Fe/H]|R,Z,\phi)$ can be 
well described (apart from local overdensities) as a sum of two components
\begin{eqnarray} 
  p(x&=&[Fe/H]|R,Z,\phi) =  \\ \nonumber
 & & [1-f_H(R,Z)]\,p_D(x|Z) + f_H(R,Z)\,p_H(x),
\end{eqnarray}
where the halo-to-disk counts ratio is simply $f_H(R,Z)=\rho_H(R,Z)/[\rho_D(R,Z)+\rho_H(R,Z)]$. 

The halo metallicity distribution, $p_H([Fe/H])$, can be modeled as a spatially
invariant Gaussian centered on 
$[Fe/H]=-1.46$ (for the Stripe 82 catalog; for the full DR6 sample, the median 
[Fe/H]=$-1.40$) and with the intrinsic (not including measurement errors) width 
$\sigma_H=0.30$ dex. For $|Z|\la10$ kpc, an upper limit on the halo radial metallicity
gradient is 0.005 dex/kpc. 

The metallicity distribution of the disk component has an rms scatter of 0.16
dex, with
the median varying as
\begin{equation} 
\label{medMetalD}
    \mu_D(Z) =  -0.78 + 0.35\,\exp(-|Z|/1.0 \, {\rm kpc}) \,\, {\rm dex},
\end{equation}
at $|Z|>0.5$ kpc. In the $|Z|=1.0-1.5$ kpc range, the median metallicity is
consistent with the measurements by Gilmore \& Wyse (1995). For $|Z|\la5$ kpc, 
an upper limit on the disk radial metallicity gradient is 0.010 dex/kpc. 
The {\it shape} of the metallicity distribution of the disk component is non-Gaussian
and can be modeled as
\begin{eqnarray} 
\label{pDmetal}
  p_D(x&=&[Fe/H]|Z)= \\ \nonumber
& & 0.37\,G[x|\mu=a(Z)+0.14, \sigma=0.11] \\ \nonumber
&+& 0.63\,G[x|\mu=a(Z), \sigma=0.21], 
\end{eqnarray} 
where the position $a(Z)$ and the median $\mu_D(Z)$ are related via $a(Z)=\mu_D(Z)-0.067$
(unless measurement errors are very large). These results represent 
powerful new constraints for the Galaxy formation and chemical evolution models 
(e.g., Tinsley 1975; Pagel \& Patchett 1975; Wyse \& Gilmore 1995; and references 
therein).
\item
The function $p_3(\mu_l, \mu_b, v_{rad}|u-g,g,g-r,l,b)$ describes proper motion
and radial velocity measurements for a given bin in the $g$ vs. $g-r$  
color-magnitude, as a function of position on the sky, and as a function
of the $u-g$ color. This function locally reflects the behavior of the velocity 
ellipsoid, but SDSS data probe sufficiently large distances to detect its spatial 
variation, as discussed in detail by B08. They find that
the detailed behavior of kinematics can also be well described (apart from 
local overdensities) as a sum of two components, disk and halo, that map well 
to components detected in spatial profiles and metallicity distribution. 
The non-rotating halo component has by and large spatially uniform 
kinematics (in an overall sense, e.g., B08 discuss several 
kinematically coherent structures), while the disk kinematics are dominated 
by a vertical ($Z$) gradient. The mean rotational velocity and the three 
velocity dispersions for disk stars can be modeled as relatively simple 
functions of the form $a+b|Z|^c$, as discussed in detail by B08 (see also
Girard et al. 2006). The shape of the rotational velocity distribution for 
disk component is non-Gaussian and can be modeled, in the $|Z|=0.8-5.0$ kpc 
range (and $R\sim8$ kpc), as
\begin{eqnarray} 
\label{pDvPhi}
  p_D(x&=&v_\Phi|Z)= \\ \nonumber 
& & 0.25\,G[x|\mu=v_n(Z), \sigma=12] \\ \nonumber
&+& 0.75\,G[x|\mu=v_n(Z)+34, \sigma=34], 
\end{eqnarray} 
where 
\begin{equation} 
\label{wZ}
                 v_n(Z) = -3+19.2\,|Z/{\rm kpc}|^{1.25} \,\,\, {\rm km/s.} 
\end{equation} 
We reiterate that the widths listed for $p_D([Fe/H]|Z)$ and $p_D(v_\Phi|Z)$ are 
{\it intrinsic} widths and {\it have to be convolved with measurement errors 
when comparing to observations.} The listed widths are measured with a relative 
accuracy of $\sim$10\%. 
\end{enumerate}

These functions provide a good description of the overall features in the
distribution of stars in the spatial-kinematic-metallicity space, as 
observed by SDSS. Qualitatively, these results are in fair agreement with 
previous work (e.g., Gilmore \& Wyse 1985; Gilmore, Wyse \& Kuijken 1989; 
Majewski 1993; Nordstr\"{o}m et al. 2004; Girard et al. 2006). Quantitatively, 
the availability of SDSS data is enabling unprecedentedly powerful and robust 
studies, not only due to its large volume, but also thanks to its accurate and diverse 
measurements. In particular, with the SDSS data, the reach of massive statistical 
studies can now be extended from $<100$ pc (the HIPPARCOS distance limit,  
e.g., Dehnen \& Binney 1998; Nordstr\"{o}m et al. 2004) to $\sim$10 kpc. 

The results presented here are only a brief illustration of the great scientific 
potential of the SDSS stellar spectroscopic database. This dataset will remain 
a cutting edge resource for a long time, because other major ongoing and upcoming 
stellar spectroscopic surveys are either shallower (e.g., RAVE, $9 < I < 12$), 
or have a significantly narrower wavelength coverage and depth (Gaia, $r\la17$).

\subsubsection{  Is there a thick disk? } 

Perhaps the most significant result of our study, in addition to detection of 
the abundant substructure in metallicity space, is that transition from the thin to 
the thick disk, seen (and originally defined by Gilmore and Reid 1983) as an 
abrupt change of slope in the log(counts) vs. $Z$ plot around $|Z|\sim1$ kpc, 
can be modeled as smooth shifts of metallicity and velocity distributions that 
do not change their shape. More
quantitatively, using the above notation, the disk metallicity and
velocity distribution can be described as 
\begin{eqnarray} 
p_D(x&=&v_\Phi \, {\rm or} \, [Fe/H] |Z)= \\ \nonumber
& & n_1(Z)\,G[x|\mu_1(Z), \sigma_1] + n_2(Z)\,G[x|\mu_2(Z), \sigma_2].
\end{eqnarray} 
Traditionally, the two components are interpreted as thin and thick disks, 
and $n_1(Z)$ and $n_2(Z)$ are constrained by stellar number counts. They are
modeled as exponential functions with scale heights of $\sim$300 pc and
$\sim$1000 pc, with $\mu_1(Z)$ and $\mu_2(Z)$, typically assumed independent 
of $Z$. This description is only mildly inconsistent with the observed
{\it marginal} metallicity and velocity distributions. However, when the 
two distributions are analyzed {\it simultaneously}, this decomposition
faces a serious difficulty. Because it combines {\it different} metallicity 
and velocity distributions for thin- and thick-disk components (the data require
offsets of 0.2 dex and 48 km/s), it predicts a strong and detectable
correlation between them. The data presented here do not display any significant
correlation, and rule out this prediction at a highly significant level 
($\sim8\sigma$).  

We find an alternative interpretation that does not imply a strong
correlation between metallicity and velocity distributions. Formally,
we find that the data can be fit with $n_1$ and $n_2$ that do not 
vary with $Z$ (eqs.~\ref{pDmetal} and \ref{pDvPhi}), while $\mu_1$ and $\mu_2$ 
are {\it coupled} and vary with $Z$ according to eqs.~\ref{medMetalD} and \ref{wZ}. 
This ability to describe the disk metallicity and velocity distributions 
using functions with universal $Z$-independent shapes has fundamental 
implications for its origin: instead of two distinct components, our data 
can be interpreted with a single disk, albeit with metallicity and velocity 
distributions more complex than a Gaussian (note that the data require 
non-Gaussian distributions even in the traditional interpretation). 
While the disk separation into thin and thick components may still be a 
useful concept to describe the fairly abrupt change of number density around 
$|Z|\sim1$ kpc (which is detected beyond doubt, see J08 for SDSS
results), the disk spatial profile may simply indicate a complex structure 
(i.e. not a single exponential function), rather than two distinct entities 
with different formation and evolution history. If this is correct,
then our results imply that {\it different processes led to the observed
metallicity and velocity distributions of disk stars,} rather than a single 
process, such as mergers or an increase of velocity dispersion due to scattering,
that simultaneously shaped both distributions.

On the other hand, it appears that stars from the solar neighborhood, 
believed to be thick-disk stars because of their {\it kinematic} behavior, 
have larger $\alpha$-element abundances, at the same $[Fe/H]$, than
do  thin-disk stars (e.g., Fuhrmann 2004; Bensby et al. 2004; 
Feltzing 2006; Reddy et al. 2006; Ram\'{i}rez et al. 2007). The thick-disk 
stars, again selected kinematically, 
appear older than the thin-disk stars (e.g., Fuhrmann 2004; Bensby et
al. 2004). Thus, it is possible that the data presented here are 
insufficient to distinguish detailed elemental and age differences, 
and that high-resolution spectroscopy is required to do so. 
If such supplemental data were available, for example, for the $\sim$20,000 
stars analyzed in Figures~\ref{Zhist} and \ref{Vphihist}, one could 
determine whether the distributions of individual $\alpha$-elements
admit a universal shape, and whether they are {\it correlated with kinematics}.
These stars are confined to several hundred sq.deg. of sky, with a sky density 
of $\sim$100 deg$^{-2}$, and those at $|Z|<4$ kpc have $g\la18$. Such an 
undertaking is within the easy reach of modern spectrographs installed on 
10m-class telescopes.  High-resolution studies of slightly brighter subsets of stars 
are planned to be undertaken with the APOGEE subsurvey, part of the proposed
next extension of the SDSS, SDSS-III.

\subsubsection{ Multidimensional Substructure } 

The samples discussed here are sufficiently large to constrain the 
global behavior of metallicity distribution and to search for anomalies.
The halo metallicity distribution is remarkably uniform. The rms 
scatter in the median value for 2 kpc$^3$ large bins of only 0.03 dex 
is consistent with expected statistical noise. The median disk metallicity
in 1 kpc$^3$ bins in the $Z=1-2$ kpc range exhibits a statistically 
significant, but still fairly small rms scatter of 0.05 dex. We detect
a vertical metallicity gradient for disk stars (0.1--0.2 dex/kpc), but 
radial gradients are limited to $\la$0.01 dex/kpc for both disk and 
halo components, outside of regions with strong substructure. 

The strongest overdensity identified in the metallicity  
space is the Monoceros stream. Its metallicity distribution is
distinct from those for both halo and disk, and has a similar width 
as the metallicity distribution of disk stars ($\sim$0.15 dex). 
Hence, recent discoveries of abundant substructure in stellar spatial 
distribution and kinematics are now extended to metallicity space.
We concur with Nordstr\"{o}m et al. (2004) that ``the Galaxy is 
a far more complicated and interesting subject than ever before''.

\subsubsection{Implications for Future Imaging Surveys }

The analysis and conclusions presented here are relevant for upcoming 
large-scale deep optical surveys such as the Dark Energy Survey (Flaugher 
et al. 2007), Pan-STARRS (Kaiser 2002) and the Large Synoptic Survey 
Telescope (Tyson 2002, LSST hereafter). Of these, only LSST plans to 
obtain data in the $u$ band\footnote{The LSST science requirements document 
is available from http://www.lsst.org/Science/lsst\_baseline.shtml}.
Over its 10-year long lifetime, the LSST survey will obtain about 
60 observations in the $u$ band of a 20,000 deg$^2$ area. 
Thanks to its large aperture, the median 5$\sigma$ depth of 
$\sim24$ (for point sources) will be significantly fainter than for 
SDSS data (22.5), and the coadded data will reach a 5$\sigma$ 
depth of $u=26$. The potential of photometric metallicity estimator 
for studying the evolution and structure of the Milky Way demonstrated 
here bodes well for LSST science mission. 

Using SDSS data, we estimate the number of stars for which LSST will provide
metallicity measurements. Based on the discussion presented in \S 2, we adopt an
error in the $u-g$ color of 0.05 mag as a practical limit for robust metallicity
studies. This color error corresponds to a metallicity error of 0.1 dex for
metal-rich stars, and 0.2 dex for metal-poor stars. The LSST data will achieve
this color accuracy for stars with $0.2 < g-r < 0.6$ if $g<23.5$. This is about
4 magnitudes deeper than the analogous limit for SDSS survey. Based on the
counts of SDSS stars, we estimate that LSST will measure metallicity accurate to
0.2 dex or better\footnote{At the bright end, LSST color errors will be $<$0.01
mag. An error of 0.01 mag in the $g-r$ color corresponds to a 50 K random error
in effective temperature, and an error of 0.01 mag in the $u-g$ color
corresponds to a random metallicity error of 0.01 dex at $[Fe/H]=-0.5$ and 0.05
dex at $[Fe/H]=-1.5$.} for at least 200 million F/G main sequence stars brighter
than $g=23.5$ (without accounting for the fact that stellar counts greatly
increase close to the Galactic plane). For these stars\footnote{The 200 million
stars from the ``metallicity'' sample will be observed over 250 times in the $g$
and $r$ bands with signal-to-noise ratios of about 20 or larger per observation
even at the faint end (and the final error in the $g-r$ color below 1\%). The
total number of stars that will be detected by LSST is of the order 10
billion.}, LSST will also provide proper motion measurements accurate to about
0.2 mas/yr at $g=21$ and 0.5 mas/yr at $g=23$ (about 10 times more accurate and
$\sim$3 magnitudes deeper than the SDSS-POSS catalog by Munn et al. used in this
work). This data set will represent a deep complement to the Gaia mission
($g\la20$; Perryman et al. 2001; Wilkinson et al. 2005), and will enable
detailed exploration of the Milky Way halo in a six-dimensional space spanned by
three spatial coordinates, two velocity components and metallicity, within a
distance limit of $\sim$100 kpc. This study can be regarded as one of first
steps in this mapping endeavor, which is bound to provide unprecedented clues
about the formation and evolution of our Galaxy. Indeed, ``these are exciting
times to study local galaxies'' (Wyse 2006).

\acknowledgements

\v{Z}. Ivezi\'{c} and B. Sesar acknowledge support by NSF grants AST-615991 
and AST-0707901, and by NSF grant AST-0551161 to LSST for design and development
activity. M. Juri\'{c} gratefully acknowledges support from the Taplin Fellowship 
and NSF grant PHY-0503584.
J. Dalcanton acknowledges NSF CAREER grant AST-02-38683. C. Allende
Prieto acknowledges support by NASA grants NAG5-13057 and NAG5-13147. T.C.
Beers, Y.S. Lee, and T. Sivarani acknowledge support from the US National
Science Foundation under grants AST 04-06784 and AST 07-07776, as well as from
grant PHY 02-16783; Physics Frontier Center/Joint Institute for Nuclear
Astrophysics (JINA). P. Re Fiorentin acknowledges partial support through the
Marie Curie Research Training Network ELSA (European Leadership in Space
Astrometry) under contract MRTN-CT-2006-033481.

Funding for the SDSS and SDSS-II has been provided by the Alfred P. Sloan
Foundation, the Participating Institutions, the National Science Foundation, the
U.S. Department of Energy, the National Aeronautics and Space Administration,
the Japanese Monbukagakusho, the Max Planck Society, and the Higher Education
Funding Council for England. The SDSS Web Site is http://www.sdss.org/.

The SDSS is managed by the Astrophysical Research Consortium for the
Participating Institutions. The Participating Institutions are the American
Museum of Natural History, Astrophysical Institute Potsdam, University of Basel,
University of Cambridge, Case Western Reserve University, University of Chicago,
Drexel University, Fermilab, the Institute for Advanced Study, the Japan
Participation Group, Johns Hopkins University, the Joint Institute for Nuclear
Astrophysics, the Kavli Institute for Particle Astrophysics and Cosmology, the
Korean Scientist Group, the Chinese Academy of Sciences (LAMOST), Los Alamos
National Laboratory, the Max-Planck-Institute for Astronomy (MPIA), the
Max-Planck-Institute for Astrophysics (MPA), New Mexico State University, Ohio
State University, University of Pittsburgh, University of Portsmouth, Princeton
University, the United States Naval Observatory, and the University of
Washington.

\appendix

\section{Photometric Parallax Relation Derived using Globular Clusters}

In Paper I, we proposed a photometric parallax relation that did not explicitly
use metallicity information, for two main reasons. First, the analysis
included stars close to the faint limit of SDSS imaging for which the accuracy
of photometric metallicity is significantly deteriorated due to increased
$u$-band noise, and, secondly, the sample also included red stars for which
metallicity is difficult to estimate. The photometric parallax relation adopted in
Paper I implicitly takes metallicity effects into account by being somewhat
shallower than a photometric parallax relation appropriate for a
single-metallicity population: nearby stars ($\la$1 kpc, or so), which are
predominantly red (due to the use of a flux-limited sample), have on average high disk-like
metallicites, while distant stars ($\sim$1-10 kpc) are predominantly blue stars
with low metallicities (at a given $g-r$ or $g-i$ color, luminosity increases
with metallicity for main-sequence stars). However, here we discuss only stars
for which photometric metallicity estimates are available and, furthermore, they
do not include very faint stars due to the flux limit ($u\la21$) imposed by
requiring proper motion information. Hence, we can explicitly account for shifts
of photometric parallax relation as a function of metallicity.  

The color-magnitude diagrams for globular clusters can be used to constrain the
photometric parallax relation and its dependence on metallicity, and to estimate
systematic errors using the residuals between the adopted relation and
individual clusters. For example, using three fiducial cluster sequences,
$M_V(B-V)$, corresponding to metallicities, $[Fe/H]$, of $-2.20$, $-0.71$ and
$+0.12$, Beers et al. (2000) spline interpolate between them to obtain $M_V$ for an
arbitrary combination of $B-V$ and $[Fe/H]$. This is the method used to compute
main-sequence distance estimates available from SDSS Data Release catalogs. 

There are several reasons to revisit the method developed by Beers et al. First,
a transformation from Johnson system to SDSS system is required to apply their
method to SDSS data. While this transformation is known to about 0.01 mag
(Ivezi\'{c} et al. 2006a), even such a small systematic error results in an
uncertainty of absolute magnitude of $\sim$0.12 mag for blue stars. Secondly, only
three fiducial color-magnitude sequences are used, and it is not clear
whether spline interpolation captures in detail the shift of the main sequence as a
function of metallicity. Thirdly, the impact of age variations on the assumed
absolute magnitudes is not quantitatively known. Furthermore, it is not known
how similar color-magnitude sequences are for different clusters with similar
metallicity. It is, therefore, desirable to determine photometric parallax
relation using a larger number of clusters, with at least some of them observed
by SDSS.

We use five globular clusters observed by SDSS, selected to have distance in the
range 7--12 kpc (using distances from Harris 1996), to constrain the {\it shape}
of the photometric parallax relation. This distance range ensures sufficient
photometric quality for stars in the color range $g-i<0.8$ ($g-r\la0.6$), where
photometric metallicity estimates are reliable. We augment this sample by data
for six additional clusters compiled by VandenBerg \& Clem (2003), which
significantly increase the sampled metallicity range and allow us to determine
the shift of photometric parallax relation as a function of metallicity. We use
additional clusters observed by SDSS and by Clem, VandenBerg \& Stetson (2008),
as well as constraints based on Hipparcos and ground-based trigonometric
parallax measurements, to test our adopted photometric parallax relation.

\subsection{   Methodology and Results    } 

\begin{deluxetable}{rrrccrrrr}
\tablenum{6} \tablecolumns{9} \tablewidth{400pt}
\tablecaption{The Globular Clusters Observed by SDSS and Used in the Photometric Parallax Analysis}
\tablehead{Name & D$^a$  & R$^b$ & $[Fe/H]_H^c$ & $[Fe/H]_{ph}^d$ & N$^e$ & $gi_{\rm min}^f$ & $gi_{\rm max}^g$ & $\Delta r^h$}
\startdata
 M 2 &  11.5 & 10.0 & -1.62 & -1.66 &  472 & 0.40 & 0.70 &  0.00 \\
 M 3 &  10.4 & 17.5 & -1.57 & -1.41 & 1279 & 0.35 & 0.80 &  0.03 \\
 M 5 &   7.5 & 17.5 & -1.27 & -1.27 & 1776 & 0.40 & 1.10 & -0.07 \\
M 13 &   7.7 & 15.0 & -1.54 & -1.65 &  829 & 0.40 & 1.00 &  0.06 \\
M 15 &  10.3 & 12.5 & -2.26 & -2.09 &  676 & 0.30 & 0.70 &  0.01 \\
\enddata
\tablenotetext{a}{Distance, in kpc, taken from Harris (1996).}
\tablenotetext{b}{Angular radius (arcmin) used for selecting cluster stars, taken from  
                     Simones, Newberg \& Cole (2008)}
\tablenotetext{c}{Metallicity, taken from Harris (1996)} 
\tablenotetext{d}{Median photometric metallicity for stars with $0.3<g-i<0.5$ and $u<21.5$} 
\tablenotetext{e}{The number of stars used for estimating $[Fe/H]_{ph}$
                  (errors are dominated by systematics)}
\tablenotetext{f}{The minimum $g-i$ color used in the analysis (determined by turn-off stars)}
\tablenotetext{g}{The maximum $g-i$ color used in the analysis (determined from $r<21.5$)}
\tablenotetext{h}{The median $r$-band offset (mag) for stars with $0.5<g-i<0.7$,
                   relative to a prediction based on
                   eqs.~\ref{GCppZ}--\ref{GCppZ2} (using distances listed in
                   the second column).}
\end{deluxetable}

\begin{figure}
\plotone{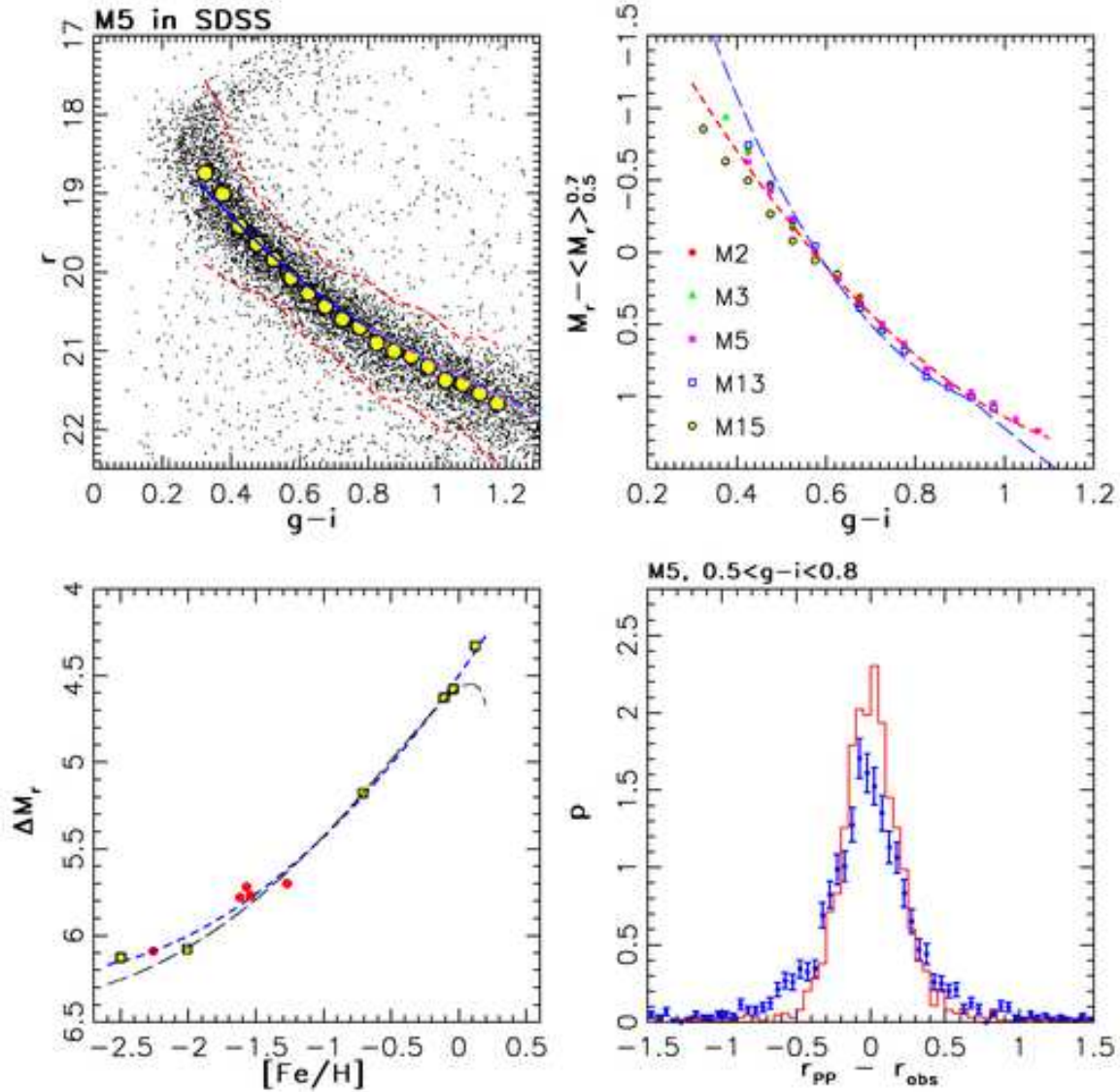}
\caption{
The top left panel shows the color-magnitude diagram for the globular cluster M5
measured by SDSS. Individual stars are displayed as small dots, while the large
dots show binned medians. The two dashed lines show the 2$\sigma$ envelope
around these medians, and the solid line is the prediction based on the adopted
photometric parallax relation (see text). The top right panel shows analogous
binned medians for five globular clusters, with each sequence rescaled by the
median magnitude for stars with $0.5 < g-i <0.7$. The short-dashed line shows a
best-fit fiducial sequence (eq.~\ref{GCppZ}). For a comparison, the long-dashed
line shows the $[Fe/H]=-2.20$ fiducial sequence from Beers et al. (2000). The
dots in the bottom left panel show the absolute magnitude offsets relative to
the fiducial relation for the five globular clusters listed in Table 6. The squares
show analogous offsets for the six globular cluster listed in Table 7. The
short-dashed line is the best unweighted linear fit to both data sets
(eq.~\ref{GCppFeH}). The thin long-dashed line is the $\Delta M_V$ vs. $[Fe/H]$
relationship from Laird, Carney \& Latham (1988), shifted to produce the same
$\Delta M_r$ at $[Fe/H]=-1.0$, as the best fit derived here. The symbols with
error bars (representing counting noise) in the bottom right panel show the
distribution of differences between $r$-band magnitudes predicted using the
adopted photometric parallax relation and the observed values. The histogram
shows the expected scatter due to photometric errors.}
\label{panels17}
\end{figure}

For clusters observed by SDSS, we select candidate cluster stars by limiting their 
angular distances from the cluster center to be less than the cluster radius 
determined by Simones, Newberg \& Cole (2008). These radii, and distance and 
metallicity data from Harris (1996), are listed in Table 6. While the faint
flux limits of SDSS imaging data limit this analysis only to relatively blue stars 
($g-i<1.0$), the color range where photometric metallicity can be determined
is fully covered.

For each cluster, we determine the median $r$-band magnitude in 0.05 mag
wide bins of the $g-i$ color. The red limit for the considered $g-i$ range is 
set by requiring $r<21.5$, and the blue end is selected to be at least 
0.05 mag redder than the vertical part of the observed sequences (turn-off
stars). The red limit ensures sufficient signal-to-noise ratios, and the
blue limit is designed to minimize the evolutionary (age) effects on the 
shape of adopted relation. That is, we deliberately construct a relation 
that corresponds to small ages first, and then study its variation with 
age using observed and model color-magnitude sequences. The adopted $g-i$ 
limits are listed in Table 6, and an example of this procedure (for M5) is 
shown in the top left panel of Figure~\ref{panels17}.  

We determine the {\it shape} of the photometric parallax relation by
{\it simultaneously} fitting data for all five clusters. To do so, we 
first shift their $r$ vs. $(g-i)$ sequences to a uniform (arbitrary) 
magnitude scale by requiring that the median $r$ magnitude for stars with 
$0.5<g-i<0.7$ is 0. These offsets depend on the cluster metallicity, as
discussed below. We then fit a parabola to all the data
points, as a function of the $g-i$ color, using and unweighted least squares 
method (a third-order polynomial is unnecessary to within $\sim$0.05 mag). 
We used the $g-i$ color because it has better signal-to-noise properties
than $g-r$ and $r-i$ colors. We did not use the so-called ``projection
on stellar locus'' technique developed in Paper I because it produces
essentially identical results for relatively bright stars considered here.
The stellar locus parametrization from Paper I can be used to express
the fiducial sequence in terms of the $g-r$ and $r-i$ colors, if needed. 

The best-fit fiducial sequence is 
\begin{equation}
\label{GCppZ}
       M_r^0(g-i) = -2.85 + 6.29 \, (g-i) -2.30 \, (g-i)^2,
\end{equation}
with $M_r^0=r-<r>=M_r-<M_r>$, valid for $0.3 < (g-i) \la 1.0$, and the
medians evaluated in the $0.5<g-i<0.7$ color range. As discernible from the 
cluster data shown in the top right panel of Figure~\ref{panels17}, individual 
clusters follow the mean relation to within 0.1 mag or better (the rms scatter 
for all data points around the best-fit relation is 0.08 mag). We compare the 
slopes of the predicted and observed sequences using the difference in
absolute magnitudes at $g-i=0.4$ and at $g-i=0.7$ (the predicted value is 
1.25 mag). The largest discrepancies of $\sim$0.1 mag are observed for M13 
(the observed sequence is steeper) and M15 (the observed sequence is
shallower). These discrepancies may be caused by a combination of metallicity 
and age effects.

We proceed by {\it assuming} that the {\it shape} of color-magnitude sequence
given by eq.~\ref{GCppZ} is a universal function independent of metallicity,
and that its {\it normalization} depends only on metallicity. While this is 
not strictly true, as we discuss below, the available data are not sufficient 
to robustly constrain the shape variation as a function of metallicity (and 
possibly other parameters, e.g. helium content; see Demarque \& McClure 1980). 

\begin{deluxetable}{rrrr}
\tablenum{7} \tablecolumns{4} \tablewidth{5in}
\tablecaption{Additional Cluster Data from VandenBerg \& Clem (2003)}
\tablehead{Name & $[Fe/H]^a$ & $M_V^b$ & $M_V^c$}
\startdata
    M 92   &  $-$2.50 &  6.30 &  6.32 \\
    M 68   &  $-$2.01 &  6.25 &  6.18 \\
   47 Tuc  &  $-$0.71 &  5.35 &  5.37 \\
 Pleiades  &  $-$0.11 &  4.80 &  4.79 \\
    M 67   &  $-$0.04 &  4.75 &  4.72 \\
  Hyades   &  $+$0.12 &  4.50 &  4.53 \\ 
\enddata
\tablenotetext{a}{Metallicity, taken from VandenBerg \& Clem (2003), except 
                  for 47 Tuc, which is taken from Beers et al. (2000)
                  (VandenBerg \& Clem adopted $[Fe/H] = -0.83$, which produces 
                     a 0.1 mag fainter $M_V$ prediction).}
\tablenotetext{b}{The absolute $V$-band magnitude for $B-V=0.60$, determined
                  with an accuracy of 0.05-0.10 mag, from figures presented in
                  VandenBerg \& Clem (2003).}
\tablenotetext{c}{The absolute $V$-band magnitude for $B-V=0.60$, determined
                  using eqs.~\ref{GCppZ}--\ref{GCppZ2}, and the SDSS to Johnson 
                  transformations from Ivezi\'{c} et al. (2006a).}

\end{deluxetable}

We place the color-magnitude sequences for each cluster on an absolute
scale using distances from Harris (1996). The offset of the measured globular 
cluster sequences relative to the best-fit fiducial sequence is a strong 
function of metallicity. We improve observational constraints on this relation 
by considering six additional clusters discussed by VandenBerg \& Clem (2003).  
We used their figures to estimate for each cluster its $M_V$ at $B-V=0.60$ 
(corresponding to $g-i=0.57$), listed in Table 7. The corresponding $M_r$ 
(i.e. the $V-r$ color) are computed using the SDSS to Johnson system
transformations from Ivezi\'{c} et al. (2007). 

The data shown in the bottom left panel of Figure~\ref{panels17}
strongly suggest a non-linear relationship (without the extended metallicity
baseline thanks to the VandenBerg \& Clem data, the five SDSS clusters would 
imply a linear relationship). The best-fit parabola is 
\begin{equation}
\label{GCppFeH}
    \Delta M_r([Fe/H]) = 4.50 -1.11\,[Fe/H] -0.18\,[Fe/H]^2,
\end{equation}
where $\Delta M_r$ is defined by
\begin{equation}
\label{GCppZ2}
        M_r(g-i,[Fe/H]) = M_r^0(g-i) + \Delta M_r([Fe/H]). 
\end{equation}
The rms scatter around the best-fit relation is 0.05 mag
for the eleven clusters used in the fit, with the maximum deviation
of 0.08 mag. This remarkably small scatter around a smooth best-fit 
function suggests that the determination of $\Delta M_r([Fe/H])$ offsets 
for individual clusters has a similar precision. Note, however,
that the overall scale of $M_r(g-i,[Fe/H])$ includes all
systematic errors inherent in cluster distances that are adopted 
from Harris (1996) compilation (including a possible covariance
with cluster metallicity). The adopted relation produces gradients
of $dM_r/d[Fe/H]=-0.57$ mag/dex at the median halo metallicity 
($[Fe/H]=-1.50$), and $-1.0$ mag/dex  at the median thin-disk 
metallicity ($[Fe/H]=-0.2$), with an offset of 1.05 mag between
these two $[Fe/H]$ values. As illustrated in the bottom left panel 
in Figure~\ref{panels17}, the best-fit relation derived here is in 
excellent agreement at $[Fe/H]<0$ with an analogous relation proposed 
by Laird, Carney \& Latham (1988).

The distributions of differences between the $r$-band magnitudes 
predicted using the above expressions and the observed values
for individual stars are consistent with expected noise due to 
photometric errors for all five clusters (see the bottom right panel 
of Figure~\ref{panels17} for an example based on M5). At the faint 
end ($r\sim21$), the expected uncertainty in $M_r$ is about 
0.3 mag (random error per star), and is dominated by random photometric 
errors in the $g-i$ color. At the bright end, the $g-i$ errors ($\sim$0.03 mag) 
contribute an $M_r$ uncertainty of $\sim$0.15 mag, and an error in 
$[Fe/H]$ of 0.1 dex results in $M_r$ error of $\la$0.1 mag. The random
errors in the $g-i$ color and photometric metallicity are by and large
uncorrelated because the $u$-band errors dominate the latter.

The SDSS cluster data discussed here are not sufficient to extend
the fiducial sequence beyond $g-i\sim1$. While not required for the 
analysis presented here, we extend for completeness the adopted relation 
using the {\it shape} of the ``bright'' relation from Paper I. Expressed 
as a function of the $g-i$ color,
\begin{equation}
\label{GCppJ08}
   M_r^0(g-i) = -1.93 + 4.39 \, (g-i) -1.73 \, (g-i)^2 + 0.452 \, (g-i)^3,
\end{equation}
valid for $(g-i) > 0.8$. We test this extension further below.

\subsection{  Testing  } 

\begin{figure}
\plotone{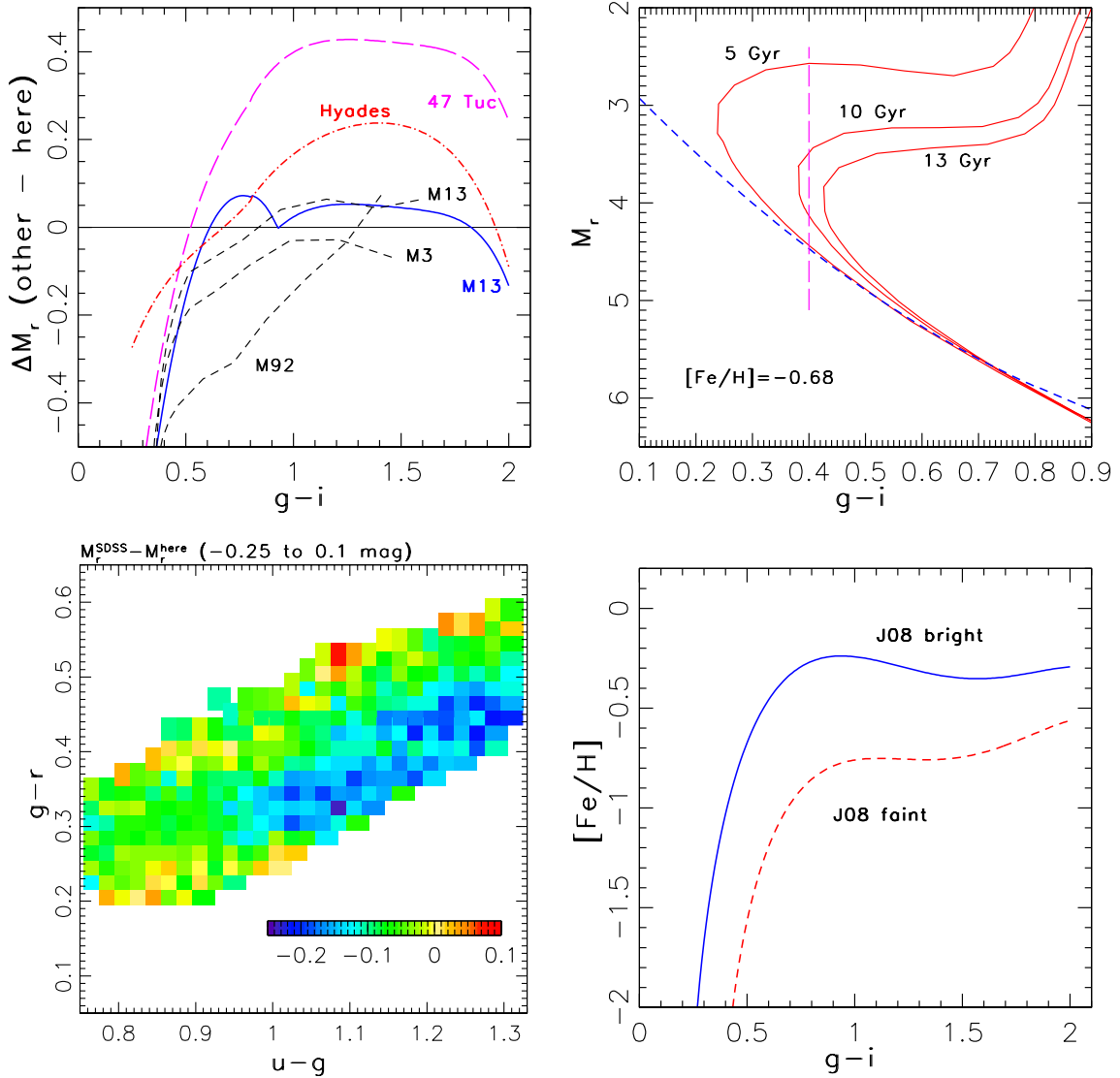}
\caption{
The top left panel shows the difference between the color-magnitude sequences
from Beers et al. (2000) for three metallicity values (solid: $[Fe/H] = -2.20$;
long-dashed: $[Fe/H] = -0.71$; dot-dashed: $[Fe/H] = +0.12$), and eqs.~\ref{GCppZ}-\ref{GCppZ2}
derived here. The three short-dashed lines shows analogous differences for the M3,
M13 and M92 sequences from Clem, VandenBerg \& Stetson (2008), as marked. The
systematic differences for blue stars are due to age effects. The solid lines in
the top right panel show $M_r$ for a Girardi et al. (2004) model with
$[Fe/H]=-0.68$, evaluated for three ages, as marked. The models are offset by
0.2 mag to brighter magnitudes, to match $M_r$ computed using eqs.~\ref{GCppZ}
and \ref{GCppFinal} (shown as the dashed line). The vertical long-dashed line
marks the turn-off color for disk stars. The dashed line is $M_r$ computed using
eqs.~\ref{GCppZ} and \ref{GCppFinal}. The bottom left panel shows the median
differences between the SDSS distance modulus for main-sequence stars
(determined using the Beers et al. sequences) and the values estimated using
eqs.~\ref{GCppZ} and \ref{GCppFinal}, color-coded as shown in the inset. The two
methods agree at the $\sim$0.1 mag level (the mean for the median difference per
pixel; the distribution rms width is $\sim$0.2 mag). The bottom right panel
shows implied metallicity, estimated using eqs.~\ref{GCppZ} and \ref{GCppFinal},
for the two photometric parallax relations proposed by Juri\'{c} et al. (2008;
solid line: ``bright'' relation; dashed line: ``faint'' relation). At the blue
end, they bracket the median halo metallicity ($[Fe/H]=-1.50$); at the red
end they sample the thin/thick disk metallicity range. }
\label{panels18}
\end{figure}

Using SDSS observations for five clusters listed in Table 6, we first
determined the median photometric metallicity for each cluster, using 
the best-fit expressions derived in this work. To avoid contamination by disk 
stars and noisy metallicity estimates, we only use stars with $0.3<g-i<0.5$ 
and $u<21.5$. Remarkably, the photometric metallicity estimates are 
consistent with the values taken from Harris (1996) to within $\sim$0.1 dex. 
This test ensures that eq.~\ref{GCppFeH} can also be used with photometric 
metallicity estimates.

We have tested eqs.~\ref{GCppZ}-\ref{GCppZ2} using an independent sample 
of clusters observed by SDSS at distances beyond our cutoff of 12 kpc 
(NGC 4147, NGC 5053, NGC 5466, NGC 5024 and Pal 5). The first four 
clusters have low metallicities ($[Fe/H] \sim -2.0$), and for Pal 5 
$[Fe/H] = -1.41$. The $r$ vs $g-i$ ridge lines predicted by eq.~\ref{GCppZ} 
agree well with the observed sequences (the data are much noisier
than for the first five nearer clusters due to their fainter 
apparent magnitudes). The only signficant discrepancy is observed 
for Pal 5, for which the predicted magnitudes are too faint by 
$\sim$0.5 mag (using a distance of 23.2 kpc).

To test the extension of photometric parallax relation to red colors, 
we use the $M_V(B-V)$ sequence for M dwarfs with the Hipparcos data, as 
compiled in Fig. 17 from VandenBerg \& Clem (2003): for $B-V$=(1.2, 1.3, 1.4), 
corresponding to $g-i$=(1.51, 1.70, 1.93), we adopt $M_V$=(7.5, 8.0, 8.5). 
Assuming that metallicity of those stars is equal to the median thin disk 
metallicity, $[Fe/H]=-0.13$ (Nordstr\"{o}m et al. 2004; Allende Prieto 
et al. 2004), we obtain $M_V$=(7.42, 7.91, 8.54). For the reddest data 
point with $V-I$=2.0, $M_V$=9.5, and we obtain $M_V$=9.47. This excellent
agreement suggests that the extension given by eq.~\ref{GCppJ08} is good to
within $\sim$0.1 mag for $g-i<2.2$. 

For redder colors ($g-i>2.0$), we compared our results with the relation
derived by Bochanski et al. (2008, in prep.), which is based on ground-based
trigonometric parallaxes for nearby stars (Golimowski et al. 2008, in prep.). 
Assuming a median metallicity of $[Fe/H]=-0.13$ for these stars, we found that 
the performance of eq.~\ref{GCppJ08} starts deteriorating around $g-i=3.0$. 
In the range ($2.0<g-i<2.8$), our relation agrees with the Bochanski et al. 
relation within 0.07 mag (rms) and $\sim$0.03 mag (median), and maximum
deviation $<$0.1 mag, evaluated on a grid with 0.01 mag steps. A linear
relation in the range $2.8<g-i<4.0$ 
\begin{equation}
\label{GCred}
                  M_r^0(g-i) = -4.40 + 3.97 \, (g-i)
\end{equation}
is a much better approximation to the observed sequence than eq.~\ref{GCppJ08} 
(but for a detailed fit please consult Bochanski et al.). Note that
for $[Fe/H]=-0.13$, this relation must be shifted by 4.64 mag to obtain 
$M_r$ (see eq.~\ref{GCppFeH}). 

As an additional test of the relation derived here, we compare it to 
color-magnitude sequences measured by Clem, VandenBerg \& Stetson (2008)
for three clusters that have turn-off colors bluer than $g-i=0.6$ (M3, M13 
and M92). Their data were obtained in the SDSS ``prime'' system, and we used 
expressions from Tucker et al. (2006) to transform those sequences onto 
the SDSS native system. For $g-i>0.5$, their sequences for M3 and M13
are in good agreement ($<0.2$ mag) with our predictions, while for
blue colors close to the turn-off color, they become progressively
brighter, as expected (see the top left panel of Figure~\ref{panels18}). 
For M92, discrepancies are larger than $\sim$0.2 mag even for red colors 
($g-i\sim1$). However, based on photometric transformations from Tucker et 
al. (2006) and Ivezi\'{c} et al. (2007), we find that the M92 sequence in 
the SDSS ``prime'' system from Clem, VandenBerg \& Stetson (2008) and the 
M92 sequence in rhw Johnson system from VandenBerg \& Clem (2003) are not 
consistent. For example, $V=20.9$ at $B-V=0.6$ taken from VandenBerg \& Clem 
implies $r=20.7$, while data listed in Table 3 from Clem, VandenBerg \&
Stetson imply $r=20.45$ at the corresponding color. We emphasize that 
the same photometric transformations result in good agreement for the 
other two clusters, and that the color-magnitude sequence for M92 from 
VandenBerg \& Clem agrees with our relation to within 0.1 mag. 

The top left panel of Figure~\ref{panels18} shows a comparison of the 
relation derived here with the three sequences from Beers et al. (2000). 
Similarly to the comparison with the Clem, VandenBerg \& Stetson sequences, 
our relation predicts fainter magnitudes for blue turn-off stars, as expected.
We emphasize that these differences are not due to errors in the color-magnitude 
sequences adopted by Beers et al., because they agree with other sources, 
e.g., with the VandenBerg \& Clem (2003) data. Rather, the differences are due 
to our design choice to exclude from fitting the parts of the clusters' 
color-magnitude sequences that are too close to their turn-off color.

Our results show that the Beers et al. spline interpolation of metallicity 
effects based on only three clusters performs remarkably well. The largest overall 
discrepancy between our photometric parallax relation and the three Beers 
et al. sequences for red colors ($g-i>0.6$) is observed for 47 Tuc: for 
$1.0 < g-i < 1.8$, the predicted $M_r$ are too bright by 0.4 mag.
Since agreement at our fiducial $g-i\sim0.6$ is satisfactory, this 
difference implies that the color-magnitude sequence is for 47 Tuc is
steeper than for other clusters discussed here. This peculiarity of 
47 Tuc has been known for some time, and may be related to its anomalous
helium content (Demarque \& McClure 1980; Hesser, Harris \& Vandenberg 
1987). We note that our relation predicts absolute magnitudes for red
stars ($B-V>1$) that are brighter by $\sim$0.3 mag than the data 
for extremely metal-rich ($[Fe/H]=+0.37$) open cluster NGC 6791 from 
VandenBerg \& Clem (2003).

\subsection{Age effects and Comparison with Models} 

By design, the photometric parallax relation derived here avoids the
increased curvature of the color-magnitude sequence close to the 
turn-off color. Its blue edge is constrained by the parts of the M3 and M15 
sequences that are {\it redwards} from their turn-off colors (see Table 
6 and the top right panel of Figure~\ref{panels17}). For stars with 
turn-off colors, the predicted absolute magnitudes can be up to $\sim$1 mag 
too faint. For example, for M5 turn-off stars selected by $0.25 < g-i < 0.35$ 
($\langle r \rangle=18.6$), the difference between predicted and observed 
$r$-band magnitudes is well described by a Gaussian distribution with a 
mean of 0.22 mag and $\sigma=0.49$ mag, implying underestimated distances 
by 11\%, on average. 

The effect of age on turn-off color and absolute magnitude, as a function
of metallicity,  can be gauged with the aid of model isochrones, e.g., such 
as those developed for SDSS photometric system by Girardi et al. (2004). 
While modeling difficulties prevent absolute normalization of such models 
to better than $\sim$0.1-0.2 mag even for hot stars (and much worse for
stars with $g-i>1$), their {\it relative} behavior, as a function of age, 
provides valuable guidance. The Girardi et al. models show that the turn-off 
color is bluer than $g-i=0.6$ even for 13 Gyr old populations and a 
metallicity at the upper end of the range relevant here ($[Fe/H]=-0.4$). 
Hence, the adopted relation is insensitive to age effects for $g-i>0.6$. 
For $g-i<0.6$, it needs to be corrected as a function of metallicity and age. 

The mean ages of halo and disk stars considered in this work can be
estimated from the blue edge of their color distributions. The number
of stars drops precipitously bluer than $g-i\sim0.25$ for the low-metallicity 
subsample ($Fe/H\la -1$, halo stars), and at $g-i\sim0.4$ for 
high-metallicity subsample (disk stars). Interestingly, the Girardi
et al. models suggest similar age for both subsamples: $\sim$10 Gyr, 
with an estimated uncertainty of $\sim$2 Gyr (due to metallicity and
color zeropoint uncertainties; we adopted 0.2 dex and 0.05 mag, 
respectively). Motivated by this result, we derive an age correction 
appropriate for stars with median halo metallicity and age of $\sim$10 Gyr
using the color-magnitude sequence for cluster M13 ($[Fe/H]=-1.54$). 
For $0.22 < g-i<0.58$
\begin{equation}
\label{GCppM13}
      \Delta M_r^{M13}(g-i) = -2.17 + 6.64 \, (g-i) -5.00 \, (g-i)^2, 
\end{equation}
which increases from 0 at the red edge to -0.95 mag at $g-i=0.22$,
and has to be added to right-hand side of eq.~\ref{GCppZ2}.

This correction for age is not strictly applicable to stars with higher
disk-like metallicity. However, the Girardi et al. models suggest
that the error is small, $<0.2$ mag for $g-i>0.45$ (i.e. 0.05 redder
than the turn-off color for disk stars), as illustrated in the top 
right panel of Figure~\ref{panels18}. For this reason, we adopt 
eq.~\ref{GCppM13} as a universal age correction for stars bluer
than $g-i<0.58$. 

Given different expressions for three color ranges (eqs.~\ref{GCppZ}, 
\ref{GCppJ08}, and \ref{GCred}) and the above age correction, for 
convenience we fit a fifth-order polynomial to a vector of $M_r$
values generated using the appropriate expressions for $0.2<g-i<4.0$, 
with a step size of 0.01 mag. Our {\it final} expression 
\begin{equation}
\label{GCppFinal}
   M_r^0(g-i) = -5.06 + 14.32 \,x -12.97 \, x^2 + 6.127 \, x^3 \\
                        -1.267 \, x^4 + 0.0967 \, x^5,
\end{equation}
where $x=(g-i)$,
reproduces individual $M_r$ values with an rms of 0.05 mag and maximum
deviation below 0.1 mag. Together with eqs.~\ref{GCppFeH} and \ref{GCppZ2},
this is the final photometric parallax relation used in this work. 

We have compared a large number of Girardi et al. models that span the 
relevant range of metallicities ($-2.3 < [Fe/H] < 0$) and ages (1--13 Gyr)
with the resulting photometric parallax relation. Model predictions are 
in good agreement (an rms of $\sim$0.1 mag) with the $M_r$ vs. $[Fe/H]$ 
dependence described by eq.~\ref{GCppZ}, but the model $M_r$ predictions are systematically 
too faint by $\sim$0.2 mag (evaluated at $g-i=0.7$). Possible explanations
for this difference are 
1) the model stars are too small by $\sim$10\%, 
2) the model $g-i$ color is too red by 0.06 mag, and
3) the model $[Fe/H]$ scale is offset relative to SDSS scale by $\sim$0.3 dex 
to larger values. A plausible combination of these effects, e.g., 
an error of 3\% in sizes, 0.02 mag in color, and 0.1 dex in metallicity,
brings data and models into agreement (the probability that all three
effects would have the same sign is 12\%).

\subsection{Comparison with SDSS Distances and J08} 

With the adopted age correction (eq.~\ref{GCppM13}), our final
expression is expected to produce very similar distances to those
published in SDSS Data Release catalogs for blue stars ($g-i<2$). 
We have confirmed that this is the case: the median offset of
implied $M_r$ evaluated in small bins of $u-g$ and $g-r$ color
(see the bottom left panel of Figure~\ref{panels18}) is -0.07 mag,
with an rms of 0.06 mag. These differences are smaller than the 
intrinsic errors of the photometric parallax method ($\sim$0.1-0.2 
mag). 

Using eqs.~\ref{GCppFeH}, \ref{GCppZ2}, and \ref{GCppFinal}, we can now 
determine ``effective'' metallicity that the two photometric parallax 
relations proposed in Paper I correspond to, as a function of the $g-i$ 
color (see the bottom right panel in Figure~\ref{panels18}). As designed, 
those two relations bracket the median halo metallicity ($[Fe/H]=-1.50$) 
at the blue end, and sample the thin/thick disk metallicity range at the 
red end.

In summary, the relations proposed here are in good agreement ($<0.1$ mag)
with the clusters M3 and M13 at the low-metallicity end for $g-i<1.5$, and 
with local stars with trigonometric parallaxes for $g-i>1.5$. At a fiducial
color $g-i=0.6$, in the middle of the color range where photometric 
metallicity can be estimated, the rms scatter around the best-fit 
$\Delta M_r$ vs. $[Fe/H]$ curve is 0.08 mag. Even in cases of known
peculiar behavior (e.g., 47 Tuc) and at the high-metallicity end 
(e.g., NGC 6791), discrepancies do not exceed 0.4 mag. Compared to the 
Beers et al. relations used by the SDSS, here we provide an estimate of 
the scatter around mean relations, a closed-form expression for the 
metallicity dependence, and extend the method's applicability farther 
into the red, to $g-i\sim4$. Given the larger 
number of globular clusters observed in SDSS system used here, as well 
as tests based on external data sets, it is likely that distance estimates 
for main-sequence stars based on the photometric parallax method (both
using relations derived here and the Beers et al. relations) do not 
suffer from systematic errors larger than $\sim$10\%. While these systematic 
distance errors are not overwhelming, they could, in principle, have an impact
on the analysis of the Milky Way kinematics. We discuss such issues further in Paper
III (B08).

\section{Additional Discussion of the Photometric Temperature Estimator}

Often, the inverse of the effective temperature is fit as a linear function 
of color (e.g. CPF). The best-fit
\begin{equation}
\label{altT}
     { 5040 \, {\rm K} \over T_{\rm eff}} = 0.532(g-r) + 0.654
\end{equation}
results in the same systematic errors and rms scatter as eq.~\ref{logT},
with the largest difference between the two relations below 50 K. 

A lower limit for the errors in estimation of the photometric effective
temperature can be readily computed using eq.~\ref{logT} and the photometric errors
in the $g-r$ color (the median value is 0.025 mag, and 0.03 mag at $g=19.5$;
these values, computed by the photometric pipeline, are reliable, as discussed in
detail by Sesar et al. 2007). This is a lower limit, because the contribution of
errors in the spectroscopic effective temperature is not included. The standard
deviation for the distribution of metallicity residuals normalized by these
errors is 1.2. Hence, one is tempted to conclude that the accuracy of the
effective temperature estimator is limited by the SDSS photometric errors. However,
this conclusion is not consistent with the behavior of the $\log(T_{\rm eff})$
vs. $g-r$ relation for a subset of 13,719 stars for which more accurate
photometry, based on $\sim$10 repeated SDSS observations, is available
(Ivezi\'{c} et al. 2007). Although for these stars the median error in the $g-r$
color is only 0.008 mag, the standard deviation for $\log(T_{\rm eff})$
residuals is not appreciably smaller (the expectation is a decrease by a factor
of three). Therefore, it is quite likely that the contribution of errors in
the spectroscopic effective temperature to the scatter of $\log(T_{\rm eff})$
residuals is not negligible. Indeed, the implied value of $\sim$100 K agrees
well with an independent estimate based on a comparison to high-resolution
spectral data, as discussed by Beers at al. 2006. The analyzed color range spans
about 15 MK spectral subtypes (from $\sim$F5 to $\sim$G9/K0, Bailer-Jones et al.
1997, 1998). Hence, the uncertainty in the photometric effective temperature
estimate of 100 K corresponds to about one spectral subtype, or equivalently,
{\it an error of one spectral subtype corresponds to a $g-r$ error of 0.02 mag}. 

A good correlation between the spectroscopic effective temperature 
and $g-r$ color extends beyond the restricted color range where
the photometric metallicity method is applicable ($0.2 < g-r < 0.4$). 
We find that everywhere in the $-0.3 < g-r < 1.3$ color range 
(roughly $-0.1 < B-V < 1.3$), the relation
\begin{equation}
\log(T_{\rm eff} / {\rm K}) = 
          3.882 - 0.316(g-r) + 0.0488(g-r)^2 + 0.0283(g-r)^3
\end{equation}
achieves systematic errors below 0.004 dex and overall rms of 0.008 dex.
The corresponding temperature range is 
4,000 -- 10,000 K. When the residuals are binned in 0.1 dex wide bins of 
metallicity and log(g), the largest median residual is 0.006 dex. 
Eq.~\ref{altT} remains valid in the $-0.3 < g-r < 0.8$ range, but
also requires non-linear terms if extended to redder colors
(or a different linear fit for the $0.8 < g-r < 1.3$ range). 

Due to the expanded $g-r$ range, the impact of metallicity and log(g) 
on $\log(T_{\rm eff})$ residuals is expected to be larger for this relation
than for eq.~\ref{logT}. Using Kurucz (1979) models, we find that 
the strongest dependence on metallicity is expected in the $0.4 < g-r < 1.2$
color range, with a gradient of $\sim$0.015 dex/dex (see also Lenz et al.
1998 for a related discussion). The measured value for SDSS sample is 
0.012 dex/dex, and implies up to $\sim$200 K offsets as metallicity varies 
from $-2.0$ to $-0.5$. The strongest dependence on log(g) is expected in the 
$-0.2 < g-r < 0.1$ color range, with a gradient of 0.02 dex/dex. The measured 
value for the SDSS sample is $\sim$0.004 dex/dex, or about five times smaller 
(150 K vs. 720 K variation, as log(g) varies by 2 dex at $g-r=0$). We do not 
understand the cause of this discrepancy.

\section{Additional Discussion of the Photometric Metallicity Estimator }

In two {\it restricted} color regions, simpler expressions than eq.~\ref{Zphotom}
can suffice. In a low-metallicity region defined by $0.8 < u-g < 1.0$ 
(and $0.2 < g-r < 0.6$, of course), the relation
\begin{equation}
\label{ZphotomLowZ}
        [Fe/H]_{ph}= 5.14(u-g) - 6.10
\end{equation}
reproduces the spectroscopic metallicity of about 27,000 stars with an rms scatter 
of 0.29 dex. We note that this is essentially the same expression as obtained 
by Ivezi\'{c} et al. (2006b), using a preliminary version of spectroscopic
parameter pipeline, {\it except for an overall shift in metallicity by 0.2 dex.}
This shift is due to the fact that SDSS stellar parameters pipeline was still
under development when the analysis of Ivezi\'{c} et al. (2006b) was performed. 

In the high effective temperature region (5800 K $< T_{\rm eff} <$ 6600 K) defined 
by $0.2 < g-r < 0.4$, the relation
\begin{equation}
\label{ZphotomHighT}
     [Fe/H]_{ph}= -21.88 + 47.39(u-g) -35.50(u-g)^2 + 9.018(u-g)^3
\end{equation}
reproduces spectroscopic metallicity of about 34,000 stars with an rms scatter
of 0.30 dex. In the range $0.8 < u-g < 1.4$ (corresponding to $-2.0 < [Fe/H] <
-0.4$) systematic errors do not exceed 0.1 dex. The systematic errors are larger
than for eq.~\ref{Zphotom} because the lines of constant metallicity in the
$g-r$ vs. $u-g$ diagram are not exactly vertical. Despite having somewhat poorer
performance, eqs.~\ref{ZphotomLowZ} and ~\ref{ZphotomHighT} are convenient when
estimating the impact of $u-g$ color error on photometric metallicity error. An
error in the $u-g$ color of 0.02 mag (typical of both systematic calibration
errors and random errors at the bright end for SDSS data) induces an error in
$[Fe/H]$ that varies from 0.02 dex at $[Fe/H]=-0.5$ ($u-g$=1.28) to 0.11 dex at
$[Fe/H]=-1.5$ ($u-g$=0.89). At $g=19.5$, the median $u-g$ error for single-epoch
SDSS data is 0.06 mag for point sources with $0.2 < g-r < 0.4$, corresponding to
median random metallicity errors of 0.10 dex for disk stars and 0.30 dex for
halo stars (for a detailed dependence of SDSS random photometric errors on
magnitude, see Sesar et al. 2007).
 
The metallicity vs. $u-g$ relation has a smaller slope at the red end (both
eq.~\ref{Zphotom} and eq.~\ref{ZphotomHighT}), and effectively introduces an
upper limit on estimated metallicity. For example, for $u-g$=1.3 and $g-r=0.4$,
$[Fe/H]_{ph}= -0.44$ (from eq.~\ref{Zphotom}, and $-0.46$ using
eq.~\ref{ZphotomHighT}). Such an upper limit is in agreement with the data
analyzed here, but we emphasize that the data set under consideration does {\it
not} include significant numbers of stars with higher metallicity. Such stars
are presumably nearby thin-disk stars, which in the $0.2 < g-r < 0.6$ range are
typically saturated in SDSS data (most SDSS data to date are obtained at high
Galactic latitudes). It is thus possible that metallicity estimates given by
both eq.~\ref{Zphotom} and eq.~\ref{ZphotomHighT} would be biased towards lower
values for stars with $[Fe/H] > -0.5$, resulting in a ``metallicity
compression''. Some evidence that this is a detectable, but not a major effect
is discussed in \S~\ref{diskEdge}, and in more detail by Lee et al. (2007b). It
will be possible to quantify this effect in detail using the data for metal-rich
stars from the ongoing SDSS spectroscopic survey of low Galactic latitudes
(SEGUE). 

Given that the $u$-band photometric errors limit the precision of photometric
metallicity estimates at the faint end, it is prudent to test whether the 
position of a star in the $r-i$ vs. $g-r$ color-color diagram could be used 
as an alternative method. We selected subsamples of stars in 0.02 mag wide
$g-r$ bins, and inspected the dependence of spectroscopic metallicity on the 
$r-i$ color in the range $-2.5 < [Fe/H] < -0.5$. The strongest correlation
between $[Fe/H]$ and $r-i$ color is observed around $g-r\sim0.4$, with a
gradient of $\Delta(r-i)/\Delta[Fe/H] \sim 0.017$ mag/dex. Hence, the 
effect of metallicity on the $r-i$ color is about 10 times smaller than 
for the $u-g$ color.  With the $r-i$ color kept fixed, we find 
$\Delta(g-r)/\Delta[Fe/H] \sim 0.04$ mag/dex. When using only the $gri$ 
bands, the photometric metallicity errors are about 0.3 dex at the bright 
end and 0.5 dex at $g=19.5$ ($<$0.1 dex and $<$0.3 dex for $ugr$ based
estimates). Therefore, the best approach for estimating photometric
metallicity using SDSS data is to use the $ugr$ bands.

\clearpage

\clearpage
\end{document}